\documentclass[]{aa}
\pdfoutput=1
\usepackage{graphicx}
\usepackage{txfonts}
\usepackage{natbib}
\usepackage{epsf}
\usepackage{rotating}
\usepackage{graphics}
\usepackage{verbatim}
\usepackage{longtable}

\def \sophie{SOPHIE}

\def \MJ{M$_{\mathrm{Jup}}$}

\def \msol{M$\mathrm{_\odot}$}
\def \kms{km\,s$^{-1}$}
\def \ms{m\,s$^{-1}$}
\def \1s{$1\,\sigma$}

\def \t0{T$_0$}

\def \starmass{M_\mathrm{s}}

\begin{document}

\title{The SOPHIE search for northern extrasolar planets\thanks{Based on observations collected with the {\it SOPHIE}  spectrograph on the 1.93-m telescope at Observatoire de Haute-Provence (CNRS), France, by the {\it SOPHIE}  Consortium  (program 07A.PNP.CONS).}}
            
\subtitle{IV. Massive companions in the planet-brown dwarf boundary}

\author{
R.~F.~D\'iaz\inst{1,2} \and
A.~Santerne\inst{3} \and
J.~Sahlmann\inst{4}\and 
G.~H\'ebrard\inst{1,2} \and 
A.~Eggenberger\inst{5}\and 
N.~C.~Santos\inst{6,7}\and
C.~Moutou\inst{3}\and 
L.~Arnold\inst{2}\and
I.~Boisse\inst{6}\and
X.~Bonfils\inst{5}\and 
F.~Bouchy\inst{1,2}\and
X.~Delfosse\inst{5}\and
M.~Desort\inst{5}
D.~Ehrenreich\inst{5}\and
T.~Forveille\inst{5}\and 
A.-M.~Lagrange\inst{5}\and
C.~Lovis\inst{4}\and
F.~Pepe\inst{4}\and
C.~Perrier\inst{5}\and 
D.~Queloz\inst{4}\and
D.~S\'egransan\inst{4}\and 
S.~Udry\inst{4}\and 
A.~Vidal-Madjar\inst{1}
}


\institute{
Institut d'Astrophysique de Paris, UMR7095 CNRS, Universit\'e Pierre \& Marie Curie, 98bis boulevard Arago, 75014 Paris, France \and 
Observatoire de Haute-Provence, CNRS/OAMP, 04870 Saint-Michel-l'Observatoire, France \and 
Laboratoire d'Astrophysique de Marseille, Universit\'e de Provence, CNRS (UMR 6110), BP 8, 13376 Marseille Cedex 12, France \and 
Observatoire de Gen\`eve,  Universit\'e de Gen\`eve, 51 Chemin des Maillettes, 1290 Sauverny, Switzerland \and 
UJF-Grenoble 1 / CNRS-INSU, Institut de PlanŽtologie et d'Astrophysique de Grenoble (IPAG) UMR 5274, Grenoble, F-38041, France \and 
Centro de Astrofisica, Universidade do Porto, Rua das Estrelas, 4150-762 Porto, Portugal \and
Departamento de F'sica e Astronomia, Faculdade de Ci\^encias, Universidade do Porto, Portugal
}

 \date{Received TBC; accepted TBC}
      
\abstract{The mass domain where massive extrasolar planets and brown dwarfs lay is still poorly understood. Indeed, not even a clear dividing line between massive planets and brown dwarfs has been established yet. This is partly due to the paucity of this kind of objects orbiting close to solar-type stars, the so-called brown dwarf desert, that hinders setting up a strong observational base to compare to models and theories of formation and evolution.}
{We search to increase the current sample of massive sub-stellar objects with precise orbital parameters, and to constrain the true mass of detected sub-stellar candidates.}
{The initial identification of sub-stellar candidates is done using precise radial velocity measurements obtained with the SOPHIE spectrograph at the 1.93-m telescope of the Haute-Provence Observatory. Subsequent characterisation of these candidates, with the principal aim of identifying stellar companions in low-inclination orbits, is done by means of different spectroscopic diagnostics, as the measurement of the bisector velocity span and the study of the correlation mask effect. With this objective, we also employed astrometric data from the Hipparcos mission and a novel method of simulating stellar cross-correlation functions.}
{Seven new objects with minimum masses between $\sim 10$ \MJ\ and $\sim 90$ \MJ\ are detected. Out of these, two are identified as low-mass stars in low-inclination orbits, and two others have masses below the theoretical deuterium-burning limit, and are therefore planetary candidates. The remaining three are brown dwarf candidates; the current upper limits for their the masses do not allow us to conclude on their nature. Additionally, we have improved on the parameters of an already-known brown dwarf (HD137510b), confirmed by astrometry.}
{}

\authorrunning{D\'iaz et al.}
\titlerunning{Massive companions in the planet-brown dwarf boundary.}

\maketitle

\section{Introduction \label{sect.intro}}

More than 15 years after the discovery of the first extrasolar planets and brown dwarfs \citep[see, for example,][]{mayorqueloz95,rebolo1995}, a number of issues remains open regarding the relation between these two types of objects. Principal among them is the problem of distinguishing members of one class from members of the other. As discussed by, for example, \citet{basribrown2006} the criteria used to classify these objects can concern their physical characteristics, and their origin, among other things. 

The common idea that brown dwarfs are "failed" stars that only fusion deuterium, and that are not able to sustain a constant luminosity throughout their lifetimes \citep[e.g.][]{burrows97,baraffe2003}, while planets are objects that are simply too light to fusion deuterium seems to imply that a definition based on physical characteristics would be the best adapted one. Traditionally, the deuterium-burning mass was taken to be 13 \MJ\ \citep{burrows2001}, but there exists an important uncertainty related to the equation of state used in these computations. Additionally, more recent models by \citet{spiegel2011} show that the deuterium-burning mass limit depends on the helium abundance, the initial deuterium abundance, and the metallicity of the model, among other things. Therefore, the authors find that the deuterium-burning mass can range from 11 \MJ\ up to 16 \MJ. Choosing a single mass as the limit between both categories may therefore lead to an oversimplification of the scenario. Additionally, since the minimum mass needed to start gravitational collapse, either in a disk or in a molecular cloud, is probably around a few Jupiter masses \citep{whitworthstamatellos2006}, and since the core-accretion models predict objects with masses as high as a few tens of Jupiter masses \citep{mordasini2009}, there surely exists an important overlap in mass for objects having different origins. For example, \citet{kirkpatrick2006} have reported the discovery of a L-type brown dwarf with a mass that could be as small as 6 \MJ. This might make the distinction by mass not completely satisfactory.

On the other hand, a criterion based on formation history would stand on the fact that brown dwarfs are thought to form as stars \citep[see the reviews by][]{luhman2007, whitworth2007}, i.e.\ by gravitational collapse and fragmentation of molecular clouds on dynamical timescales, while planets are widely believed to form from accretion of a solid core in a protoplanetary disk, a process believed to be much slower \citep{pollack96,alibert2005}. 
This could in principle allow us to distinguish ones from the others, but this is not without a few drawbacks. First of all, it is practically impossible to trace unambiguously the history of an object back to its origins. Additionally, there is not a complete consensus about the mechanisms ruling formation processes, neither for brown dwarfs, nor for massive planets. For example, massive planets are also proposed to be able to form by disk collapse \citep{boss1997} or by protostellar cloud fragmentation followed inward migration by disk capture \citep{font-ribera2009}. A definition based on formation history might then prove provisional.

To produce a consistent picture of brown dwarfs and extrasolar planets, a solid observational base for formation and evolution models is crucial. Constraining the multiplicity properties of these objects, such as frequency, separation, and mass ratio distributions, should permit distinguishing between different formation models. 

In this regard, most of the known extrasolar planets have been discovered by high-precision radial velocity (RV) surveys \citep[see exoplanet.eu, and][]{schneider2011}. Due to the large reflex motion they induce in the stars they monitor, these surveys should easily discover brown dwarfs companions and properly characterise their orbital elements, and frequency. However, the brown dwarf desert \citep{marcybutler2000,gretherlineweaver2006}, i.e.\ the paucity of massive sub-stellar companions to solar-type stars in close orbits, has hindered this endeavour, resulting in only few tens of such objects known to date \citep{sozzetti2010, sahlmann2011}.  Indeed, the frequency of close-in brown dwarf companions around sun-like stars has been recently determined to be below 0.6\% \citep{sahlmann2011}. Additionally, the RV method provides a lower limit for the mass only, which means that a fraction of the brown dwarf candidates detected by RV surveys are actually low-mass stars in low-inclination orbits. As a consequence, studies of massive planets and brown dwarf properties are faced with a small and uncertain sample of objects, and therefore all efforts to either increase the number of these objects or to determine their true mass accurately should prove valuable. 

Different authors have attempted to measure the true mass of RV candidates using astrometric measurements \citep{halbwachs2000,zuckermazeh2001,sahlmann2011,reffert2011}. Although most of these analyses fail to yield direct measurements of the mass, and are limited by the precision of their measurements to upper limits to the mass, they have proven efficient in discarding relatively massive stellar companions in low-inclination orbits.

In this paper, we report the detection of 7 new objects with minimum masses between 10 and 90 \MJ, orbiting main-sequence stars of spectral types F, G and K. The detections is based on precise radial velocity measurements obtained as part of the sub-progamme 2 of the \sophie\ search for northern extrasolar planets \citep{bouchy2009}. We have already announced three of these objects, albeit with preliminary parameters \citep{diaz2011}. Additionally, we improve on the orbital parameters of an already known brown dwarf, HD137510 b. Finally, our attempts to measure the true mass of these candidates are also described. 

In Sect. \ref{sect.observations} we describe the observations and the data reduction, in Sect. \ref{sect.dataanalysis} we describe the methods used to analyse the data, fit the orbits, and estimate the stellar parameters of the primaries. We also describe the techniques employed to constrain the true nature of the companions, including the analysis of astrometric data from the Hipparcos mission. In Sect.~\ref{sec.results} we report the results obtained for each object, which are roughly classified using the mass-based criterion described above into planetary candidates, brown dwarf candidates and stellar-mass objects. The results are then discussed in Sect.~\ref{sect.discussion}. Finally, in Sect.~\ref{sec.summary} we summarise the results and present our conclusions.

\section{SOPHIE Radial Velocities \label{sect.observations}}

\begin{table*}[t]
\caption{Target characteristics and summary of observation log. \label{table.obslog}}
\begin{center}
\begin{tabular}{crrcccccccc}
\hline
\hline
Target & \multicolumn{2}{c}{Coordinates} &m$_\mathrm{v}$& Sp. Type & Dist.\tablefootmark{a} &\multicolumn{4}{c}{Observations}&Sect.\\
Name&R.A.&DEC& &  &  & First Obs.\tablefootmark{b}&Last Obs.\tablefootmark{b}&Time Span&N&\\
&(hh:mm:ss)&($^\circ$:$^\prime$:$^{\prime\prime}$)&&&(pc)&&&(days)&&\\
\hline
\object{HD14651}	&02:22:00.9	&+04:44:48.3	&8.3	&G6	&39.6	&2006-11-04	&2010-09-14	&1410.1	&15	&\ref{sec.browndwarfs}\\
\object{HD16702}	&02:42:58.7	&+58:11:52.1	&8.3	&G0	&45.2  	&2007-09-22	&2010-11-02	&1137.9	&22	&\ref{sec.stars}\\
\object{HD22781}	&03:40:49.5	&+31:49:34.7	&8.8	&K1	&31.8   	&2006-11-12	&2010-09-05	&1392.1	&33	&\ref{sec.planets}\\
\object{HD30246}	&04:46:30.4	&+15:28:19.4	&8.3	&G1	&51.5    	&2006-11-12	&2010-10-18	&1436.9	&25	&\ref{sec.browndwarfs}\\
\object{HD92320}	&10:40:56.9	&+59:20:33.0	&8.4	&G3	&41.5   	&2006-12-21	&2010-11-27	&1437.1	&16	&\ref{sec.browndwarfs}\\
\object{HD137510}	&15:25:53.3	&+19:28:50.6	&6.3	&F9	&41.8  	&2008-03-24	&2010-08-27	&886.7	&13	&\ref{sec.browndwarfs}\\
\object{HD156279}	&17:12:23.2	&+63:21:07.5	&8.1	&G6	&35.4	&2010-05-14	&2011-01-23	&254.2	&15	&\ref{sec.planets}\\ 
\object{HD167215}	&18:12:59.4	&+28:15:27.4	&8.1	&F7	&78.6	&2009-07-29	&2010-10-09	&437.0	&14	&\ref{sec.stars}\\
\hline
\end{tabular}
\end{center}
\tablefoot{
\tablefoottext{a}{Distance from Hipparcos catalogue \citep{hipparcos}.}
\tablefoottext{b}{Universal Time date at beginning of observation.}
}
\end{table*}

\subsection{Observations\label{sect.observations.velocimetry}}

Radial velocity measurements were performed with \sophie, a high-resolution \emph{echelle} 
spectrograph fiber-fed from the Cassegrain focus of the 1.93-m telescope at the Haute-Provence Observatory (OHP). It is installed in a temperature-stabilised environment and the dispersive elements are kept at constant pressure in order to provide high-precision radial velocities \citep{perruchot2008}. \sophie\ spectra cover most of the visible wavelength range in 39 spectral orders.

Observations were performed in high-resolution mode which produces spectra with resolving power of $\lambda/\Delta
\lambda\sim75\,000$ at 550~nm. Two optical fibers are used in this mode: one of them records the stellar spectrum while the other can be used to simultaneously calibrate the spectrum with a Thorium Argon lamp (\emph{thosimult} mode) or to record the background sky spectrum during the observations (\emph{objAB} mode). Most of the observations reported here were obtained in the \emph{objAB} mode, since the stability of the instrument is high enough to allow for calibrations to be obtained every 2 hours without degrading the sought velocity precision\footnote{The same is not true for other \sophie\ progammes, which require measurements with the highest possible precision \citep[see][]{bouchy2009}.}. The observations performed in \emph{thosimult} mode date from the first season of observations with \sophie\ when its stability had not yet been confirmed. Moreover in the \emph{objAB} mode, the sky fiber can be used to correct for scattered light contaminating the star fiber \citep[e.g.][]{barge2008,pollacco2008,santerne2011}.

The targets reported in this paper are listed in Table~\ref{table.obslog}. We include the date of the first and last observation of each target, the time span of observations and the number of measurements obtained. In the last column of Table~\ref{table.obslog} we indicate the Section in which the results of each target are presented. The exposure time was varied in order to reach a S/N of 50 per resolving element under varying weather conditions. The exposure time and S/N of each individual observation are reported in Table~\ref{table.RV}, available on-line.

\subsection{Data reduction \label{sect.datareduc}}
The spectra were reduced and extracted using the \sophie\ pipeline \citep{bouchy2009}, which performs all the steps from the raw images to the wavelength-calibrated 2D spectra. These are then correlated using numerical binary masks corresponding to different spectral types (G2, K5 and M4). The cross-correlation function (CCF) averaged over all 39 spectral orders is then fit with a gaussian function to obtain the RV, the  full-width at half-maximum (FWHM), and contrast relative to the baseline. Additionally, the bisector velocity span of the CCF is computed in the way described by \citet{queloz2001} for further analysis, as well as the radial velocity computed considering only the blue or red halves of each spectral order, which we call $RV_\mathrm{blue}$, and $RV_\mathrm{red}$, respectively (see \S\ref{sect.seeing}). For reference, in Table~\ref{table.CCF} we report for all targets studied in this paper the numerical mask closest to the measured spectral type (\S\ref{sect.stpar}), and the median values of the CCFs parameters: RV, FWHM and Contrast. The errorbars reported correspond to the standard deviation of each magnitude.

The complete radial velocity datasets obtained with the mask closest to the spectral type of each target are reported in Table~\ref{table.RV}, where the uncertainties include the photon noise error, estimated with the method by \citet{bouchy2001}, the error in the wavelength calibration ($\sim 2$ \ms), and an additional systematic error of 3 \ms, which was added quadratically to the other sources of uncertainty. The systematic errors arise from imperfect guiding, the so-called seeing effect \citep[][see also \S\ref{sect.seeing} of the present paper]{boisse2011b}, uncorrected velocity drifts of the spectrograph, and other still unidentified sources.



In April 2007, \sophie\ was upgraded with a bandpass filter to reduce the contamination of the stellar spectra by Argon lines coming from the ThAr lamp during observations in \emph{thosimult} mode. For three stars of our sample some observations were performed in \emph{thosimult} before the upgrade: HD14651, HD22781, and HD30246. These observations are identified in Table~\ref{table.RV}, and they exhibit a clear contamination in the red part of the spectra. To try to mitigate the effects of this contamination¤, the 6 reddest spectral orders were not considered for the computation of the average CCF and the radial velocity. To maintain the uniformity of the reduction, the remaining spectra of each star were also reduced in the same way. The resulting velocities are slightly different from those obtained using all spectral orders for these same spectra, but no systematic shift exists between both observation modes. In general, the orbital fits performed using the corrected velocities exhibit a reduced scatter. The remaining observations were performed in \emph{objAB} mode.

\section{Data Analysis \label{sect.dataanalysis}}
\subsection{Stellar Parameters \label{sect.stpar}}
The CCF obtained (\S\ref{sect.observations.velocimetry}) has been used to obtain an estimate of the stellar metallicity ([Fe/H]) and the projected stellar rotational velocity ($v\,\sin I$, where $I$ is the inclination of the rotation axis of the star with respect to the line of sight) using the calibrations described in detail by \citet{boisse2010}. The estimated uncertainty due to these calibrations is 1 $\mathrm{km\,s^{-1}}$ on $v\,\sin I$  and about 0.09 dex on [Fe/H]. Additionally, a more detailed spectral analysis was carried out using the \sophie\ spectra obtained in \emph{objAB} mode, and not contaminated by moonlight. The method \citep{santos2004} provides the effective temperature $T_\mathrm{eff}$, the surface gravity $\log\,g$ and the metallicity [Fe/H] of the observed stars. To obtain the stellar masses, we interpolated the isochrones by \citet{girardi2000}, using the method described by \citet{dasilva2006}. The values derived from this analysis are reported in Table~\ref{table.stellarparam}. The uncertainties for the stellar masses range from 2.6\% to 5.8\%. However, these are the formal uncertainties, which do not take into account the systematic uncertainties in the models, and we therefore decided to adopt a conservative uncertainty of 10\% on the stellar mass.  The spectral types reported in Table~\ref{table.obslog} are obtained from interpolating the tables from \citet{cox2000} to the values of $T_\mathrm{eff}$ obtained. We note that in all cases, the spectroscopic determination of the metallicity agrees with the method based on the CCF \citep{boisse2010} to within 1-$\sigma$.

\setcounter{table}{2}
\begin{table}[t]
  \centering 
  \caption{Cross-correlation function parameters. Median values are reported; the error intervals correspond to the standard deviation of each magnitude.}
\label{table.CCF}
\begin{tabular}{lcccc}
\hline
\hline
Target 					& Mask & RV & FWHM & Contrast \\
						&	&[km s$^{-1}$]&[km s$^{-1}$]&[\%]\\
\hline
HD14651\tablefootmark{a}	&G2&52.2&$7.53\pm0.04$&$43.1\pm0.6$\\
HD16702					&G2&5.5&$7.58\pm0.02$&$35.8\pm0.4$\\
HD22781 \tablefootmark{a}	&K5&8.2&$6.60\pm0.03$&$34.3\pm1.7$\\
HD30246\tablefootmark{a}	&G2&41.2&$9.34\pm0.14$&$35.1\pm2.4$\\
HD92320					&G2&1.1&$7.70\pm0.02$&$38.3\pm0.5$\\
HD137510				&G2&-6.7&$12.14\pm0.06$&$29.1\pm0.4$\\
HD156279				&G2&-20.5&$7.64\pm0.01$&$47.1\pm0.5$\\
HD167215				&G2&-44.6 &$9.74\pm0.04$&$22.2\pm0.7$ \\
\hline
\end{tabular}
\tablefoot{
\tablefoottext{a}{Only spectra acquired in \emph{objAB} mode are considered for the contrast and the FWHM.}
}
\end{table}

\begin{table*}[t]
  \centering 
  \caption{Stellar parameters of target stars. \label{table.stellarparam}}
  \label{table_parameters}
\begin{tabular}{lccccc}
\hline
\hline
Parameters							&HD14651		&HD16702		&HD22781		&HD30246		\\
\hline
Effective temperature [K]					&$5491\pm26$		&$5908\pm25$		&$5027\pm50$		&$5833\pm44$		\\
$v \sin I$ [\kms]\tablefootmark{a}			&3.07			&2.56			&1.73			&4.96			\\
\hspace{0pt}[Fe/H]						&$-0.04\pm0.06$	&$-0.12\pm0.06$	&$-0.37\pm0.12$	&$0.17\pm0.10$	\\
$\log\,g$								&$4.45\pm0.03$	&$4.46\pm0.03$	&$4.60\pm0.02$	&$4.39\pm0.04$	\\
$M_\textrm{s}$ [M$_\odot$]\tablefootmark{b}	&$0.96\pm0.03$	&$0.98\pm0.04$	&$0.75\pm0.03$	&$1.05\pm0.04$	\\
Age [Gyr]								&$[0.2-6.6]$		&$[0.1-4.6]$		&$[0.6-8.9]$		&$[0.2-3.7]$		\\
Age$_{R^\prime_\mathrm{HK}}$ [Gyr]\tablefootmark{c}		&$>5.0$			&$[0.6-1.1]$		&$>4.6$			&$[0.4-0.9]$		\\
$\log R^\prime_\mathrm{HK}$				&$-5.02\pm0.11$	&$-4.54\pm0.05$	&$-5.02\pm0.14$	&$-4.50\pm0.06$	\\
&\\
\hline
\hline
Parameters							&HD92320		&HD137510		&HD156279		&HD167215		\\
\hline
Effective temperature [K]					&$5664\pm24$		&$6131\pm50$		&$5453\pm40$		&$6351\pm68$		\\
$v \sin I$ [\kms]\tablefootmark{a}			&3.13			&7.42			&2.51			&4.94			\\
\hspace{0pt}[Fe/H]						&$-0.10\pm0.06$	&$0.38\pm0.13$	&$0.14\pm0.01$	&$-0.21\pm0.14$	\\
$\log\,g$								&$4.48\pm0.03$	&$4.02\pm0.04$	&$4.46\pm0.03$	&$4.19\pm0.07$	\\
$M_\textrm{s}$ [M$_\odot$]\tablefootmark{b}	&$0.92\pm0.04$	&$1.36\pm0.04$	&$0.93\pm0.04$	&$1.15\pm0.07$	\\
Age [Gyr]								&$[0.2-6.6]$		&$[2.3-3.0]$		&$[0.4-7.7]$		&$[2.0-5.2]$		\\
Age$_{R^\prime_\mathrm{HK}}$ [Gyr]\tablefootmark{c}		&$[2.2-4.7]$		&$>5.4$			&$>5.4$			&$>8.0$		\\
$\log R^\prime_\mathrm{HK}$				&$-4.80\pm0.08$	&$-5.09\pm0.18$	&$-5.01\pm0.06$	&$-5.25\pm0.19$	\\
&&&\\
\hline
\end{tabular}
\tablefoot{
\tablefoottext{a}{Estimated from CCF as described in Sect.~\ref{sect.stpar}. Uncertainty is estimated  to about 1 \kms\ in $v \sin I$ \citep[see also][]{boisse2010}.}
\tablefoottext{b}{Formal errors issued from isochrone fitting are reported. A more realistic 10\% was used for the rest of the analysis.}
\tablefoottext{c}{Age estimated from the age-activity relation by \citet{mamajekhillenbrand2008}.}
}

\end{table*}


\subsection{Stellar activiy}
We computed the $\log R^\prime_{HK}$ activity index \citep{noyes84}  on each spectrum of each target using the method described by \citet{boisse2010}. In Table~\ref{table.stellarparam} we report the median value of the index for each target and its standard deviation\footnote{The standard deviation of $\log R^\prime_{HK}$ was computed using the median absolute deviation, which is less sensitive to the presence of outliers.}, computed using the spectra obtained in \emph{objAB} mode only, since the contamination by the ThAr lamp present in the rest of the spectra is expected to change the  depth of the spectral lines. The obtained values were used to estimate the age of the host stars using the relation by \citet{mamajekhillenbrand2008}. We report the results in Table~\ref{table.stellarparam}. The activity-determined ages are in agreement with the values obtained from isochrone fitting, except for HD137510 and HD167215. These stars are the two least active stars in the sample, and their $\log R^\prime_{HK}$ are either at the limit of or outside the range of applicability of the \citet{mamajekhillenbrand2008} formula, which is valid only for  $\log R^\prime_{HK}>-5.1$.

It is known that stellar activity is a source of uncertainty in the RV measurements \citep[e.g.][]{queloz2001,boisse2011}. In some cases \citep{boisse2010, melo2007}, the RV jitter produced by stellar activity has been corrected, with the result of improved determinations of the orbital and planetary parameters. This is done by decorrelating the residuals of the orbital fit with the bisector velocity span. For all our sample stars, we checked if the residuals of the best-fit Keplerian orbit exhibited any dependence with the measured bisector span. In no case a correlation is found with a significance above 2 $\sigma$. In the case of HD30246, the most active star in our sample ($\log R^\prime_\mathrm{HK} = -4.5$), a slight hint of negative correlation is seen, with a fit slope of $b = -0.13 \pm 0.08$. On the other hand, HD16702 has a similar activity level ($\log R^\prime_\mathrm{HK} = -4.54$), but the residuals do not exhibit any correlation with activity index $R^\prime_\mathrm{HK}$. However, as will be discussed in \S\ref{sec.stars}, the companion to HD16702 is most likely a low-mass star, which might contaminate the determination of $R^\prime_\mathrm{HK}$. We note that the stars studied by \citet{boisse2010} (HD189733) and \citet{melo2007} (HD102195) exhibit activity levels similar to that of HD30246 and HD16702, but their data is much more precise, which might permit detecting subtler trends.

\subsection{Moon contamination}
Whenever present, moonlight scattered in the Earth's atmosphere can pollute the observed stellar spectrum. If the Moon's RV is close to the velocity of the observed star, moonlight can bias the measurement of the velocity. Prior to performing a keplerian fit to the data, we have verified that none of the RV measurements were biased by this effect.  For an observation to be polluted, two conditions must be fulfilled: in the first place, the scattered moonlight must contribute significantly to the light entering the spectrograph's fiber. For the observations in \emph{objAB} mode, this can be addressed using fiber B. We considered conservatively that the moonlight level is enough to pollute a stellar spectrum if both the S/N in fiber B was larger than 3 and the contrast of the corresponding CCF was larger than 1\%, i.e. if the CCF peak of the Moon is detected in the sky fiber, even if marginally. The second condition is that the Moon RV must be close to the stellar velocity. To study this, we considered the Baricentric Earth Radial Velocity (BERV) on the day of the observation, and in the direction of the target, which is within $\sim 1$ \kms\ of the Moon Radial Velocity. The Moon RV is considered to be close to the measured RV if the BERV was less than $\sim 7\sigma$ away from the measured stellar peak, where $\sigma$ is the width of the the CCF.  For those cases in which the second fiber was dedicated to obtain the spectrum of a ThAr lamp simultaneously, for which no sky monitoring is possible, the presence of the moon was assumed whenever its illuminated fraction was larger than 40\%. Besides this, the same criterion based on the BERV was applied.

With these criteria, none of the reported velocities is polluted, due mainly to the fact that the targets are bright. However, for limiting cases, to confirm that the effect of moonlight pollution was negligible, we performed the Moon correction as is described by \citet{barge2008} and \citet{pollacco2008} for SOPHIE and by \citet{bonomo2010} for HARPS. In all cases, we found that the corrected velocities were in excellent agreement with the uncorrected measurements, which were therefore retained.

\subsection{Keplerian Orbits \label{sect.keplerianorbits}}
The radial velocity measurements of all the targets exhibit variations greatly in excess of the measurement precision. The RVs were fit using a single Keplerian orbit model. A genetic algorithm was employed to pick up the starting point for the Levenberg-Marquardt method. The best-fit parameters are reported in Table~\ref{table.parameters}, together with the 68\% confidence interval, obtained from 5000 Monte Carlo simulations. In most cases, the histograms of the obtained parameters have a single peak, and gaussian-like appearance, and therefore the errors reported can be taken as the 1-$\sigma$ intervals. In a few cases, however, double-peaked or highly asymmetric  distributions were found. These cases are studied in detail for each object in \S\ref{sec.results}. In Figure~\ref{fig.RV} we plot the RV curves for all stars observed together with the best-fit model and the residuals, after correction of the seeing effect (see below). 

\subsubsection{Correction of the seeing effect \label{sect.seeing}}
Radial velocities obtained with \sophie\ exhibit a systematic effect thought to be caused by an incomplete radial scrambling in the optical fiber coupled with aberrations in the optical system, which lead to a variation in the illumination of the spectrograph's pupil. The consequence of this effect is an artificial variation in the measured velocity when the point spread function changes, as a consequence of seeing variations, focus variations, etc. This effect induces RV variations with an amplitude of about 20 m s$^{-1}$ and constituted the main limitation on the spectrograph's precision until very recently\footnote{This has been solved by a recent upgrade of the fiber link \citep{perruchot2011}.}. 


\citet{boisse2011b} present a detailed description of this effect and provide a way to diagnose and correct it.  We follow the method described by these authors, but using a different variable to decorrelate the data.  Since simulations of the optical path of \sophie\ have shown that variations in the pupil's illumination produce a differential effect between the red and blue parts of each order \citep{boisse2011b}, we have used the difference between the velocities measured in each half of the CCD (see \S\ref{sect.observations.velocimetry}) as an indicator of this effect. We define the pupil's illumination proxy $\delta_{RV}$ as
\begin{displaymath}
\delta_{RV} = RV_\mathrm{blue} - RV_\mathrm{red}\;\;,
\end{displaymath}
where $RV_\mathrm{blue}$ and $RV_\mathrm{red}$ indicate the velocities measured using the blue and red part of the orders, respectively. 

Not all stars in our sample exhibit a clear correlation between the residuals and $\delta_{RV}$. This is also the case when the estimator from \citet{boisse2011b} is used, but the reasons for this differential behaviour are currently under study. For the moment, we have decided to perform the correction only on the stars for which the Pearson's correlation coefficient between $\delta_{RV}$ and the fit residuals is greater than 0.5. In this case the residuals are decorrelated and added back to the fit curve. The resulting data set is then used to find a new solution to the 1-Keplerian fit, which is shown in Figure~\ref{fig.RV}. The data of only three stars fulfil this criterion: HD16702, HD92320 and HD137510, for which the correlations between residuals and $\delta_{RV}$ are shown in Fig.~\ref{fig.resplots}. For the rest of the sample, the correlation is not significant, and the data reported do not include any correction. 

\subsection{Resolving the nature of the companions}

The high minimum mass of most of the reported companions imply a non-negligible probability that the objects are actually low-mass stars in low-inclination orbits. To narrow down this possibility and, when possible, conclude on the nature of the detected objects, we have carried out a series of tests and simulations using the available RV data and astrometric observations from the Hipparcos satellite \citep{hipparcos}.

\subsubsection{Bisector Analysis and CCF Simulations \label{sect.bisector}}
If the companion object is relatively luminous, we expect it to pollute the peak of primary star in the CCF, and to produce a variable distortion of its shape, which results in a variation of its bisector velocity span throughout the object's orbit. We have searched for significant variations in the bisector velocity span, as well as correlations between the bisector velocity span and RV measurements. The detection of such correlation indicates that the velocity signal is the product --at least in part-- of a deformation of the stellar CCF. This can either mean that the velocity signal is in fact a product of stellar activity --which would be unprecedented for the large RV amplitudes observed in these targets, and that could easily be identified using other diagnostics such as the correlation between RV and the $\log R^\prime_{HK}$ index--, or that the companion is contributing non-negligibly to the observed CCF. For all the stars of our sample, except for HD16702 (see \S\ref{sec.results} below), the bisector velocity span observations have negligible variability and do not exhibit any significant correlation with radial velocity measurements.

In the context of this paper, the lack of correlation between the bisector velocity span and the radial velocities can be used to constrain the mass of the companion by means of CCF simulations. To exploit this possibility, we assume that both the host and the companion objects are dwarf stars, and that each contributes its own peak to the observed CCF. We model each CCF as a gaussian function. The parameters of the primary CCF set to the median values of the observed quantities. The area and FWHM of the secondary peak are varied according to the modeled mass ratio and rotational velocity, assuming the same metallicity as the primary star, and using the calibrations from \citet{boisse2010}. For each observed orbital phase, the primary and secondary CCF are simulated at the expected position in RV space, and are added after normalisation using the flux ratio of the stars. Here we assume that the measured RV corresponds approximately well to the RV of the primary star, which holds for small mass ratios. For each simulated orbital phase, the RV, bisector velocity span, FWHM and contrast of the joint CCF are computed similarly to how is done for the real data, and the results are compared to the observations. The value of the $\chi^2$ statistics is computed over a grid of masses and rotational broadening values for the secondary, and confidence regions are obtained from intersecting the $\chi^2$ surface with constant planes. In general, we have found that the bisector velocity span provides strong constrains on the mass of the secondary, resulting in many cases in lower upper limits than those issued from astrometric measurements. The results from these simulations are reported for each target in \S\ref{sec.results}.

\subsubsection{Mask Effect}
Similar to the bisector analysis, the dependence of the amplitude of the RV variations with the numerical mask used to measure them can reveal the contribution of a secondary stellar object to the flux entering the spectrograph \citep{bouchy2009b, santos2002}. 


When analysing our data in search of this effect, we proceed in the following way: first, we compute the best Keplerian fit as described in \S\ref{sect.keplerianorbits}, using the best-matching numerical mask (Table \ref{table.CCF}). Then we adjust the RV measurements obtained with other masks, fixing all orbital parameters, except for the semi-amplitude $K$ and the center-of-mass velocity $\gamma$, which is expected to change from mask to mask. Since the fit is linear on these two parameters, the errors are obtained directly from the covariance matrix. For all our targets, except for HD16702 (see \S\ref{sec.stars}), the amplitudes obtained in all three masks are compatible.

\subsubsection{Analysis of the Hipparcos astrometry}
All stars listed in Table~\ref{table.obslog} were catalogued by the Hipparcos astrometry satellite \citep{hipparcos}. We used the new Hipparcos reduction \citep{vanleeuwen2007} to search for signatures of orbital motion in the Intermediate Astrometric Data (IAD). The analysis was performed as described in \citet{sahlmann2011}, where a detailed description of the method can be found. Using the orbital parameters given by the RV analysis (Table~\ref{table.parameters}), the IAD was fitted with a seven-parameter model, where the free parameters are the inclination $i$, the longitude of the ascending node $\Omega$, the parallax $\varpi$, and offsets to the coordinates ($\Delta \alpha^{\star}$, $\Delta \delta$) and proper motions ($\Delta \mu_{\alpha^\star}$, $\Delta \mu_{\delta}$). A two-dimensional grid in $i$ and $\Omega$ was searched for its global $\chi^2$-minimum with a standard nonlinear minimisation procedure. The statistical significance of the derived astrometric orbit was determined with a permutation test employing 1000 pseudo orbits.   Uncertainties in the solution parameters were derived by Monte Carlo simulations. This method has proven to be reliable in detecting orbital signatures in the Hipparcos IAD \citep{sahlmann2011, sahlmann2011b}. Where the astrometric signal was not detected in the data, i.e. the derived significance is low, the Hipparcos observations were used to set an upper  limit to the companion mass following the method described by \cite{sahlmann2011, sahlmann2011b}.


\begin{table}\caption{Parameters of the Hipparcos astrometric observations}
\label{tab:HIPparams} 
\centering  
\begin{tabular}{l c c c c c} 	
\hline\hline %
Target & HIP & Sol. & $N_\mathrm{Orb}$\tablefootmark{b} & $\sigma_{\Lambda}$\tablefootmark{c}  & $N_\mathrm{Hip}$\tablefootmark{d} \\  
       &     & type\tablefootmark{a} &           & (mas)                 &           \\  
\hline 
HD14651 & 11028 & 5 & 13.4 & 3.4 & 59  \\ 
HD16702 & 12685 & 5 & 15.0 & 5.3 & 140  \\ 
HD22781 & 17187 & 5 & 1.7 & 4.0 & 69  \\ 
HD30246 & 22203 & 5 & 0.8 & 3.5 & 56  \\ 
HD92320 & 52278 & 5 & 8.1 & 6.4 & 233  \\ 
HD137510 & 75535 & 5 & 1.4 & 1.9 & 101  \\ 
HD156279 & 84171 & 5 & 8.7 & 3.5 & 110  \\ 
HD167215 & 89270 & 7 & 1.8 & 5.3 & 144  \\ 
\hline
\end{tabular} 
\tablefoot{
\tablefoottext{a}{'5': standard five-parameter astrometric solution; '7': solution including first derivatives of the proper motion.}
\tablefoottext{b}{Portion of the orbit covered by the Hipparcos measurements.}
\tablefoottext{c}{Median precision.}
\tablefoottext{b}{Number of observations.}
}
\end{table}
Table~\ref{tab:HIPparams} lists the target identifiers and the basic parameters of the Hipparcos observations relevant for the astrometric analysis. The solution type indicates the astrometric model adopted by the new reduction. For the standard five-parameter solution it is '5', whereas it is '7' when the model included proper-motion derivatives of first order to obtain a reasonable fit, which is the case for HD167215. The parameter $N_\mathrm{Orb}$ represents the number of orbital periods covered by the Hipparcos observation timespan and $N_\mathrm{Hip}$ is the number of astrometric measurements with a median precision of $\sigma_{\Lambda}$. Outliers in the IAD had to be removed because they can substantially alter the outcome of the astrometric analysis. We eliminated one outlier for HD30246 and HD92320 and two outliers for HD156279.


An orbit signature is detected for HD16702, with a 2.5-$\sigma$ significance, which provides a measurement of the real mass of the companions and places it in the stellar-mass range (see \S\ref{sec.stars}). For the rest of the targets, no significant orbit is detected, but upper mass limits are derived. These are reported on a target basis in \S\ref{sec.results}. The only exception is HD30246, for which not even an upper limit can be reliably derived from Hipparcos data (see \S\ref{sec.browndwarfs}).

\begin{table*}[t]
\centering 
  \caption{Fitted orbits and planetary parameters. \label{table.parameters}}

\begin{tabular}{lcccccc}
\hline
\hline
											&HD22781\tablefootmark{\dagger}			&HD156279	&HD14651			&HD30246\\
\multicolumn{4}{l}{\emph{Orbital Parameters}}\\
\hline
$P$ 				[days]						& $528.07 \pm 0.14$	&$131.05 \pm 0.54$				& $79.4179 \pm 0.0021$ 	&$990.7 \pm 5.6$			\\
$e$											& $0.8191 \pm 0.0023$	&$0.708 \pm 0.018$				& $0.4751 \pm 0.0010$ 	&$0.838 \pm 0.081$			\\
$\omega$ 		[$^{\circ}$]						&$-44.08 \pm 0.56$		&$-95.8 \pm 1.7$				& $-151.92 \pm 0.21$ 	&$-65 ^{+ 18} _{- 13}$		\\
$K$				[\ms]							&$726.4 \pm 7.1$		&$578 \pm 20$					& $2595.0 \pm 3.6$		&$2035 ^{+ 1600} _{- 560}$	\\
$T_0$ (periastron)	[BJD-2\,454\,000]				&$881.40 \pm 0.12$		&$1525.59 \pm 0.18$			& $537.607 \pm 0.045$	&$1314.0 ^{+ 9.3} _{- 7.2}$	\\
$V_0$ 			[\kms]						&$8.2508 \pm 0.0042$	&$-20.6432 \pm 0.0050$			& $52.7244 \pm 0.0030$	&$41.91 ^{+ 1.05} _{- 0.19}$	\\
$d_1$			[\ms/yr]						&$14.2761 \pm 0.0019$	&--							& -- 					&--						\\
$d_2$			[\ms/yr$^2$]					& $-3.10954$			&--							& -- 					&--						\\
%
\multicolumn{4}{l}{\emph{Fit Parameters}}\\
\hline
Typical photon noise   [\ms]						&$2.61$				&2.83						&$3.17$				&$4.35$					\\
Typical uncertainty\tablefootmark{a}   [\ms]			&$4.41$				&$4.26$						&$4.64$				&$5.67$					\\
$\sigma_\mathrm{O-C}$		[\ms]					&$8.36$				&9.08						&$5.26$				&$14.09$					\\
%
\multicolumn{4}{l}{\emph{Derived Parameters}}\\
\hline
$f(M_\textrm{s},M_\textrm{p})$ [$10^{-6}$ M$_\odot$] 	&$3.956 \pm 0.066$		&$0.926 \pm 0.034$				&$97.95 \pm 0.41$ 		&$132 ^{+ 204} _{- 41}$ 		\\
$a$				[AU]							&$1.167 \pm 0.039$		&$0.495 \pm 0.017$				&$0.361\pm0.013$		&$2.012\pm0.069$			\\\
Min. astrometric signal [$10^{-5}$ arcsec]				&$22.62 \pm 0.26$		&$5.68\pm0.31$				&$36.71\pm0.03$		&$48.8^{+ 9.7} _{-4.9}$	\\
$M_\textrm{c} \sin i$	[\MJ]							&$13.65 \pm 0.97$		&$9.71 \pm 0.66$				&$47.0\pm3.4$			&$55.1 ^{+ 20.3} _{- 8.2}$		\\
$M_{c,\mathrm{up-lim}}$ [M$_\odot$]				&[0.23--0.31]			&[0.35--0.41]					&[0.28--0.4]			&[0.28--0.37]\\
$M_{c}$ [M$_\odot$]\tablefootmark{b}				&--					&--							&--					&--						\\
\\
\hline
\hline
											&HD92320			&HD137510						&HD16702			&HD167215				\\
\multicolumn{4}{l}{\emph{Orbital Parameters}}\\
\hline
$P$ 				[days]						&$145.402 \pm 0.013$ 	&$801.30 \pm 0.45$					& $72.8322\pm0.0023$	&$632 ^{+ 260} _{- 79}$		\\
$e$											&$0.3226 \pm 0.0014$ 	&$0.3985 \pm 0.0073$				& $0.1373\pm0.0017$	&$0.37 ^{+ 0.13} _{- 0.06}$	\\		
$\omega$ 		[$^{\circ}$]						&$-22.99 \pm 0.54$ 		&$31.2 \pm 1.1$					& $-102.44\pm 0.56$	&$140 \pm 15$				\\
$K$				[\ms]							&$2574.3 \pm 6.6$ 		&$531.5 \pm 7.0$					& $2423.7\pm 3.0$		&$2143 ^{+ 550} _{- 370}$	\\				
$T_0$ (periastron)	[BJD-2\,454\,000]				&$798.17 \pm 0.18$ 	&$187.8 \pm 2.0$					& $983.45\pm0.12$		&$1748 ^{+ 270} _{- 82}$		\\
$V_0$ 			[\kms]						&$-0.1418 \pm 0.0045$ 	&$-6.4157 \pm 0.0081$\tablefootmark{c} & $4.5602\pm0.0022$	&$-45.17 ^{+ 0.30} _{- 0.09}$	\\
$d_1$			[\ms/yr]						&-- 					&--								& -- 					&--						\\
$d_2$			[\ms/yr$^2$]					&-- 					&--								& -- 					&--						\\
%
\multicolumn{4}{l}{\emph{Fit Parameters}}\\
\hline
Typical photon noise [\ms]   						&3.31				&3.85							&$3.32$				&7.51					\\
Typical uncertainty\tablefootmark{a} [\ms]				&$4.69$				&$5.32$							&$4.78$				&$8.16$					\\
$\sigma_\mathrm{O-C}$		[\ms]					&9.07				&10.76\tablefootmark{d}				&$7.85$				&7.92					\\
\multicolumn{4}{l}{\emph{Derived Parameters}}\\
\hline
$f(M_\textrm{s},M_\textrm{p})$ [$10^{-6}$ M$_\odot$] 	&$217.9 \pm 1.7$ 		&$9.61 \pm 0.30$					&$104.41\pm0.39$		&$520 ^{+ 660} _{- 250}$		\\
$a$				[AU]							&$0.536 \pm 0.018$ 	&$1.880 \pm 0.064$					&$0.344\pm0.012$		&$1.56 ^{+ 0.40} _{- 0.14}$	\\
Min. astrometric signal [$10^{-5}$ arcsec]				&$75.2\pm0.2$			&$86.3\pm0.9$						&$34.59\pm0.06$		&$144 ^{+ 106} _{- 33}$	\\
$M_\textrm{c} \sin i$	[\MJ]							&$59.4 \pm 4.1$ 		&$27.3 \pm 1.9$					&$48.7\pm3.4$			&$92 ^{+ 29} _{- 18}$		\\
$M_{c,\mathrm{up-lim}}$ [M$_\odot$]				&[0.26--0.34]			&0.057							&--					&--						\\
$M_{c}$ [M$_\odot$]	\tablefootmark{b}				&--					&[0.019--0.057]						&[0.35--0.45]			&[0.2--0.29]				\\
\hline
\end{tabular}

\tablefoot{
\tablefoottext{\dagger}{A second companion is probably present in the system (see \S\ref{sec.planets}). The orbital parameters for HD22781 b  might therefore change slightly once the two-Keplerian model is used.}
\tablefoottext{a}{Including wavelength calibration error and 3 \ms\ of systematic error.}
\tablefoottext{b}{99.7\%-confidence interval.}
\tablefoottext{c}{Systemic radial velocity of SOPHIE data. The values obtained for the other instruments are: $0.1587 \pm 0.0082$ \kms, $0.1352\pm 0.083$ \kms and  $0.045 \pm 0.011$, for McD, TLS and Lick data, respectively.}
\tablefoottext{d}{Dispersion for SOPHIE data. The values for other data sets are: 13 \ms, 20 \ms, and 16 \ms for McD, TLS and Lick data, respectively.}
}
\end{table*}

\begin{figure*} 
\begin{center}
\includegraphics[width=0.5\textwidth]{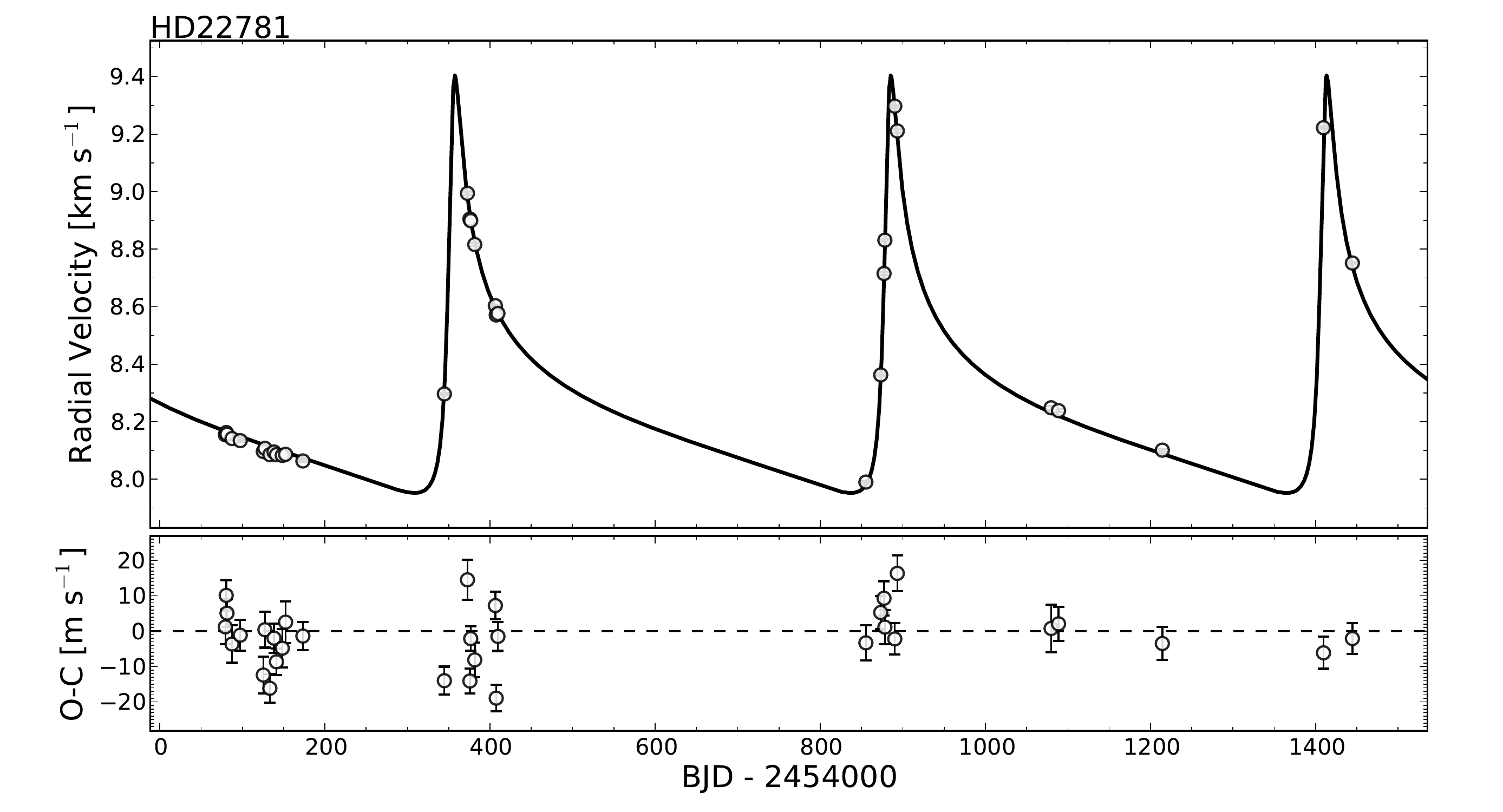}\includegraphics[width=0.5\textwidth]{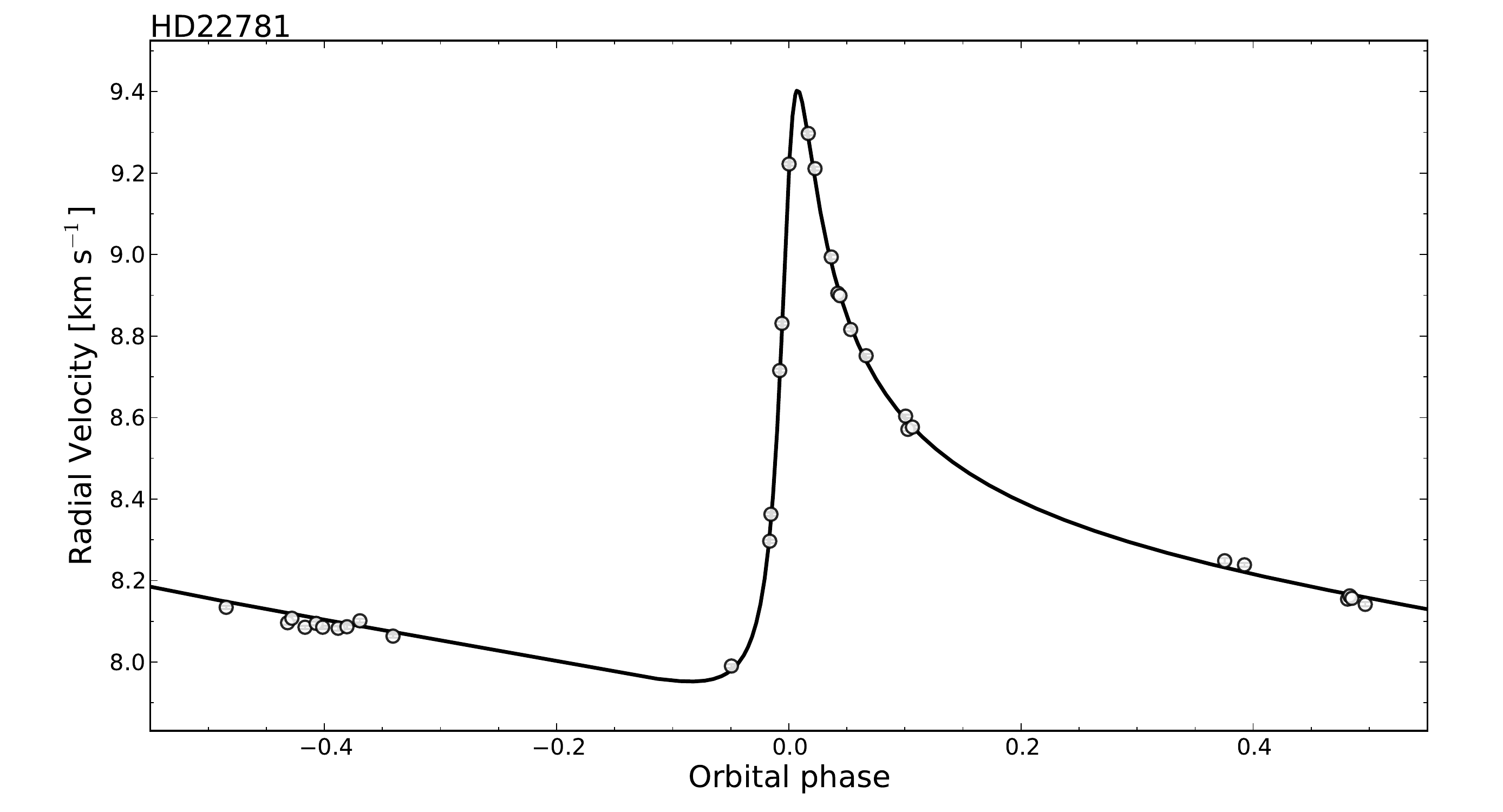}
\includegraphics[width=0.5\textwidth]{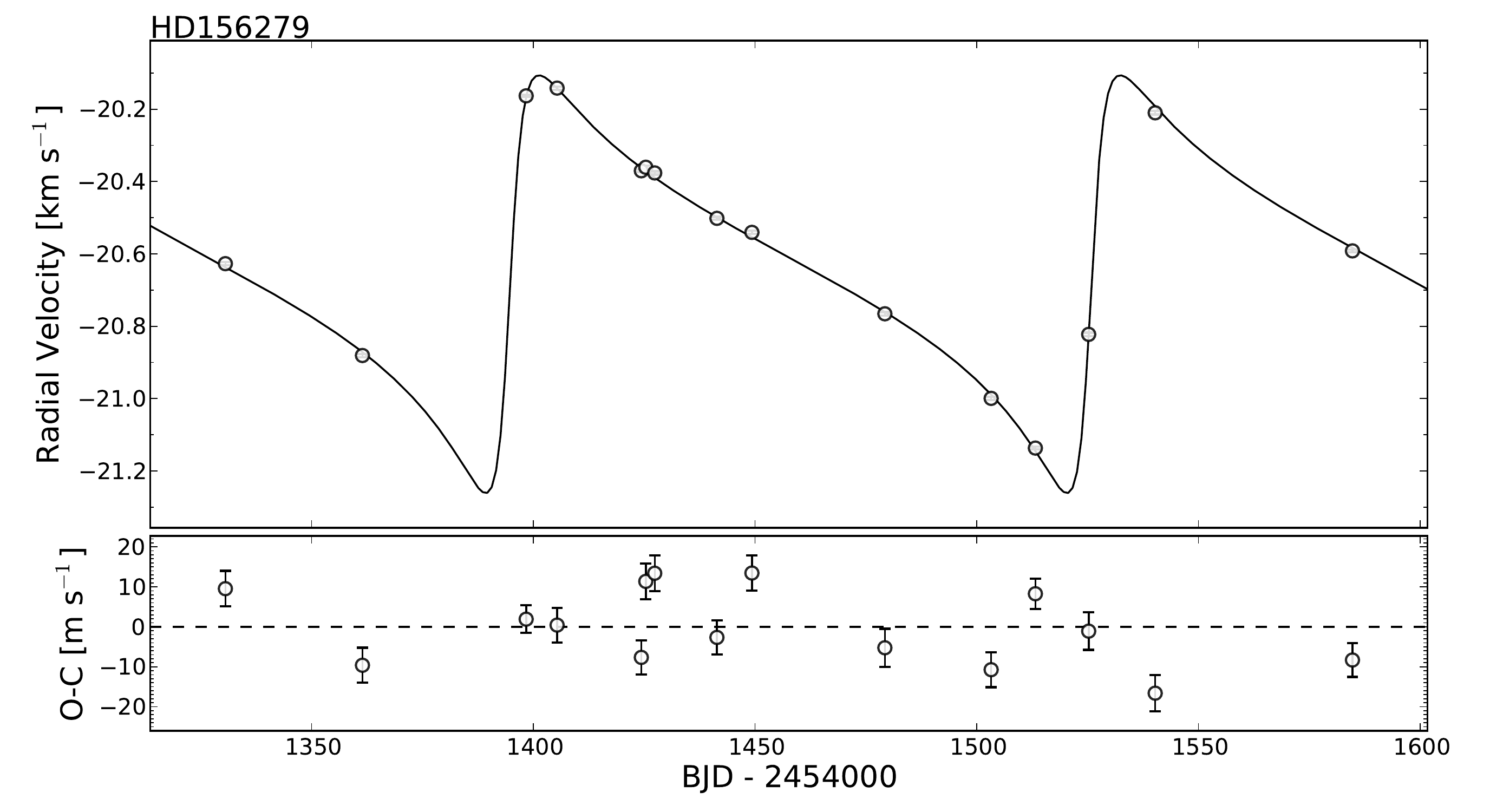}\includegraphics[width=0.5\textwidth]{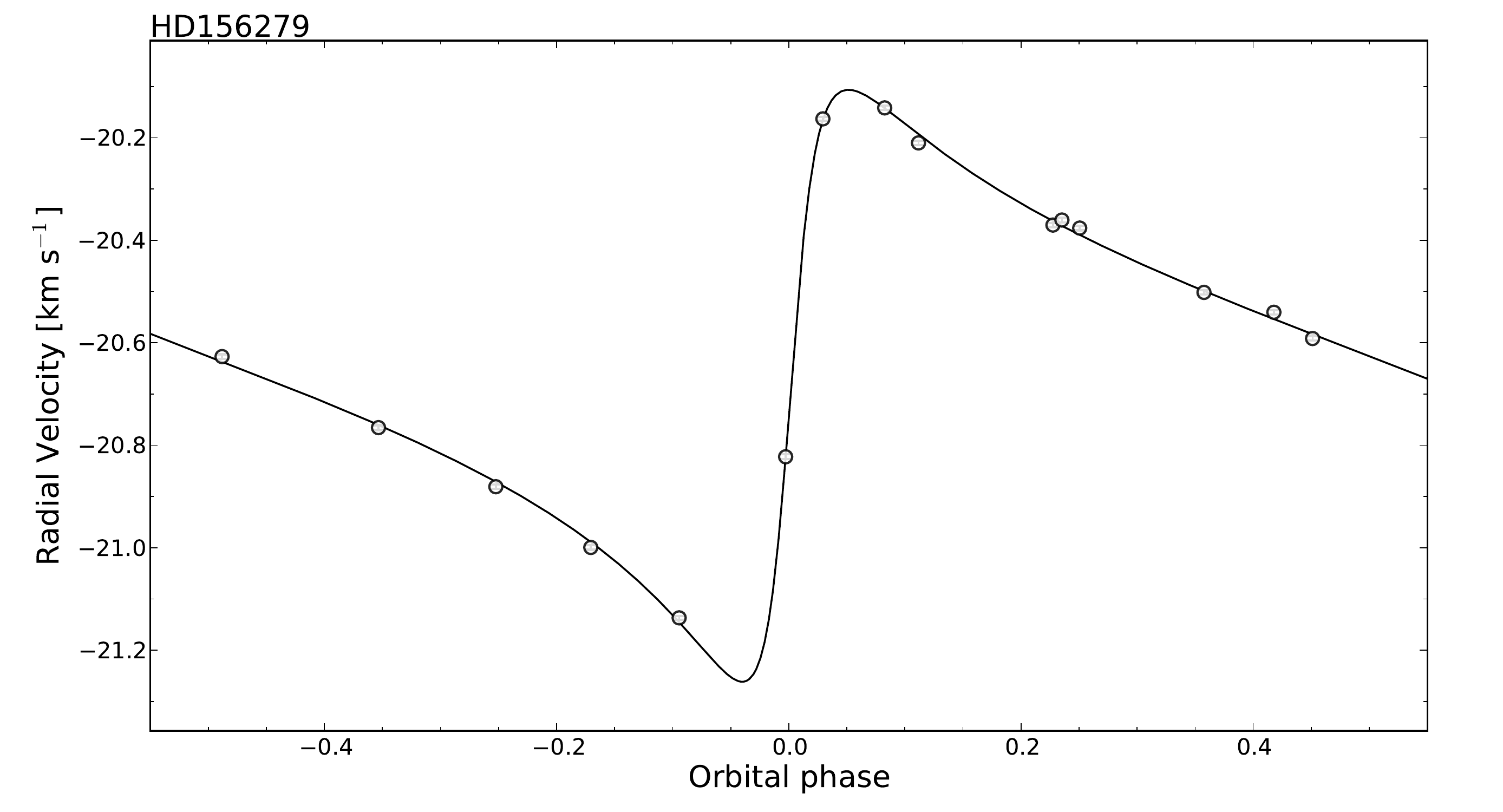}
\includegraphics[width=0.5\textwidth]{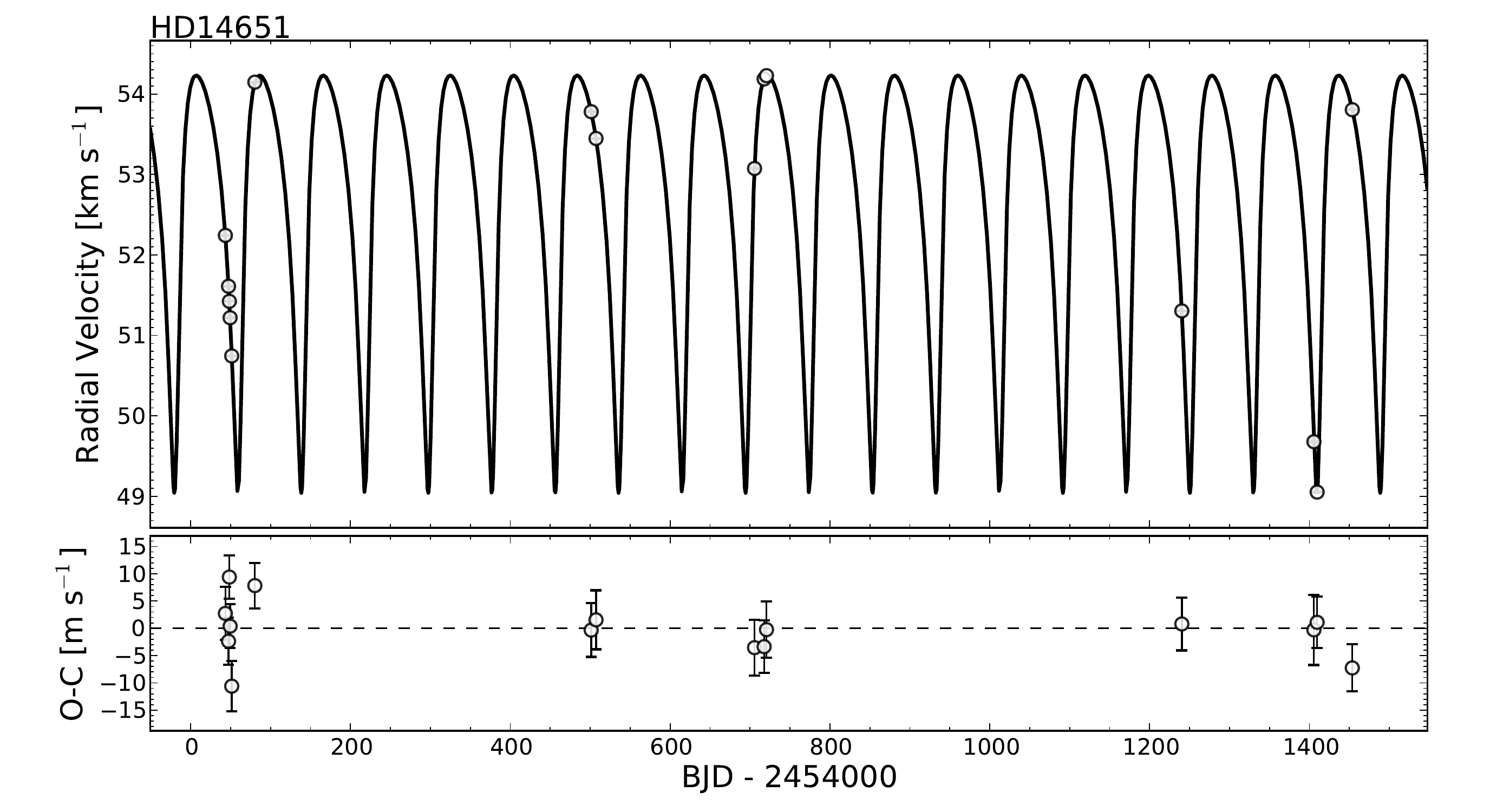}\includegraphics[width=0.5\textwidth]{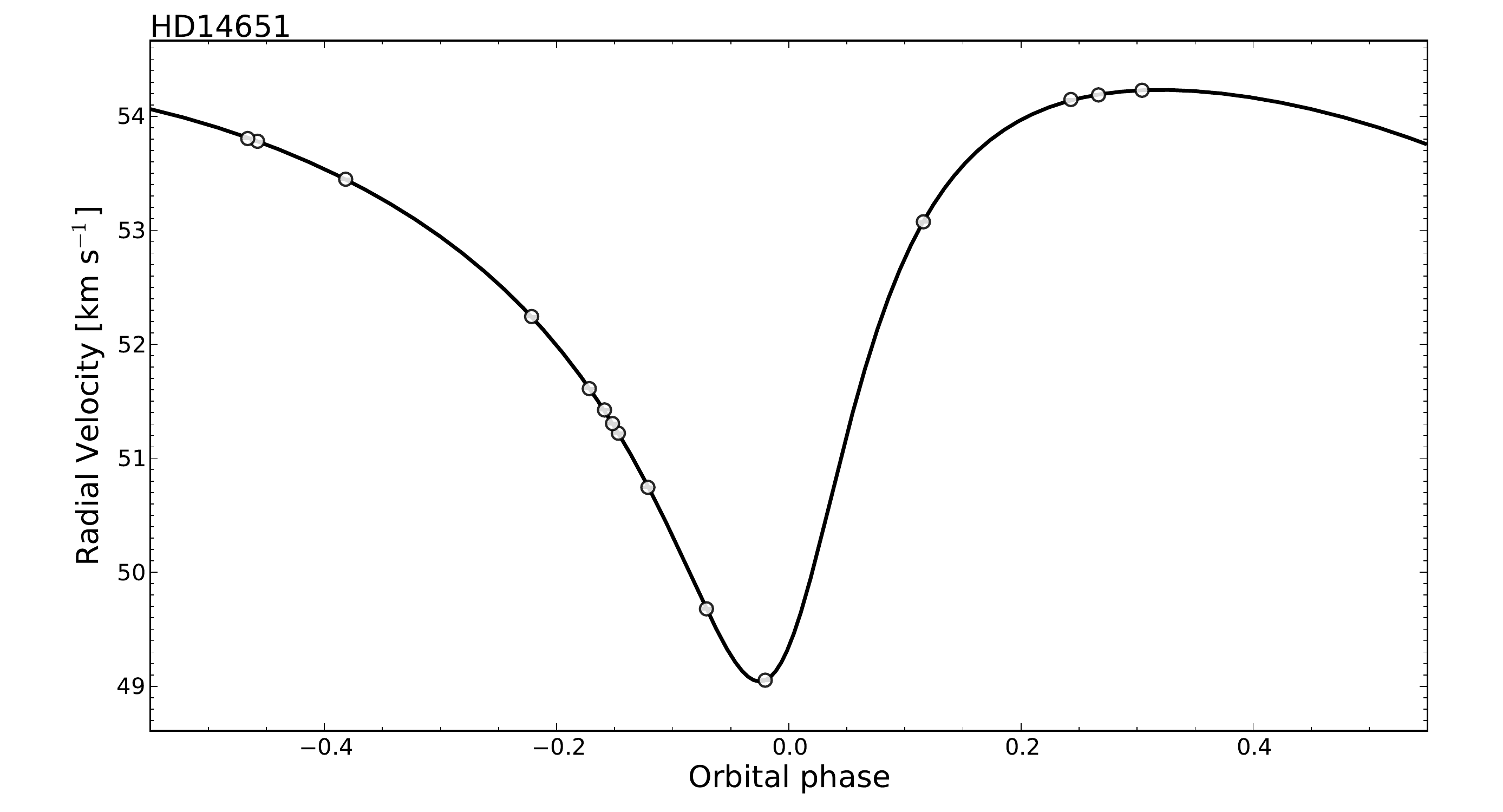}
\includegraphics[width=0.5\textwidth]{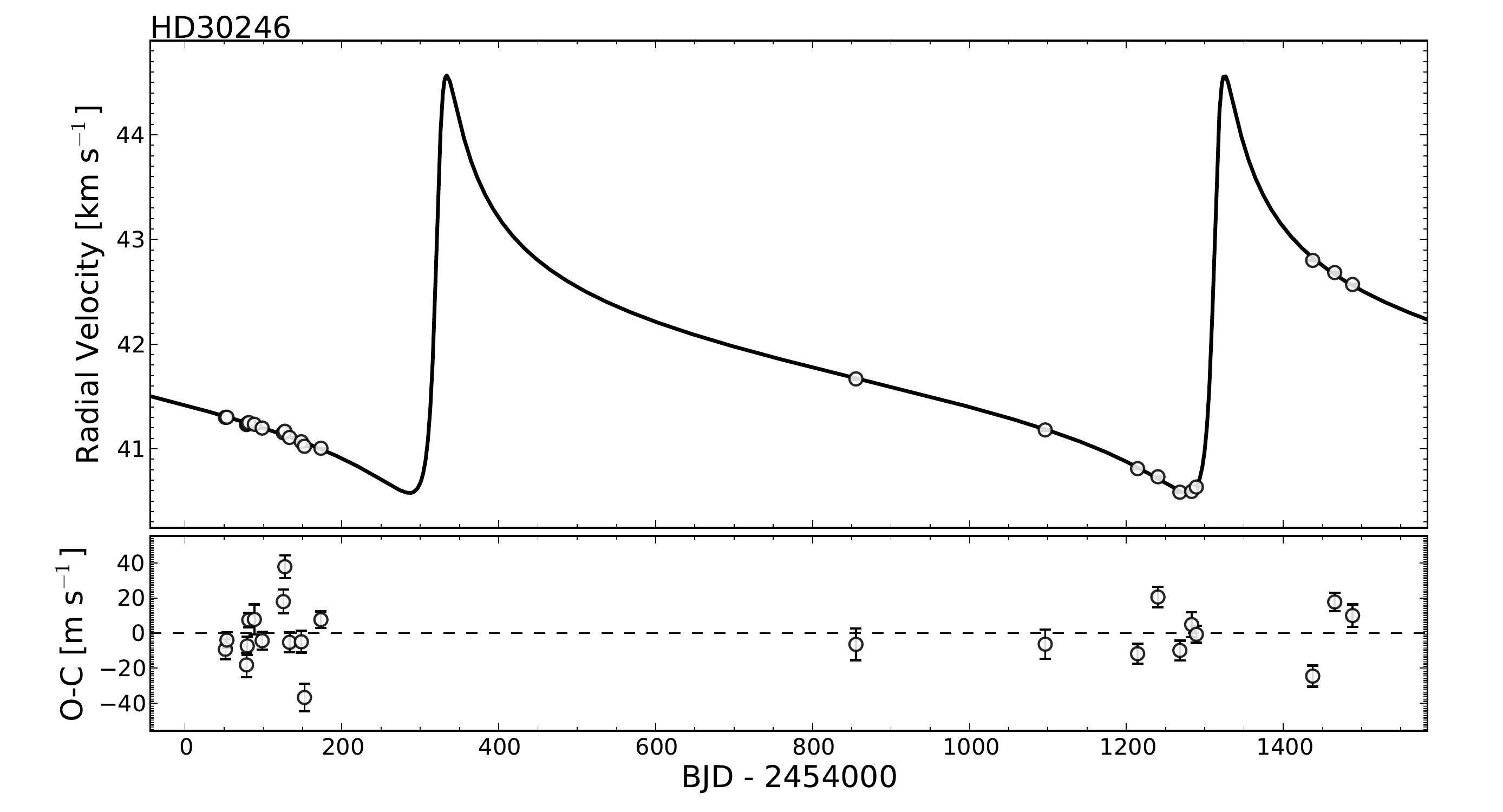}\includegraphics[width=0.5\textwidth]{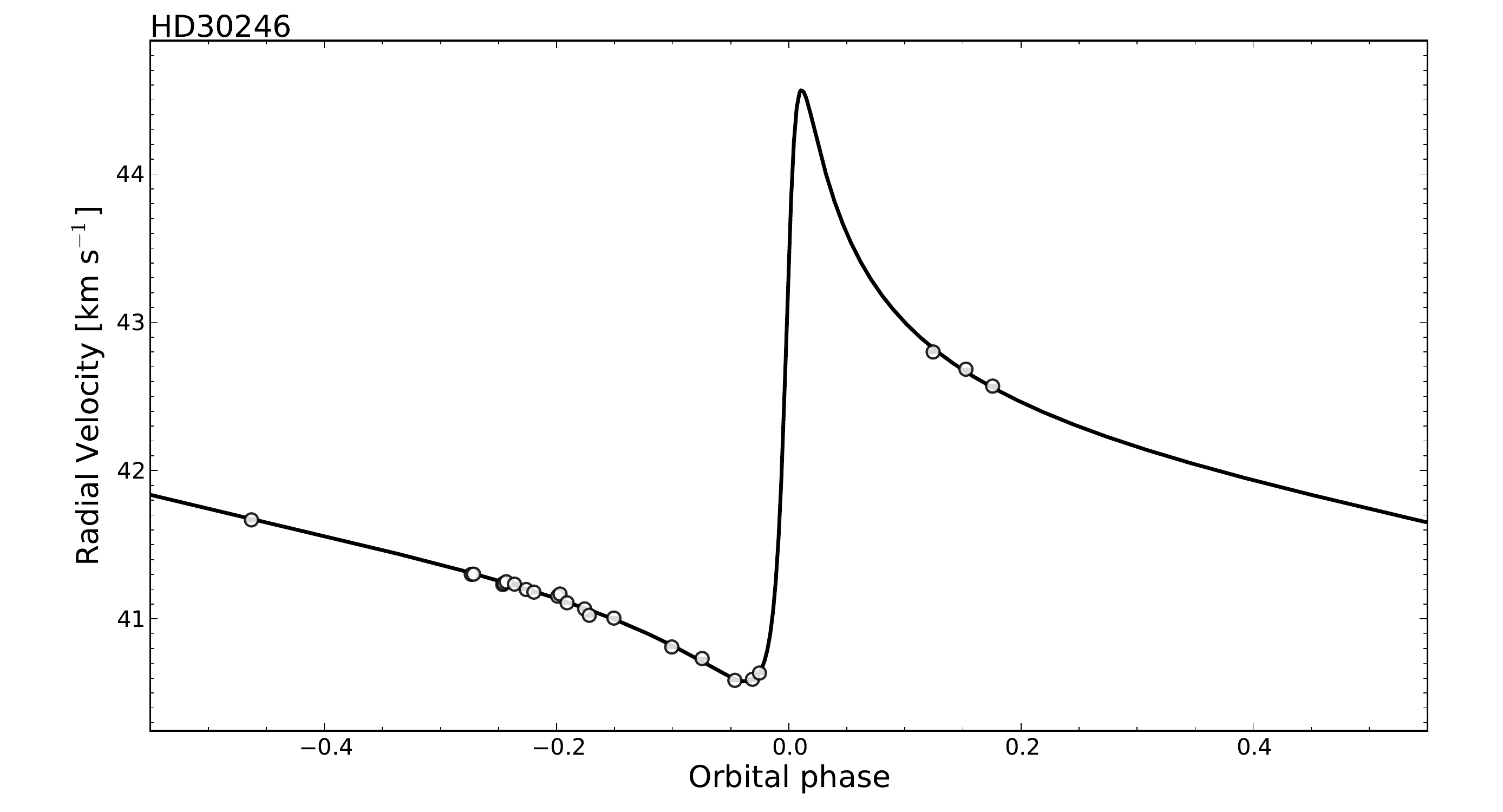}
\caption{Radial velocity measurements obtained with SOPHIE for all objects studied in this paper. In the left column, the measurements are plot as a function of time, while on the right column, they have been folded to the orbital phase. The solid curve represents the best fit to the data and the lower panels on the left column show the residuals to the fit.}
\end{center}
\end{figure*}
\setcounter{figure}{0}
\begin{figure*} 
\begin{center}
\includegraphics[width=0.5\textwidth]{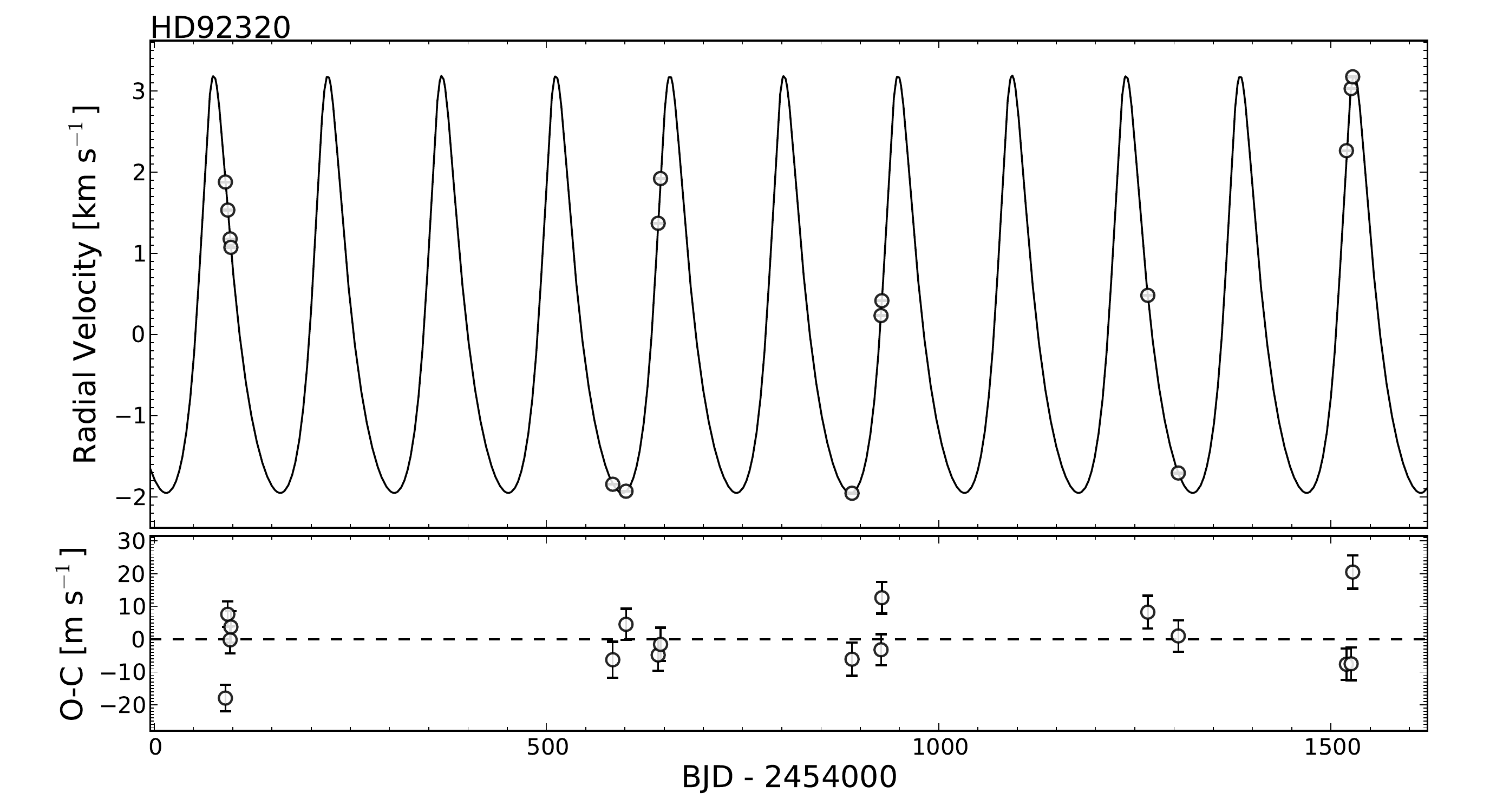}\includegraphics[width=0.5\textwidth]{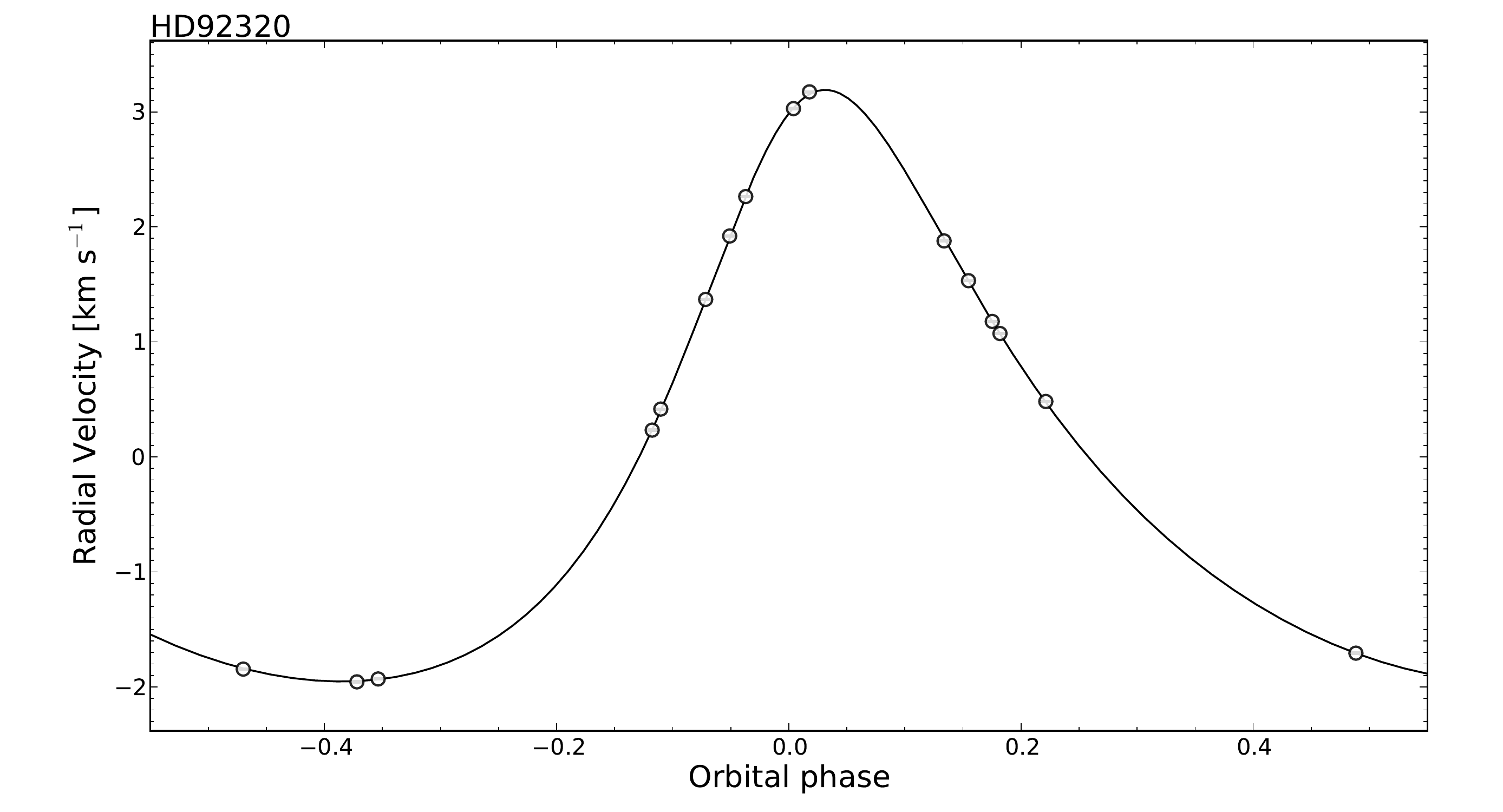}
\includegraphics[width=0.5\textwidth]{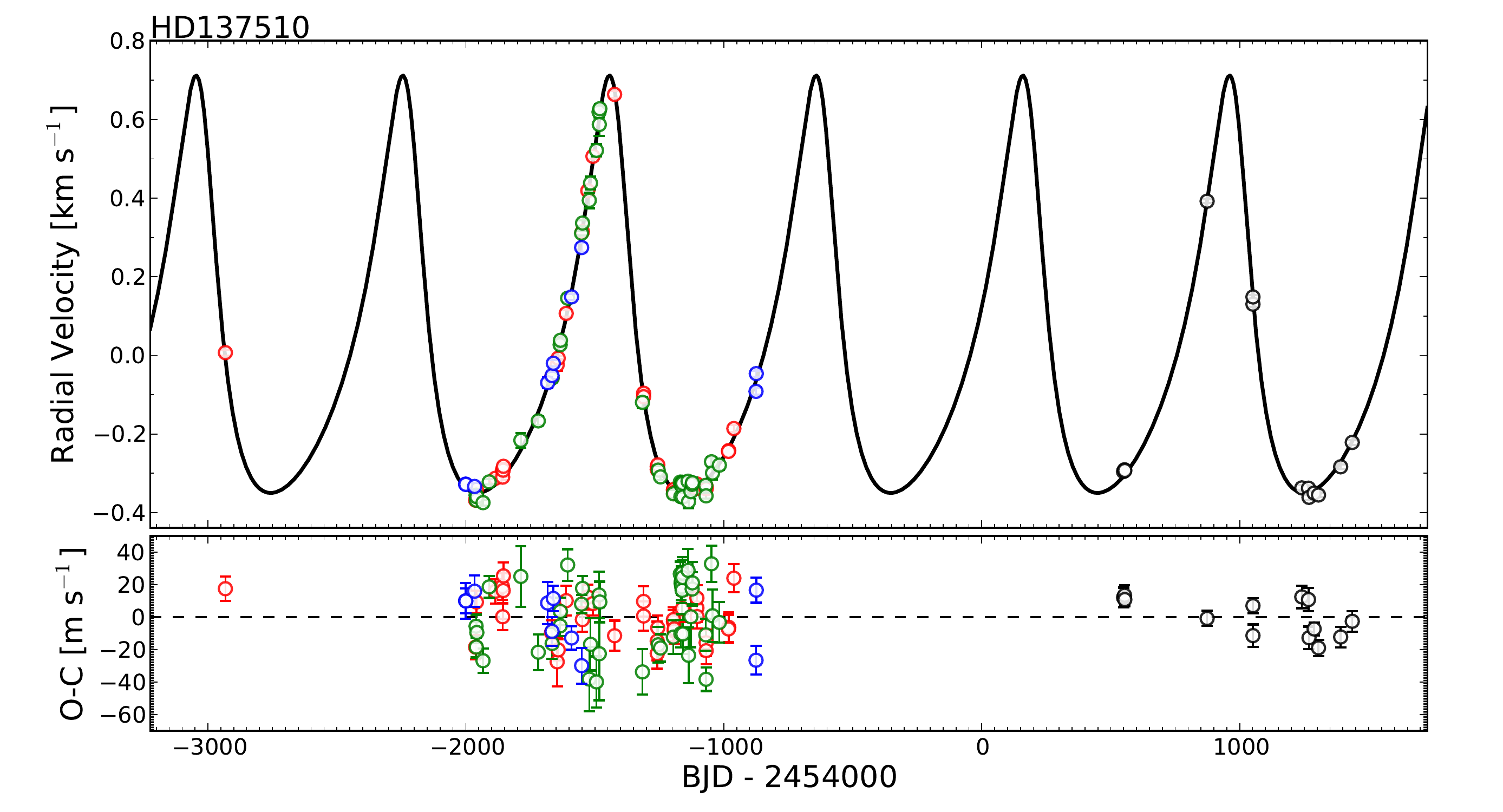}\includegraphics[width=0.5\textwidth]{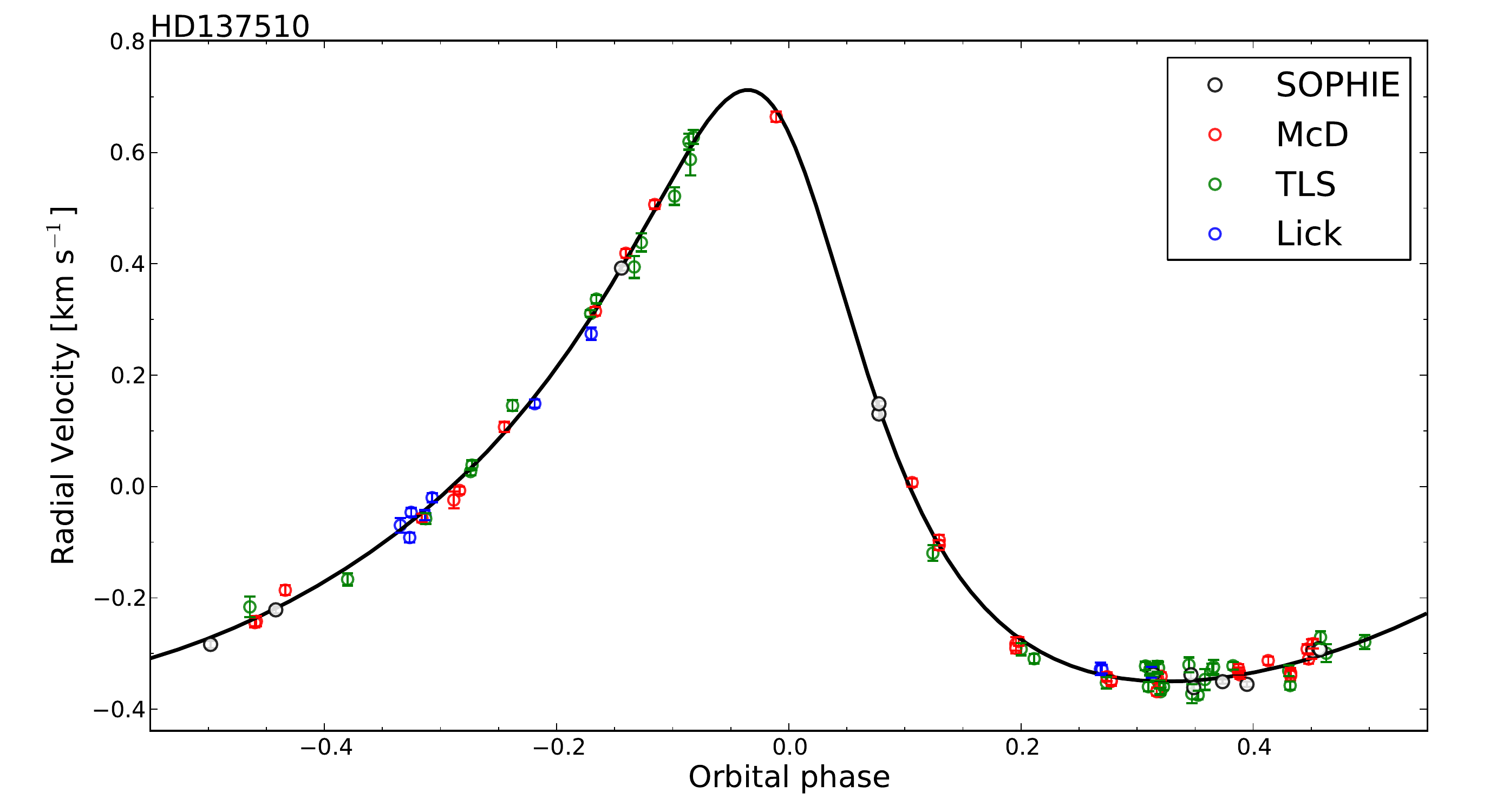}
\includegraphics[width=0.5\textwidth]{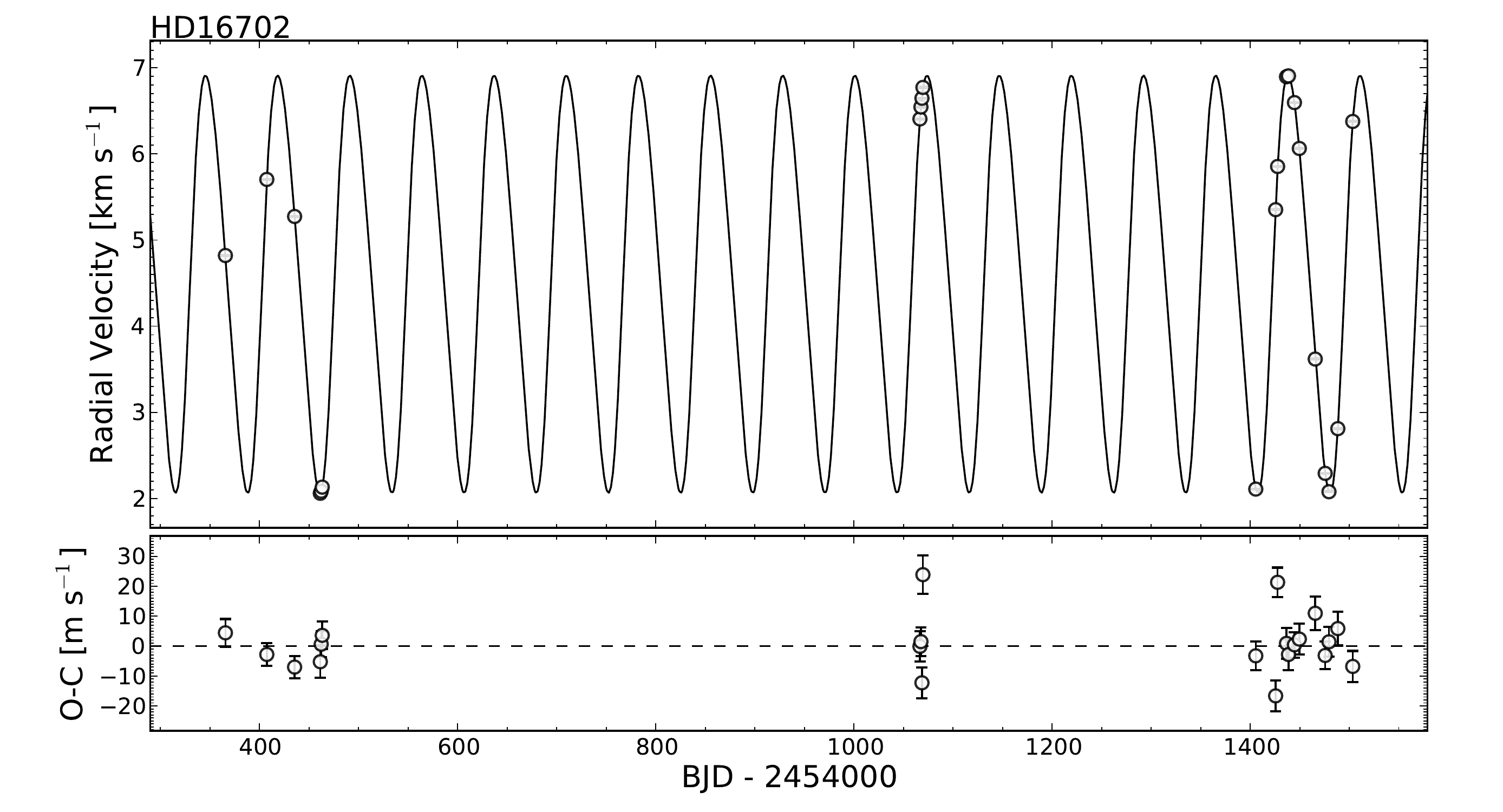}\includegraphics[width=0.5\textwidth]{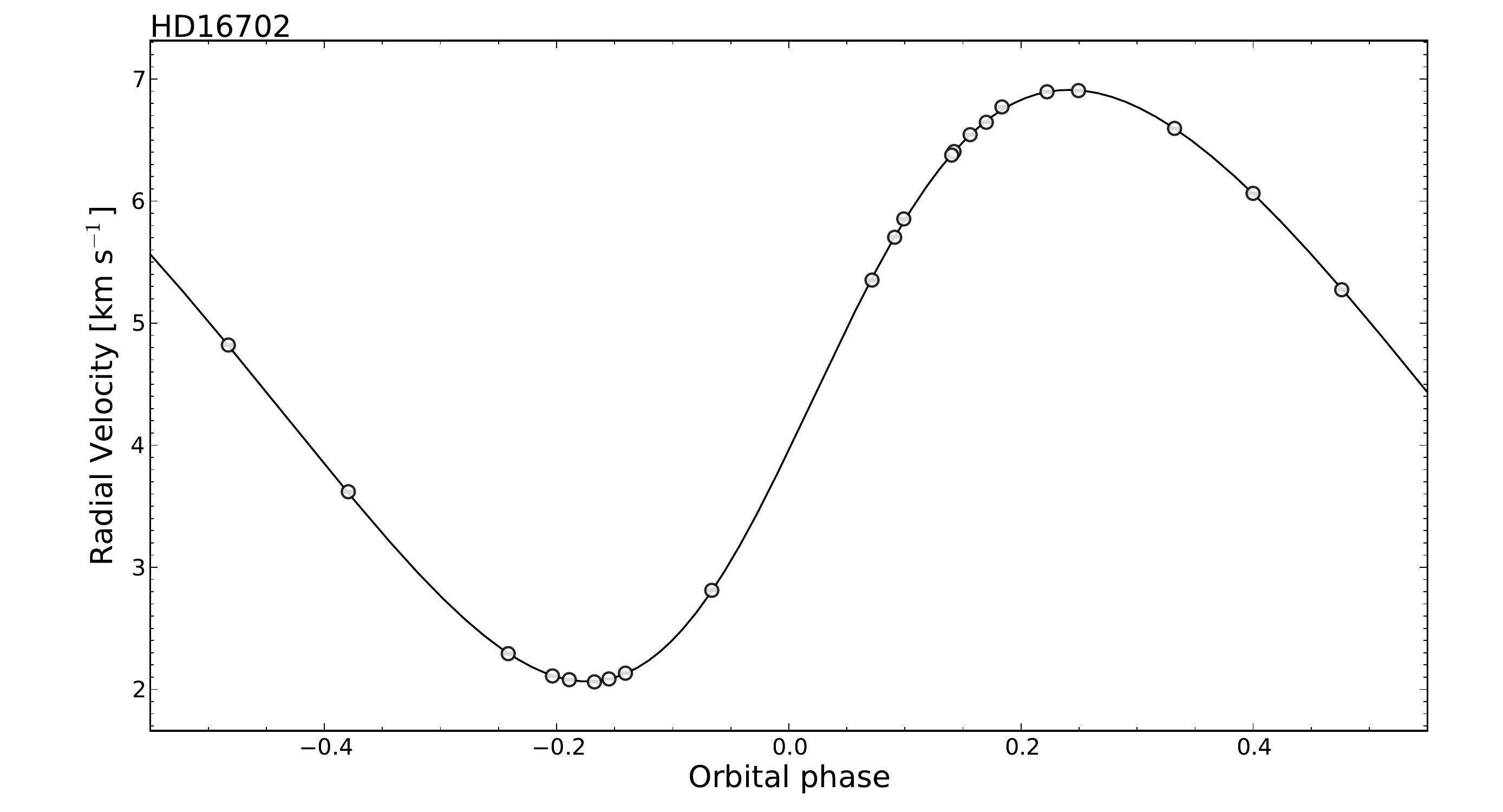}
\includegraphics[width=0.5\textwidth]{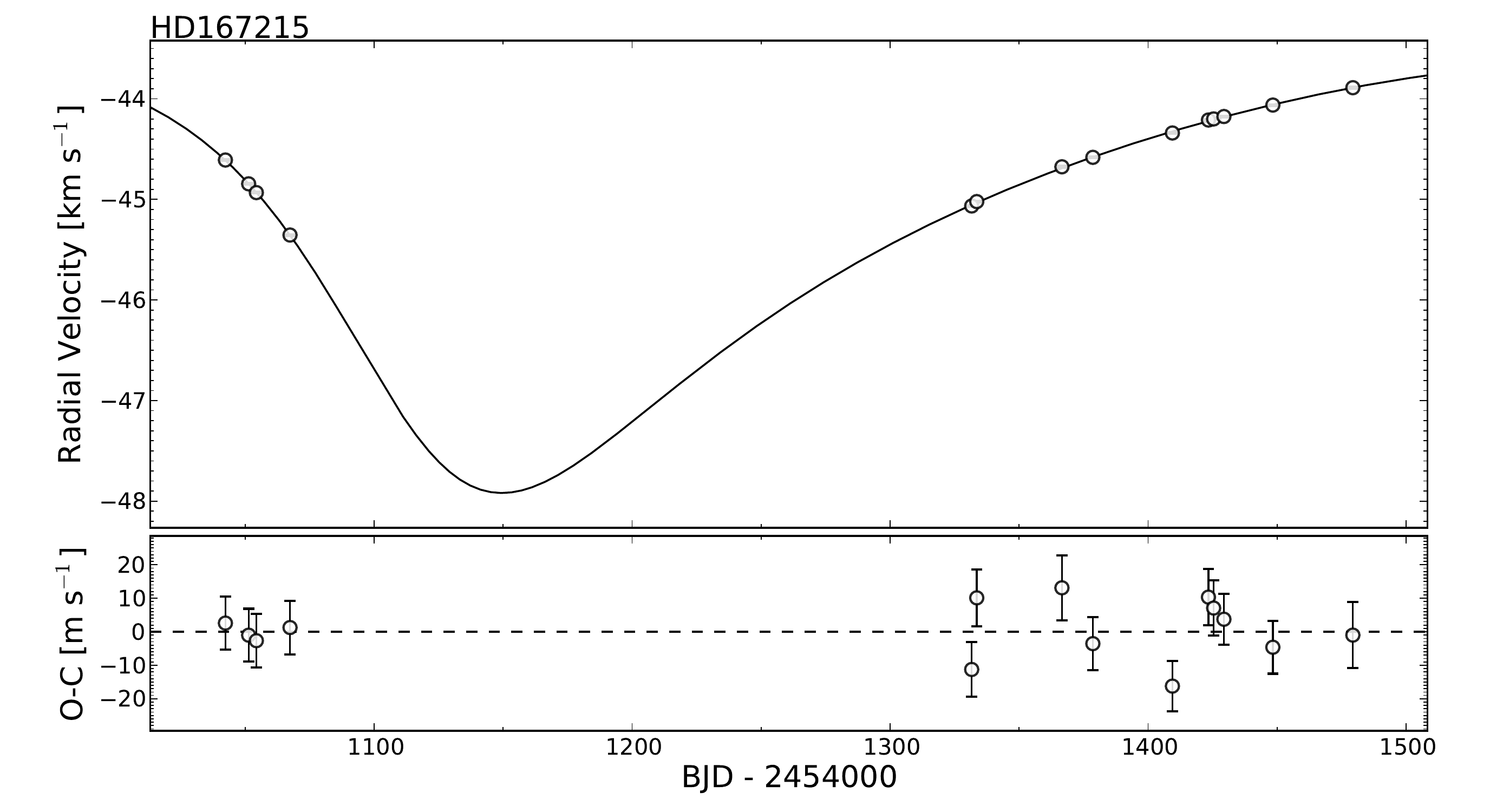}\includegraphics[width=0.5\textwidth]{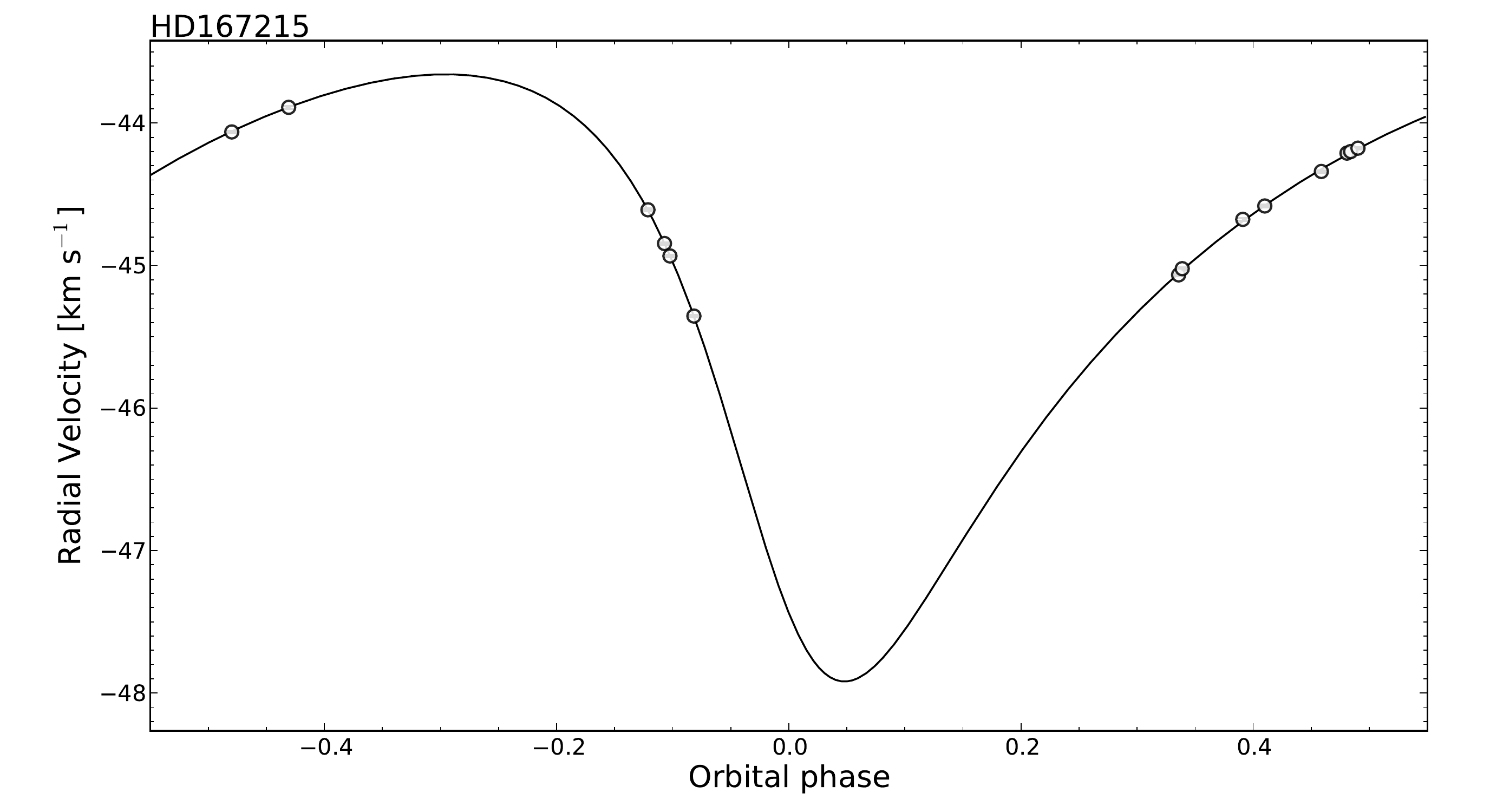}
\caption{Cont. For HD137510, the red, green and blue points correspond to data from McDonald, TLS, and Lick, respectively.  \label{fig.RV}}
\end{center}
\end{figure*}


\section{Results \label{sec.results}}

In this section we discuss the results for each target. We have classified the reported companions roughly according to their minimum mass and evidence for or against them being stellar companions.

\subsection{Planetary candidates \label{sec.planets}}


\subsubsection*{HD22781}
HD22781 (HIP 17187) is a V=8.8 magnitude K0 star located at d=33 parsecs from the Sun, with a stellar mass $\starmass=0.75$ M$_\odot$.  The RV measurements obtained with \sophie\ (Fig.~\ref{fig.RV}) reveal the presence of a sub-stellar companion with minimum mass $M_c \sin i =  14$ \MJ\ in a highly eccentric ($e=0.82$), 528-day period orbit.  Due to the effective temperature of the target, the velocities used to obtained the Keplerian fit are those obtained using the K5 mask.

\begin{figure}[t] 

\begin{center}
\includegraphics[width=\columnwidth, trim=0 0 0 1cm, clip=true]{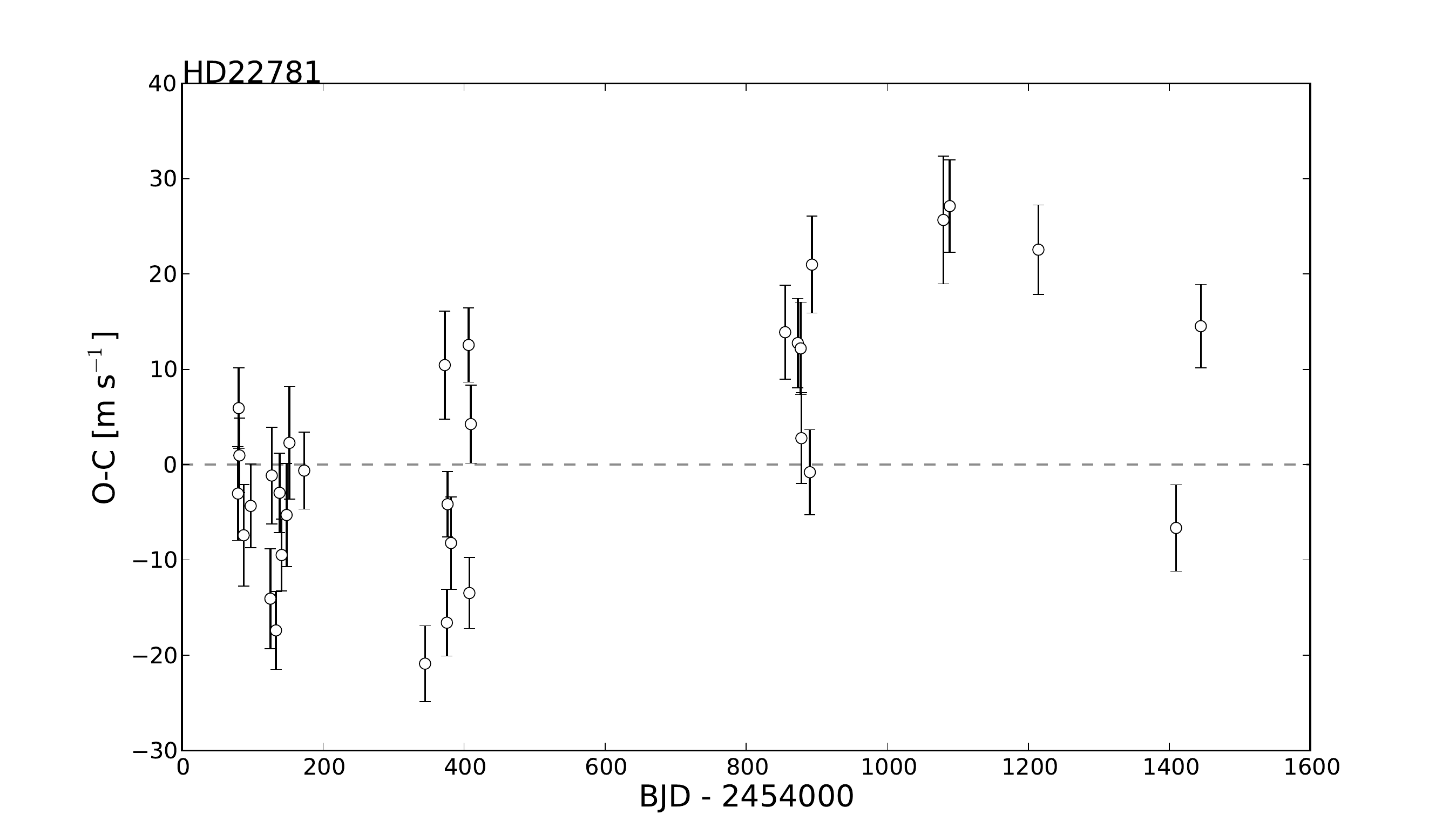}
\caption{Residuals of a single-Keplerian fit to the radial velocities of HD22781, not including any additional drift.\label{fig.22781res}}
\end{center}
\end{figure}

The residuals of a single-Keplerian fit exhibit a scatter of around 12 \ms, largely in excess of the typical measurement error, and too large to be explained by the effect of stellar jitter, estimated to be about 1 \ms\ ($\left<\log R^\prime_{HK}\right>=-5.0$). Furthermore, the residuals seem to exhibit a trend, with a maximum around $\mathrm{BJD}=2\,455\,100$ and a subsequent decrease (see Fig.~\ref{fig.22781res}). In view of this, the RV measurements were also fit using a Keplerian orbit plus a quadratic drift. The best-fit parameters of this solution are reported in Table~\ref{table.parameters}. We note that the parameters corresponding to the Keplerian orbit are consistent at the 1-$\sigma$ level with those derived from a single-Keplerian fit. Concerning the parabolic drift, we can obtain a rough estimate of the possible body producing such a drift by assuming that its orbital period and the RV semi-amplitude correspond roughly to twice the time between the zeros of the parabola, and to its maximum, respectively. In this way the fit coefficients imply a companion with minimum mass around 1.0~\MJ\ in a 9.2-year orbit. A two-Keplerian fit, on the other hand, produces a fit with half this period and a slightly smaller  minimum mass (0.8 \MJ) mass for the second companion, and reduces the scatter in the residuals to about 7 \ms, which is comparable to the scatter level of the other targets. However, the fit semi-amplitude has a significance below 2-$\sigma$. We therefore decided to keep the solution with a quadratic drift, and note that this system warrants further observations in order to confirm the presence of a second orbiting body. We therefore warn the reader that the orbital elements reported in Table~\ref{table.parameters} should be taken as preliminary.


The bisector velocity span of these observations is compatible with a constant (reduced $\chi^2$ smaller than unity) and a linear regression of the bisector as a function of RV  yields a flat slope ($b <  0.012$, at 3-$\sigma$ confidence level). These results hold for all spectroscopic masks employed, reinforcing thus the case for a sub-stellar companion. Additionally, no significant difference is observed in the semi-amplitudes obtained using different masks. The astrometric orbit derived from Hipparcos data has a low significance of $22~\%$, but we can set an upper limit of $0.31\,M_{\sun}$ for the mass of the companion, which is increased by the low number of 69 Hipparcos measurements with moderate precision and the high eccentricity of the system. If the secondary is a slow-rotating star, this constraint can be slightly improved by means of the CCF simulations described in \S\ref{sect.bisector}, which give an upper limit of 0.23 M$_\odot$. In the case of a fast rotating star, the upper limit is similar to that from astrometry.

We conclude that the companion to HD22781 is most likely a sub-stellar object, which could have formed either by core-accretion \citep{mordasini2009} or by gravitational collapse. The term "superplanet" used by \citet{udry2002} for objects of similiar mass seems adequate (see \S\ref{sect.discussion}). The detection of a potential second companion --albeit with low significance-- also points to the planetary nature of HD22781 b, although it is known that circumbinary planets exist \citep{doyle2011}. The dynamical stability of this scenario could be promptly tested by numerical simulations, but this is outside the scope of this paper.  In any case, this system warrants additional RV observations over a longer time period.


\begin{figure}
\begin{center}
\includegraphics[width=\columnwidth, trim=0 0 0 1cm, clip=true]{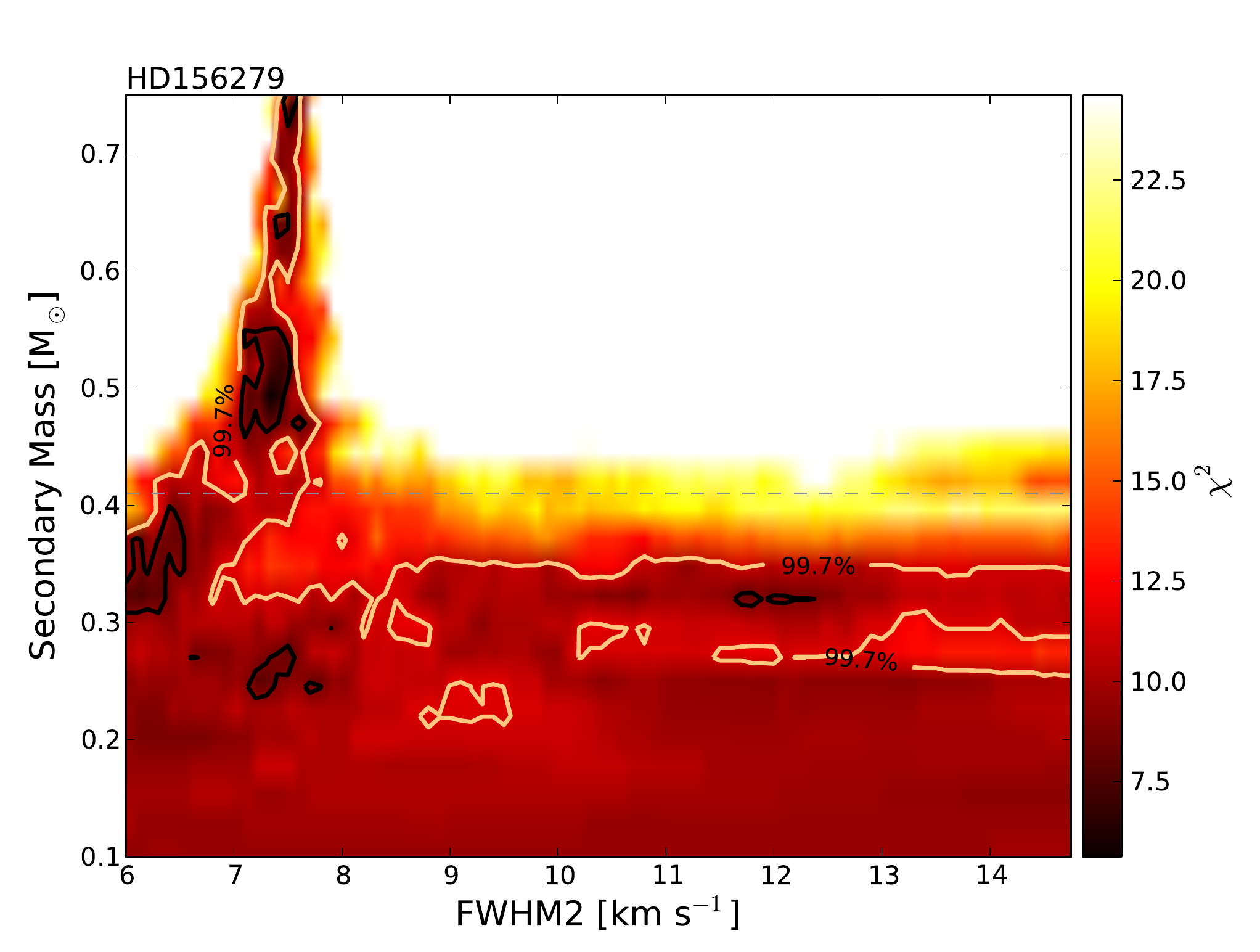}
\caption{Confidence regions for the mass of the secondary and the FWHM of its CCF, issued from numerical simulations of HD156279. The dashed line is the upper limit derived from the astrometric analysis of Hipparcos data. At the 3-$\sigma$ level, the mass of the secondary is constrained by these simulations to $\lesssim 0.35$ M$_\odot$. Astrometric data discard the high-mass family with FWHM $\sim 7$ \kms\ permitted by the bisector data.  \label{fig.blends156279}}
\end{center}
\end{figure}

\subsection*{HD156279}
\object{HD156279} (HIP 84171) is a G6 dwarf star with a mass of $0.93$ \msol. It is located 35 pc  away from the Sun and has a V-band magnitude of 8.08. The RV variations observed with SOPHIE imply the presence of a planetary-mass companion ($M_c \sin i = 9.71 \pm 0.66$ \MJ)  in a 130-day eccentric orbit. The best-fit orbit is shown in Fig.~\ref{fig.RV} and the parameters are reported in Table~\ref{table.parameters}.


Bisector velocity span measurements for this star do not exhibit any significant variation in time, nor any dependence with RV, and present a scatter of 6.5 \ms. The CCF simulations (Fig.~\ref{fig.blends156279}) put an upper limit of roughly 0.35 M$_\odot$, although solutions with higher mass are possible for rotational velocities around the resolution limit of SOPHIE (FWHM $\approx$ 6-7 \kms, depending on spectral type).  However, this family of solutions is discarded by Hipparcos measurements, for which the derived upper mass is $0.41\,M_{\sun}$. Additionally, no mask effect is detected at a significant level, which also supports the hypothesis of a sub-stellar object orbiting HD156279.

HD156279 b is the least massive object reported in this paper, and may have been formed as a planet. Its minimum mass is below the smallest deuterium-burning mass determinations \citep{spiegel2011}. HD156279 b is therefore established as a planetary candidate.

\subsection{Brown dwarf candidates \label{sec.browndwarfs}}

\subsubsection*{HD14651}

HD14651 (HIP 11028) is a V=8.3 magnitude star, with a mass $\starmass = 0.955$ \msol, measured as described in Sect.~\ref{sect.stpar} The Keplerian fit to the RV measurements obtained with \sophie\ (Fig.~\ref{fig.RV}) indicates that this star possesses a companion of minimum mass $M_c \sin i =  47$ M$_\mathrm{J}$ in a 79.4-day orbit, with an eccentricity $e = 0.48$. 

Evidence that the companion is indeed a sub-stellar object comes from the fact that the amplitude of the orbital fit does not depend on the numerical mask used to compute the CCF, and from the bisector analysis, which is compatible with a constant (reduced $\chi^2=0.6$), and does not exhibit any correlation with the RV measurements. Indeed, the Pearson and Spearman correlation coefficients between the RV and the bisector velocity span are -0.11 and -0.14, respectively, and the slope obtained by linear regression is consistent with zero at the 1-$\sigma$ level. Additionally, we do not detect an orbital signature in the Hipparcos astrometry, the significance being $17~\%$. We can set an upper limit of 0.72 M$_{\sun}$ for the mass of the companion, which is high partly due to the low number of Hipparcos measurements ($N_\mathrm{Hip}=59$) with moderate precision and the considerable eccentricity of the system. On the other hand, the CCF simulations constrain the mass of the secondary to be smaller than 0.4 M$_\odot$ if the star is rotating with projected velocity $v \sin I \approx 10$ \kms, and smaller than 0.28 M$_\odot$ if $v \sin I < 3$ \kms.

We conclude that the companion to HD14651 is a \textit{bona fide} brown dwarf candidate, that requires further observations to pinpoint its true nature. For example, astrometric observations from the future GAIA mission, or near-infrared spectroscopy that could detect the peak of the companion, and therefore measure the mass ratio between primary and secondary (see \S\ref{sect.discussion}).


\subsubsection*{HD30246}
HD30246 (HIP 22203) is a V=8.3 magnitude G1 star located at 51 parsecs from the Sun, with a measured mass $\starmass = 1.047$ M$_\odot$.  The RV measurements obtained with \sophie\ exhibit a peak-to-peak variation of $2$ \kms, that can be explained by the presence of a  $M_c \sin i =  56$ M$_\mathrm{J}$ companion in a 990-day orbit, with an eccentricity $e = 0.85$. HD30246 belongs to the Hyades cluster \citep{griffin1988} and has been the object of several studies, discussed in part below. It is also the most active star in the sample, compatible with its probably young age \citep{holmberg2009}.

\begin{figure*} 
\begin{center}
\includegraphics[width=0.85\columnwidth, trim=0 0 0 1cm, clip=true]{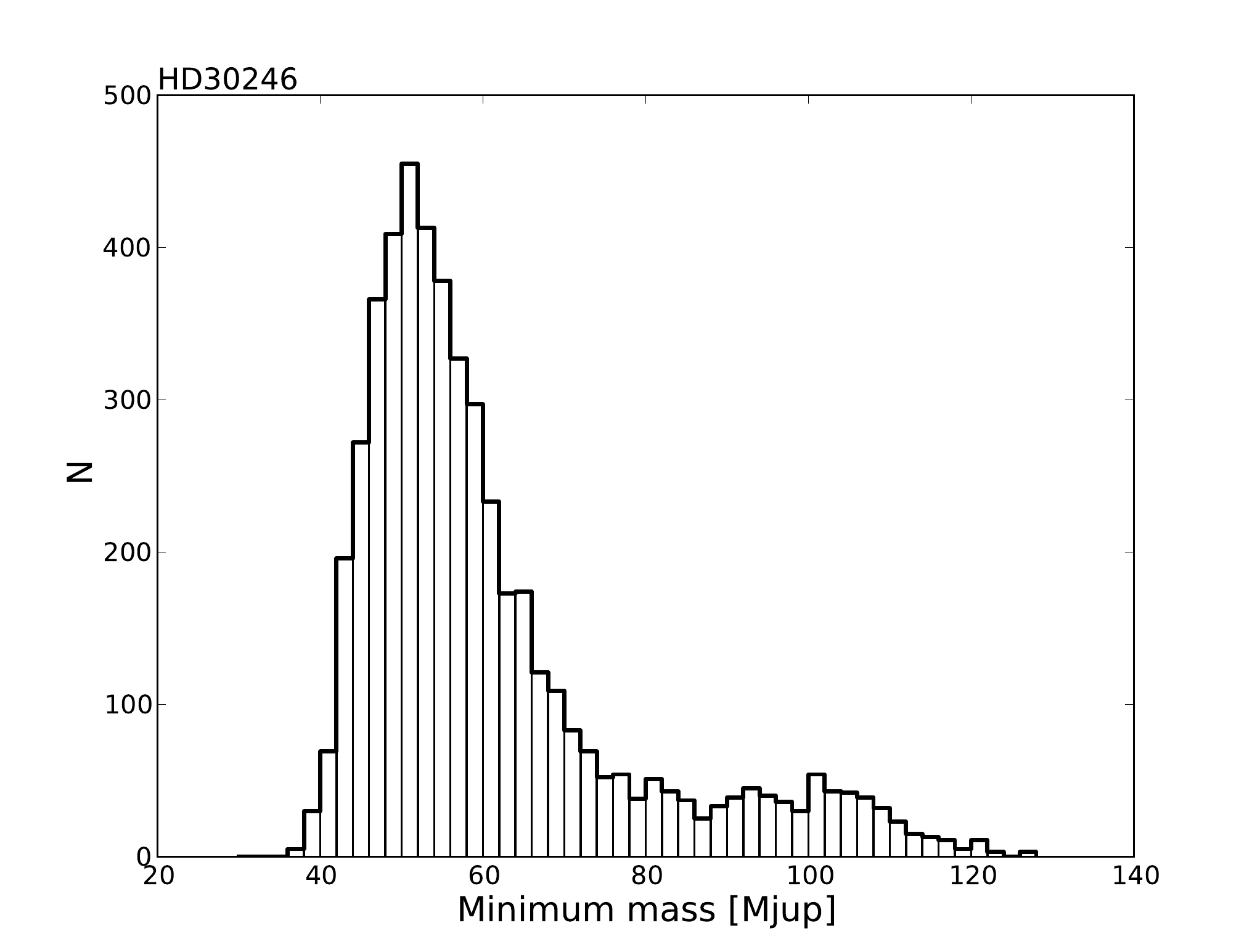}
\includegraphics[width=0.85\columnwidth, trim=0 0 0 1cm, clip=true]{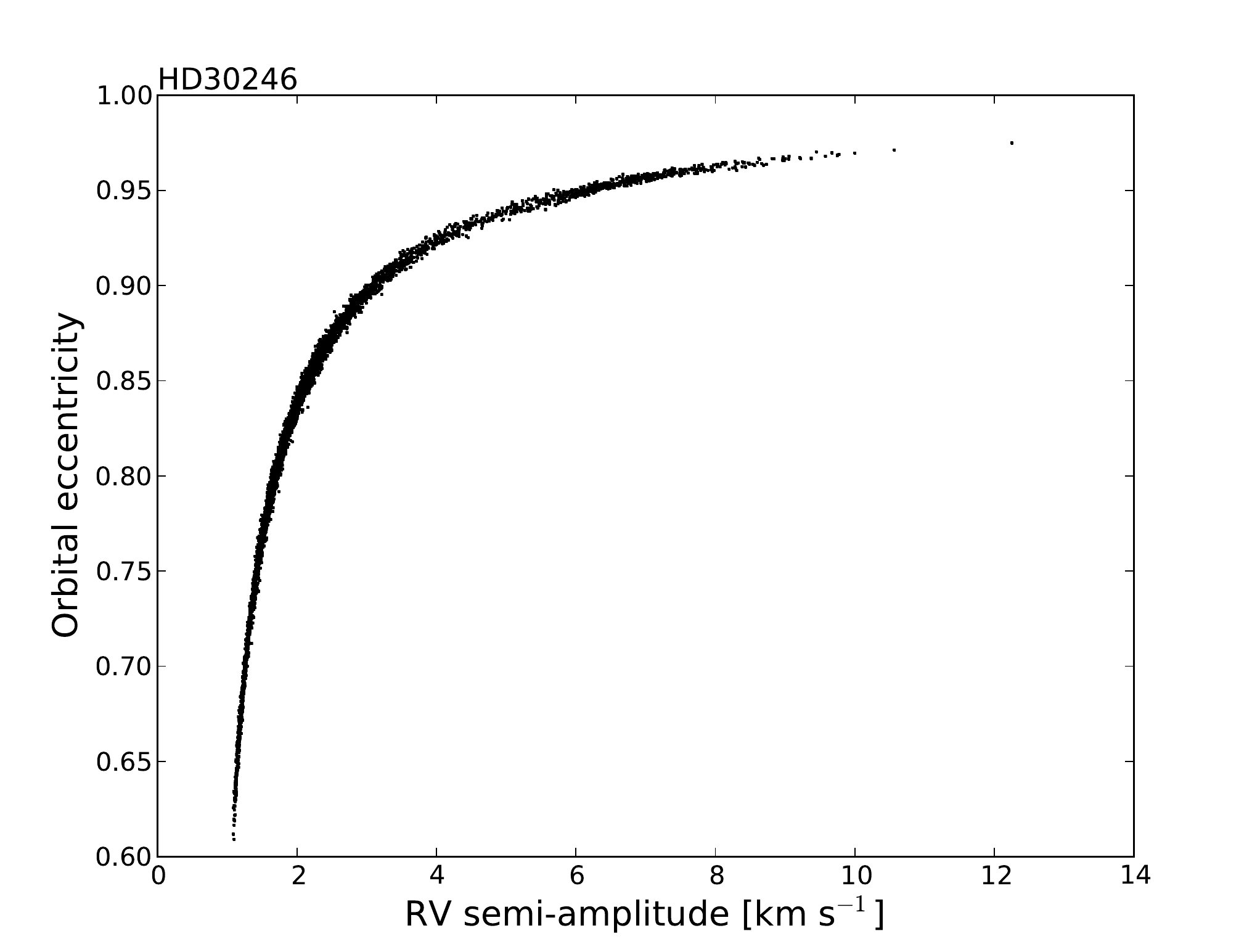}
\caption{\emph{Left:} Minimum mass distribution for HD30246, obtained from 5000 Monte Carlo simulations. The bimodal behaviour is clearly seen, and allows for stellar-mass solutions. \emph{Right}: Correlation between orbital eccentricity and RV semi-amplitude from the same Monte Carlo simulations, showing that high-mass solutions correspond to very eccentric orbits. \label{fig.bimodal30246}}
\end{center}
\end{figure*}

For this target, observations do not cover one of the extrema of the RV curve (see Fig.~\ref{fig.RV}). As a consequence, the Monte Carlo distribution of the minimum mass of the companion has a long high-mass tail, and a slightly bimodal behaviour, allowing for a family of solutions corresponding to a low-mass star (Fig.~\ref{fig.bimodal30246}). The distributions of the semi-amplitude $K$ and the orbital eccentricity $e$ exhibit a similar feature, with the high-mass solutions corresponding to high-eccentricity orbits. Since the value of the $\chi^2$ statistics is not systematically higher for any value of the eccentricity or the semi-amplitude $K$, the identification of the correct solution is hindered. We note, however, that less than 14\% out of the 5000 Monte Carlo realisations of the data produce an orbital solution with minimum mass above 80 \MJ. The best fit orbital solution is shown in Fig.~\ref{fig.RV}, and the obtained parameters  obtained using mask G2 are reported in table~\ref{table.parameters}.

\begin{figure}
\begin{center}
\includegraphics[width=\columnwidth, trim=0 0 0 1cm, clip=true]{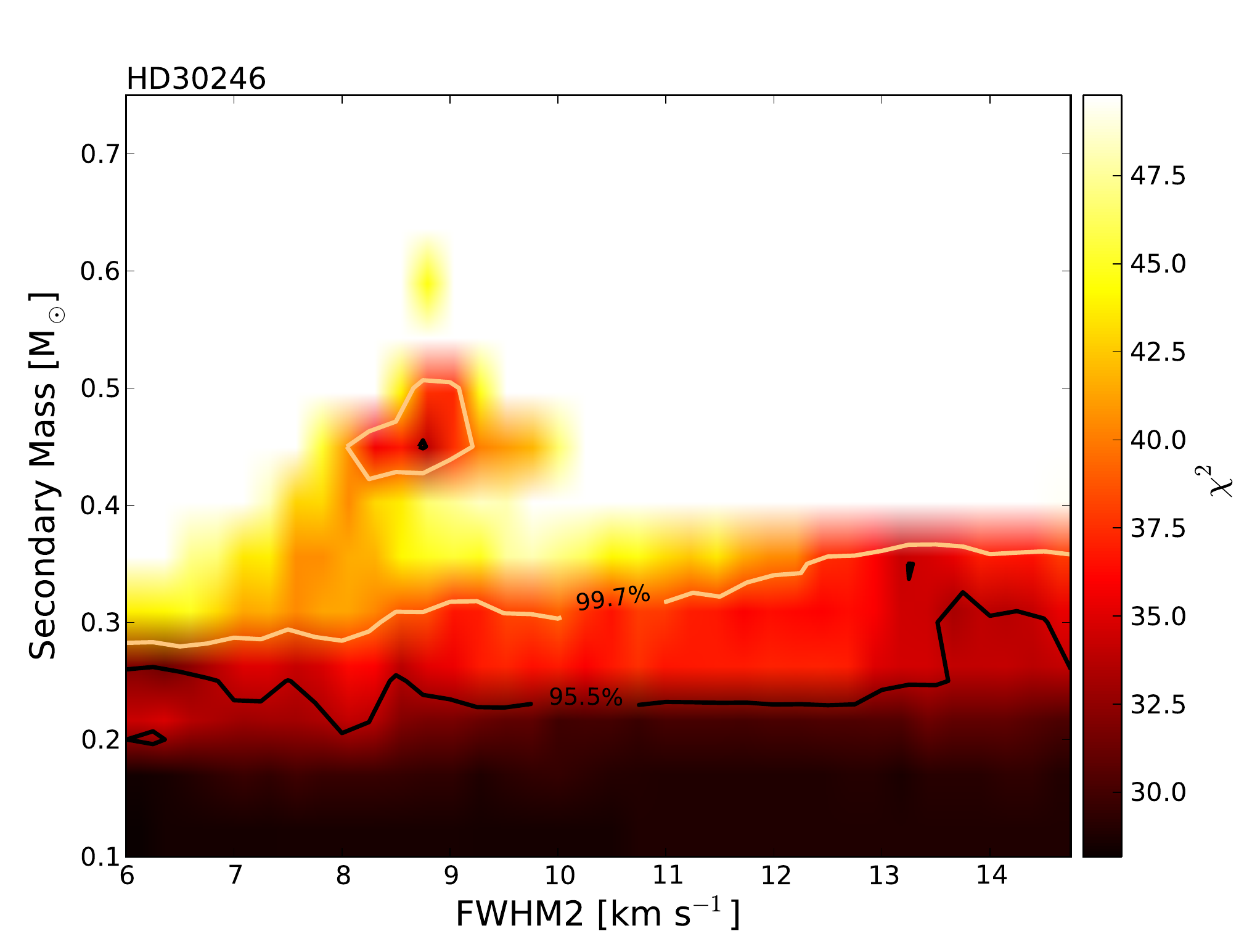}
\caption{Confidence regions for the mass of the secondary and the FWHM of its CCF, issued from numerical simulations of HD30246. At the 3-$\sigma$ level, in addition to the upper limit at around 0.33 M$_\odot$, there exists a solution with $\sim 0.45$ M$_\odot$. \label{fig.blendsim30246}}
\end{center}
\end{figure}

The bisector velocity span is compatible with a constant, exhibiting a dispersion of around 16 \ms. Also, no relation is detected with the RV above the 1.8-$\sigma$ level ($b = -0.0059 \pm 0.0034$), and the amplitudes of the RV variation measured with different masks are in agreement ($K_{G2}=1995.1\pm4.1$ \ms, $K_{K5}=1996.8\pm4.6$ \ms, $K_{M4}= 2005\pm16$ \ms). Additionally the large scatter in the fit residuals (around 13 \ms) can be partially explained by the stellar jitter of the host star, which is estimated to be around 10 \ms, given the median value of activity index $\left<\log R^\prime_\mathrm{HK}\right>= -4.50$ \citep{santos2000}. Also, the analysis of the Hipparcos astrometry provide orbits with low significance (87.6~\%, or $1.5\,\sigma$), meaning that the orbital signature is not detected. However, because the orbit is not covered and in addition is highly eccentric ($e\sim0.8$), we refrain from deriving an upper limit to the companion mass\footnote{Even when taking the obtained solutions at face value, the derived mass limits of $\sim$$50 - 400\,M_J$ would not allow us to establish the nature of the companion.}. On the other hand, the CCF simulations described in \S\ref{sect.bisector} allow us to set an upper limit at the 3-$\sigma$ level of 0.28 M$_\odot$ (respectively, 0.37 M$_\odot$) for simulations in the slow-rotating (respectively, fast-rotating) extreme of the explored parameter range. Additionally, a possible solution is found for M$_\mathrm{c}$ in the interval $[0.42-0.51] $ M$_\odot$, with rotational velocity $v\sin I$ between 3 \kms and 4.5 \kms, at the 3-$\sigma$ level (see Fig.~\ref{fig.blendsim30246}).

HD30246 has been reported to be a spectroscopic binary \citep{griffin1988}, based on 13 RV measurements with a precision of around 700 \ms. Additionally, signs of the spectrum of the secondary were found in 4 infrared spectra obtained with the NASA Infrared Telescope Facility by \citet{bendersimon2008}. However, according to these authors, the detection of a secondary peak in the CCFs of this target is challenging, and leads to poorly-constrained velocities for the secondary (see \S5.3 and Figure 4 of their paper). The derived mass ratio $q = 0.06 \pm 0.04$ implies a mass $M_\mathrm{p}	= (66 \pm 45)$ M$_\mathrm{Jup}$ for the companion, when the stellar mass derived in this paper is considered, $M_\mathrm{s} = 1.05 \pm 0.11$ M$_\odot$\footnote{A 10\% error in the stellar mass is used, rather than the formal error reported in Table~\ref{table.stellarparam}}. This is in agreement with our determination for the minimum mass of the companion. Additionally, the orbital parameters of HD30246 b, obtained by Stefanik \& Latham and reported by \citet{bendersimon2008}, agree with the parameters we derived from SOPHIE data. Furthermore, K-band speckle interferometry \citep{patience1998} provides a limiting magnitude difference for companions at 0.05" from HD30246 of $\Delta K = 4.6$. Considering the $V-K$ relations for main-sequence stars \citep{cox2000} and the mass-luminosity relations from \citet{delfosse2000}, this translates roughly to a maximum mass of 0.19 $M_\odot$ for companions to HD30246 at this orbital distance, which is slightly larger than the distance obtained from the Keplerian fit to the data, $a = 0.04''$.

We conclude that the initial claim of binarity by \citet{griffin1988} warrants revision, and that the remaining facts support a sub-stellar candidate companion to HD30246. The upper limit for the mass is however in the stellar regime, and therefore further astrometric observations will be needed to decide unambiguously on the nature of this object, as well as a continuing RV monitoring, in order to better constrain the orbital parameters.

\begin{figure*}[t]
\begin{center}
\includegraphics[width=0.7\columnwidth, trim=0 0 0 1cm, clip=true]{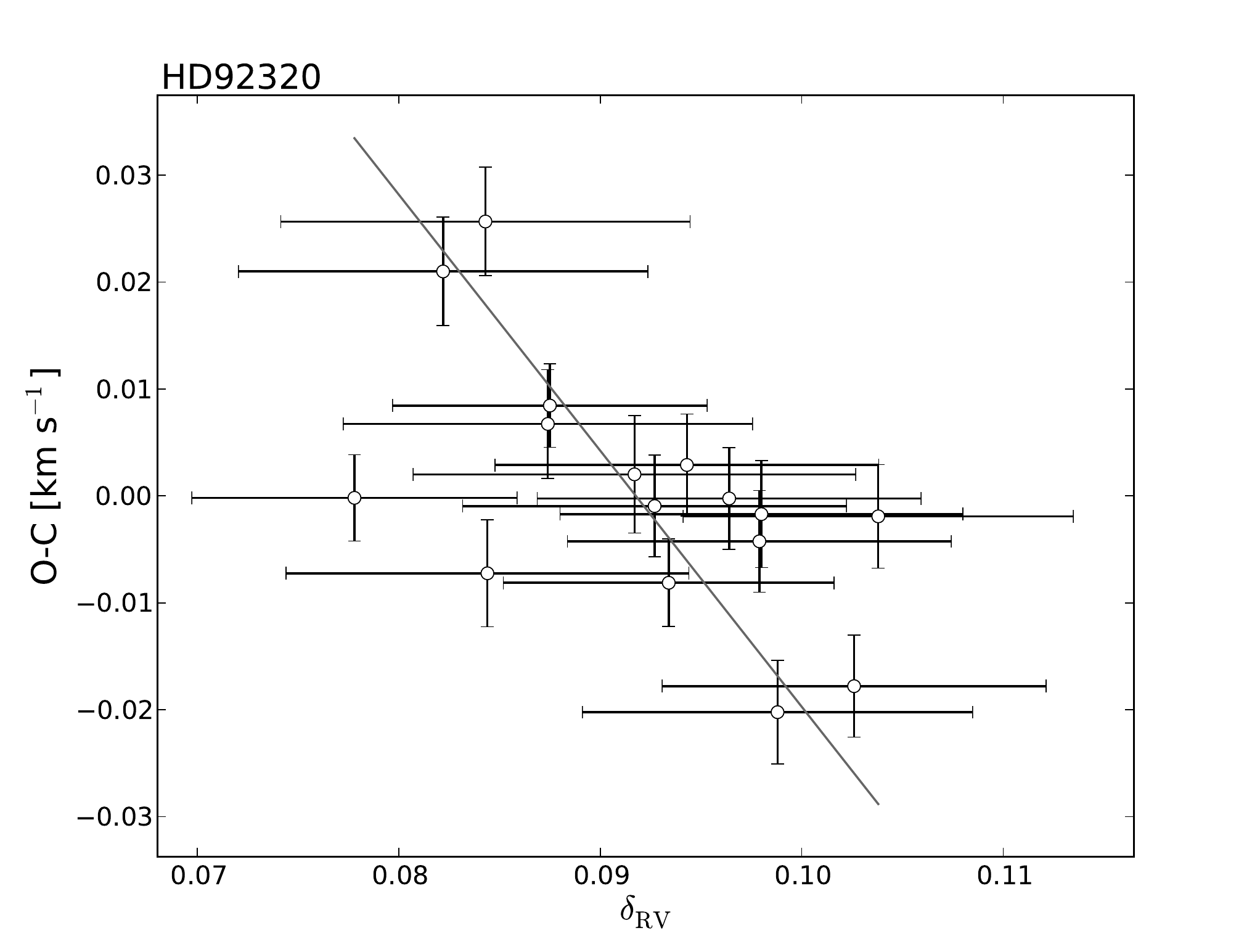}\includegraphics[width=0.7\columnwidth, trim=0 0 0 1cm, clip=true]{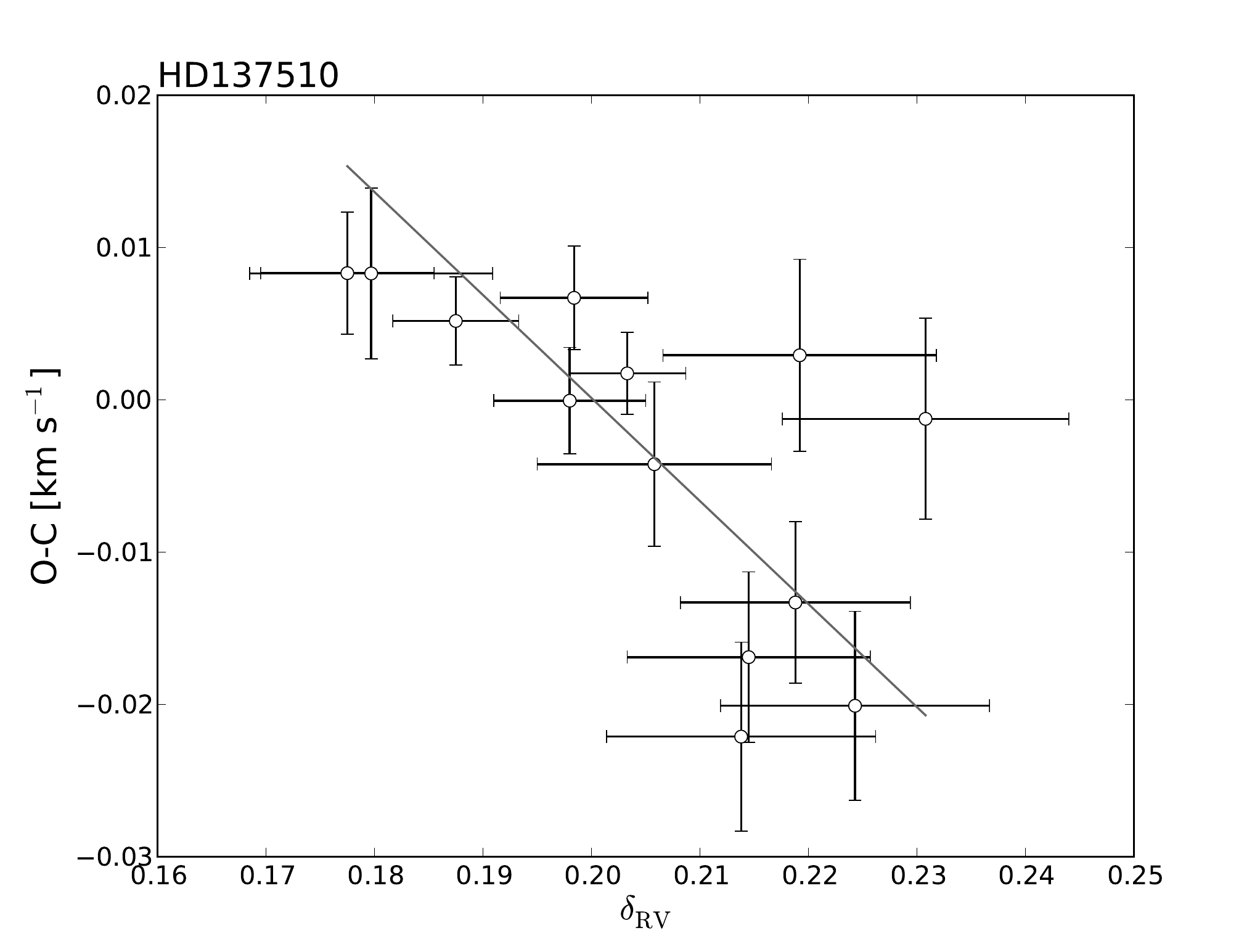}\includegraphics[width=0.7\columnwidth, trim=0 0 0 1cm, clip=true]{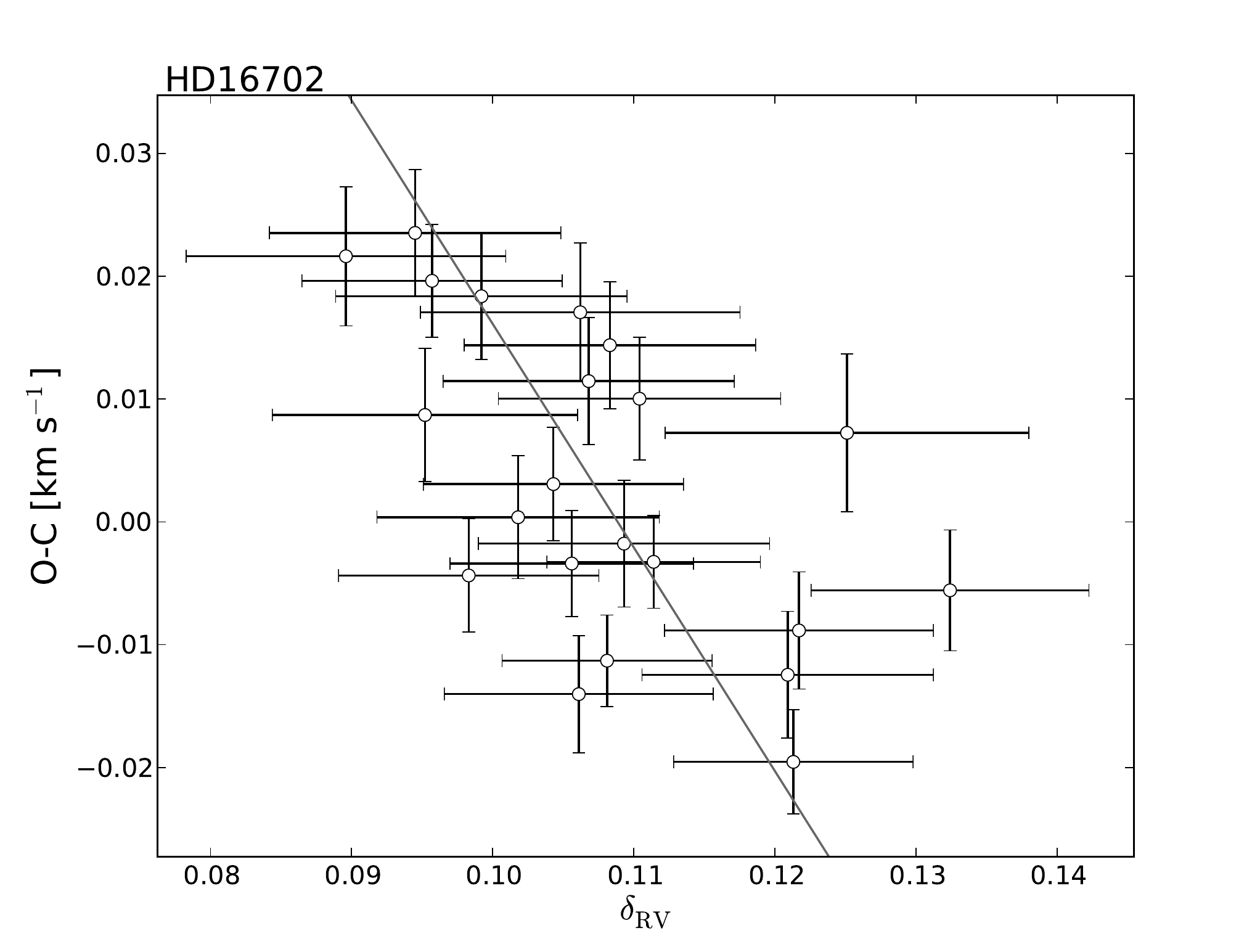}
\caption{Residuals of the Keplerian orbital fit to the uncorrected data as a function of pupil illumination indicator $\delta_{RV}$ for the three stars in our sample showing a significant ($r>0.5$) correlation.  The line is a fit to the data, and was used to correct the radial velocities, which leads to a reduced scatter. \label{fig.resplots}}
\end{center}
\end{figure*}


\subsection*{HD92320}
\object{HD92320} (HIP 52278) is a V=8.38 magnitude star located at 41 parsecs from the Sun. The 16 RV measurements obtained since late 2006 are fitted by a 145-day Keplerian orbit with a semi-amplitude of about 2.6 \kms, and an eccentricity of 0.32. The spectral analysis coupled to the comparison with evolutionary models yields a stellar mass of $\starmass=0.916$ M$_\odot$, and a companion minimum mass $M_c \sin i =  59.5$ \MJ.

The Keplerian fit residuals exhibit a weak correlation with pupil illumination proxy $\delta_{RV}$ (see \S\ref{sect.seeing}), with Pearson and Spearman correlation coefficients -0.60 and -0.66, respectively.  The trend is given mainly by the four points with largest absolute residuals (see Fig.~\ref{fig.resplots}). These measurements do not exhibit unusual signal-to-noise ratios, CCF contrast or FWHM. We have therefore concluded that the relation is real and consequently, the detrended measurements were used to obtain the orbital fit plotted in Fig.~\ref{fig.RV} and the parameters reported in table~\ref{table.parameters}. 

The lack of correlation between bisector velocity span and RV\footnote{The Pearson's and Spearman's coefficients are -0.083 and -0.144, respectively}, together with the absence of mask effect argues for a substellar companion as the source of the periodic RV variations. Additionally, we did not detect an orbital signature in the Hipparcos data, but we could set an upper limit of $0.34\,M_{\sun}$ for the mass of the companion. The CCF simulations described in \S\ref{sect.bisector} provide a similar upper limit to the mass: at the 99.7\% confidence level, the upper limit to the true mass of the companion ranges from 0.26 M$_\odot$, for slow rotators below the detection limit up to 0.35 M$_\odot$ for the fastest rotating stars simulated ($v \sin I = 10$ \kms). The companion to HD92320 is therefore established as a brown dwarf candidate that merits further observations. GAIA measurements should easily detect the reflex motion induced by this object (\S\ref{sect.discussion}), and near-infrared spectroscopy might permit detecting the spectrum of the secondary.


\subsection*{HD137510}


\object{HD137510} (HIP 75535) is a high-metallicity $1.36$-\msol\ sub-giant late-F star, located 42 pc away from the Sun. Its V-band magnitude is 6.26, making it the brightest star in our sample. A brown dwarf companion candidate with a minimum mass of 26 \MJ\ was discovered by \citet{endl2004}, based on observations acquired at the McDonald Observatory (McD) in Texas, USA and at Th\"uringer Landessternwarte Tautenburg (TLS), in Germany. Later, \citet{butler2006} published an independent orbital solution based on 10 RV measurements obtained at Lick Observatory, which led to a slightly less massive object of 22.7 \MJ, in a mildly less eccentric orbit. Thirteen new measurements of this star were obtained with SOPHIE, which permit improving the system parameters. Indeed, the determination using the combined datasets has reduced uncertainties, particularly in the orbital period and in the eccentricity.

The SOPHIE observations were corrected as described in \S\ref{sect.seeing} for the seeing effect. The Pearson's and Spearman's coefficients between the residuals and $\delta_\mathrm{RV}$ are -0.67 and -0.70, respectively (Fig.~\ref{fig.resplots}). The dispersion of the residuals of the Keplerian fit to the SOPHIE data is improved by this procedure: from around 8.6 \ms\ for the uncorrected data to 4.6 \ms. The fit to the combined data of all four instruments is also improved, albeit less drastically: the scatter of the residuals is reduced from 15.2 \ms to 14.4 \ms. In Table~\ref{table.parameters} we report the parameters obtained from the combined fit of all four datasets, and the best fit curve is shown in Fig.~\ref{fig.RV}. 

Concerning the stellar parameters, we find an effective temperature $\mathrm{T_{eff}}=6130 \pm 50$ K, in good agreement with the value by \citet{lebre99}, and about 160 K (3.1-$\sigma$) and 230 K (2.8-$\sigma$) larger than the value reported by \citet{valentifischer2005} and \citet{endl2004}, respectively. These discrepancies might be due to unaccounted systematic errors in the different methods and models used. The metallicity found as described in Sect.~\ref{sect.stpar} ([Fe/H] = $0.38\pm0.13$) is in agreement with previous determinations, and agrees also with the metallicity value obtained from the CCF as described in \S\ref{sect.stpar}.

The scatter in the bisector velocity span measurements is below 16 \ms, and produced mainly by one point, which if left out reduces the scatter to around 12 \ms, consistent with the typical error of the RV data ($\sim$5 \ms). Additionally, no correlation with the RV measurements above the 1.5-$\sigma$ level is present, and there is no significant difference in the RV amplitudes measured through different numerical masks. The results of the CCF simulations are compatible with a substellar object at the 3-$\sigma$ level, yet another solution with $M_c \approx 0.5$ M$_\odot$ is also possible. This last solution is discarded by the astrometric analysis. Indeed, the substellar nature of HD137510 b has already been established by \cite{sahlmann2011} and later confirmed by \cite{reffert2011}. \cite{sahlmann2011} derived an upper companion-mass limit of $64.4$ \MJ\ using Hipparcos astrometry and the spectroscopic elements of \cite{butler2006}. On the basis of the updated RV solution presented here, we find a similar result: the astrometric signature is not detected, but because of the updated orbital elements and the slightly lower stellar mass, we obtain an updated upper limit of $59.5$ \MJ\ for the mass of the companion, which confirms that HD137510 b is a brown dwarf companion orbiting an F-star.

\subsection{Stellar companions \label{sec.stars}}


\subsubsection*{HD16702}


\begin{figure}
\begin{center}
\includegraphics[width=\columnwidth, trim=0 0 0 1cm, clip=true]{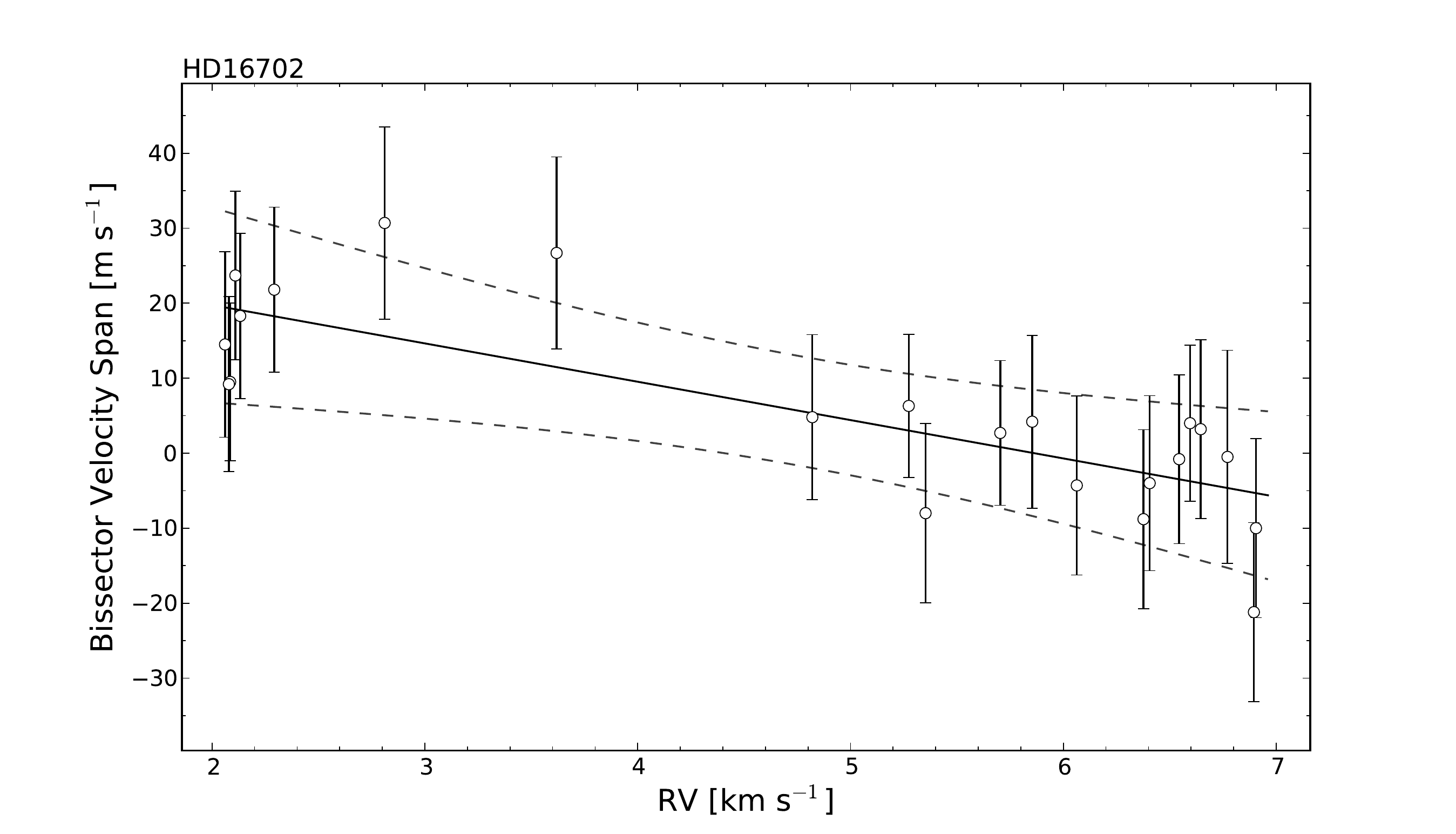}
\caption{Bissector velocity span and radial velocities for HD16702 (corrected for the seeing effect). The solid line represents the best fit to the data and the dashed lines indicate the 3-$\sigma$ confidence region. A correlation is detected to 4-$\sigma$ significance, which  hints to the presence of an additional blended star in the fiber of the spectrograph. \label{fig.bissector16702}}
\end{center}
\end{figure}

HD16702 (HIP 12685) is a V=8.3 magnitude G0 star located at 45 parsecs from the Sun. The orbital fit to the RV data (Fig.~\ref{fig.RV}) indicates that this star possesses a companion  in a 72.8-day orbit, with an eccentricity $e = 0.14$ and minimum mass $M_c\sin i =  48.7$ \MJ, given a stellar mass $\starmass = 0.98$ M$_\odot$.

The residuals of the keplerian fit exhibit a clear correlation with $\delta_{RV}$ (see Figure.~\ref{fig.resplots}), which allowed us to correct the data, as described in \S\ref{sect.seeing}. When this is done, the scatter  in the fit residuals is reduced from 12 \ms\ to the final value of $\approx$ 8 \ms. The final fit parameters are listed in Table~\ref{table.parameters}, and the best-fit curve is plotted in Fig.~\ref{fig.RV}.

The bisector analysis for this target shows a significant correlation between bisector velocity span and radial velocity, shown in Fig.~\ref{fig.bissector16702}. A linear regression to the data shows that the slope $b$ is significantly different from zero, $b=-5.1\pm1.3$, i.e.\ a detection at the 4-$\sigma$ level. Additionally, the amplitude of the RV variations depends on the spectral mask used to compute the CCF. Indeed, we find $K_{G2}=2423.7\pm3.0$ \ms, $K_{K5}=2412.0\pm5.2$ \ms, and $K_{M4} = 2344.0\pm9.5$ \ms, which implies a 8-sigma difference between mask G2 and mask M4. These facts indicate that the companion around HD16702 might be a star in a low inclination orbit. The spectra obtained would then consist of the addition of the spectrum of the solar-like primary star and the secondary star, probably a low-mass star, moving in anti-phase in velocity space. When correlated with the M4 mask, the relative weight given to the spectrum of the secondary is increased, and the resulting CCF is then skewed towards the velocity of the secondary, which produces a decrease in the measured amplitude of the RV variations.

\begin{figure}\begin{center} 
\includegraphics[width= 0.8\linewidth, trim= 0 0cm 0 0cm,clip=true]{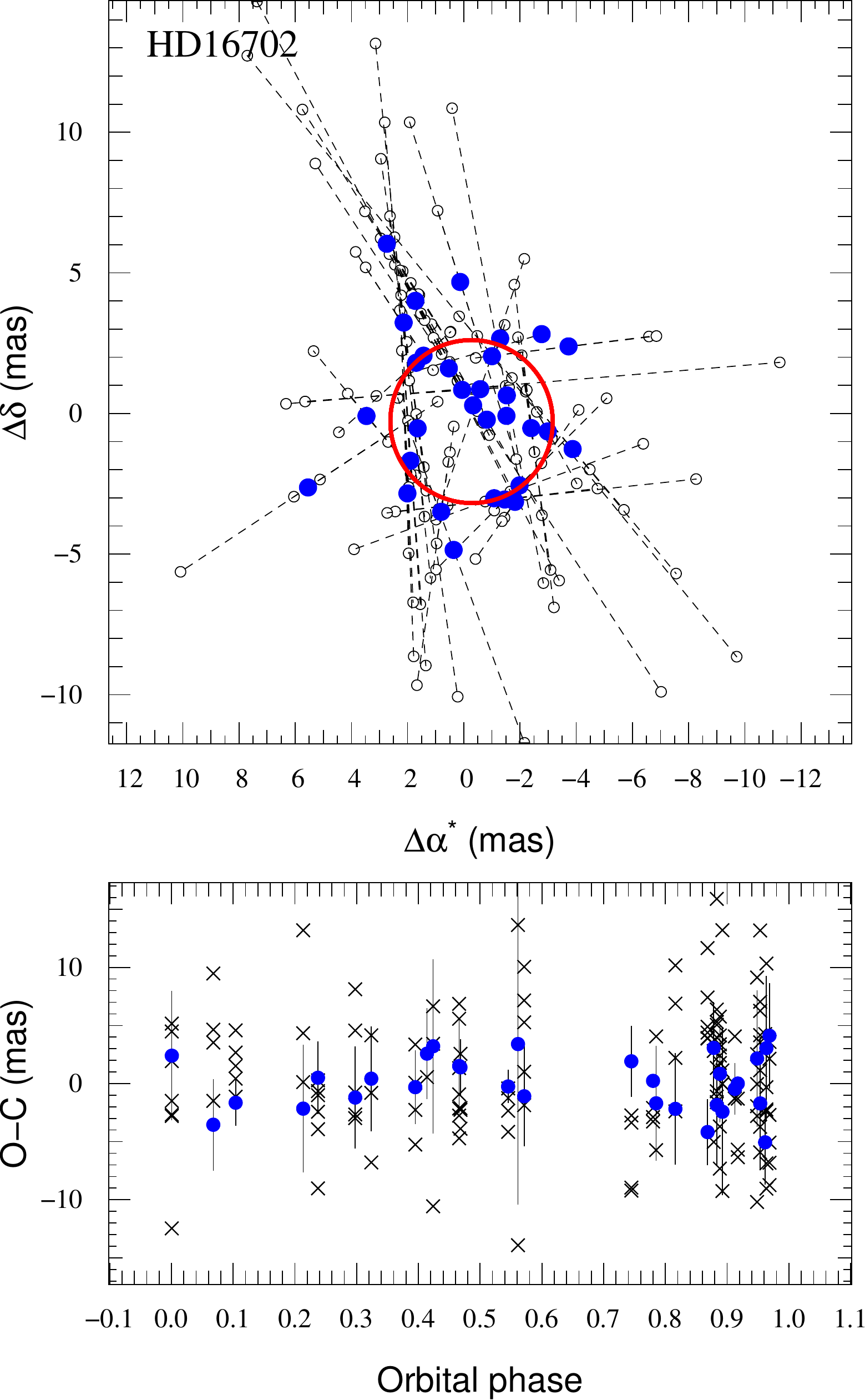}
 \caption{\emph{Top:} Astrometric orbit of HD16702 projected on the sky. North is up and east is left. The solid red line shows the model orbit, which is orientated counter-clockwise, and open circles mark the individual Hipparcos measurements used for the astrometric fit. Blue filled circles show normal points for each satellite orbit number and are used for display only. \emph{Bottom:} O-C residuals for the normal points of the orbital solution (filled blue circles) and the five-parameter model without companion (crosses).}
 \label{fig:orbit}
 \end{center} \end{figure}

The astrometric analysis and CCF simulations confirm this scenario. We detect the astrometric signal in the Hipparcos data with a significance of $2.5\,\sigma$ (98.7~\%) and measure a semimajor axis of $2.9 \pm 0.6$~mas. We see the system almost face-on with an orbital inclination of $i = 6.6 ^{+1.6}_{-1.1}$ degrees, corresponding to a companion mass of $M_c = 0.55 \pm 0.14$ M$_{\sun}$ and we measure the longitude of the ascending node of $\Omega=146^{+20}_{-7}$~$\degr$. The coordinate and proper motion offsets are $\Delta \alpha^{\star} = -0.4\pm0.7$ mas, $\Delta \delta = 0.0\pm0.5$ mas, $\Delta \mu_{\alpha^\star} = -1.8\pm 1.2$ mas yr$^{-1}$, $\Delta \mu_{\delta} = 0.6\pm0.8$ mas yr$^{-1}$, and the updated parallax is $\varpi = 20.9\pm0.8$ mas, which is compatible with the original value. The sky-projected orbit of HD16702 and the Hipparcos measurements are shown in Fig.~\ref{fig:orbit}. The values of $2.5\,\sigma$ for the significance and of 6.5 for the astrometric S/N for this orbit indicate that we are close to the limit of the detection capabilities of Hipparcos. However, our astrometric analysis rules out the possibility of the companion being a sub-stellar object. Indeed, at the 3-$\sigma$ level, the lower and upper companion mass limits are $0.19$ M$_\odot$ and $1.05$ M$_{\sun}$, respectively.


\begin{figure}[t] 
\begin{center}
\includegraphics[width=\columnwidth, trim=0 0cm 0 1cm, clip=true]{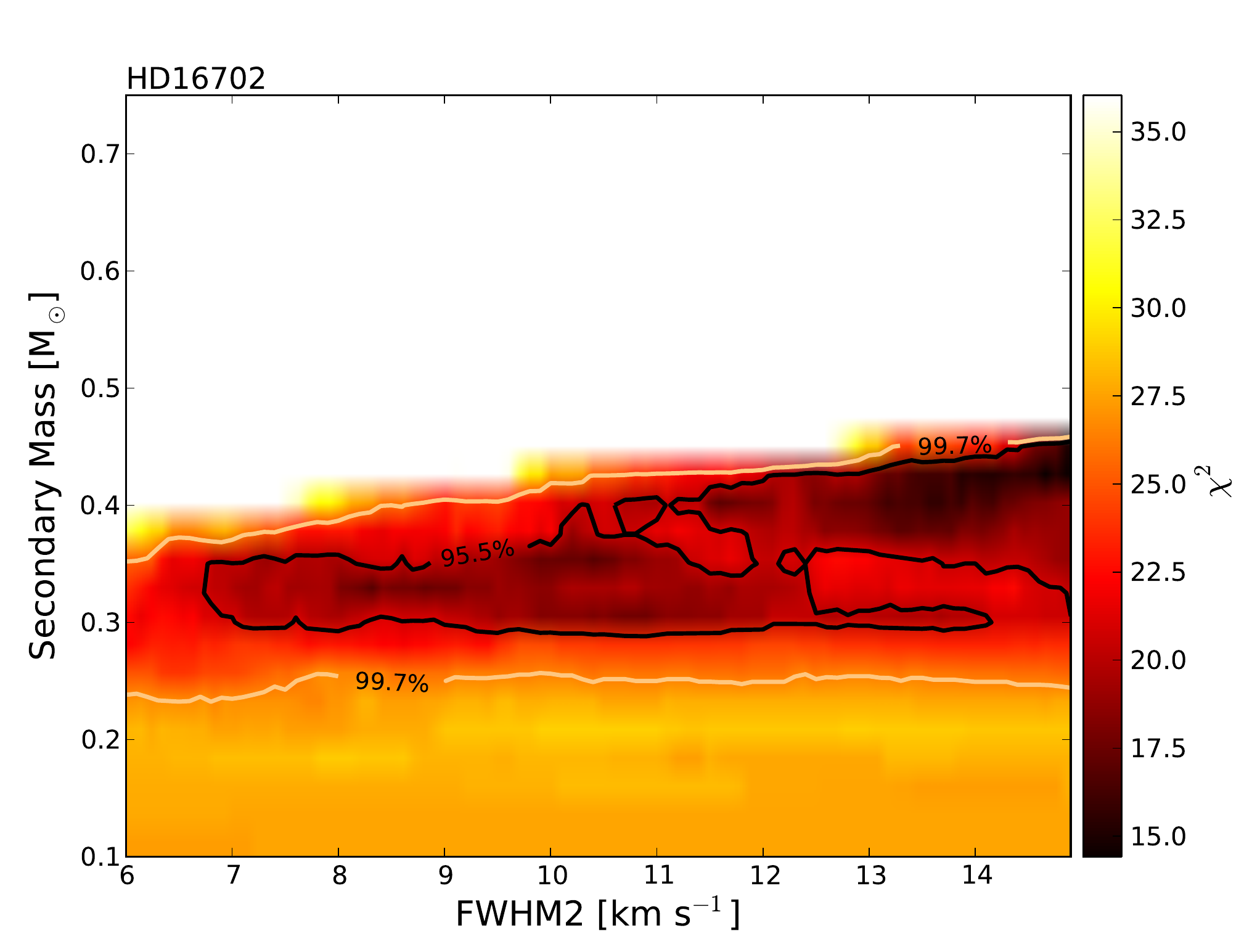}
\caption{Confidence regions for the mass of the secondary and the FWHM of its CCF, issued from numerical simulations of HD16702. At the 3-$\sigma$ level, a strong constraint can be put to the mass of the secondary. \label{fig.blendsim16702}}
\end{center}
\end{figure}

The CCF simulations permit further constraining the companion mass. In Fig.~\ref{fig.blendsim16702} we show the confidence regions obtained from the $\chi^2$ surface, as described in \S\ref{sect.bisector}. We deduce that the maximum allowed value, at 99.7 \% confidence level, range from 0.35 M$_\odot$ to 0.45 M$_\odot$, depending on the FWHM of the secondary. Interestingly, due to the significant correlation between bisector and radial velocity shown in Fig.~\ref{fig.bissector16702}, also a lower limit of about 0.25 M$_\odot$ can be put to the mass of the secondary. On the other hand, no strong constraints can be put on the rotational velocity of the secondary. We note that the best-fit  point has $\chi^2=14.4$, comparable with the number of observations (N=22), which means that the confidence regions are a rigorous  way to estimate the mass of the secondary. 

HD16702 is therefore a binary star, with a mass ratio ranging from $q\approx0.25$ to $q\approx [0.35 - 0.46]$. See \S\ref{sect.discussion} for further discussion on this object.
 

\subsection*{HD167215}
The spectral analysis of HD167215 (HIP 89270) shows that this object is a low-metallicity $1.15$-\msol\ late-F star. It is located 79 pc away the Sun, and the RV measurements obtained with SOPHIE, although inconclusive due to their incomplete phase coverage, point at a 600-day period variation with an amplitude of 2~\kms.

\begin{figure*} 
\begin{center}
\includegraphics[width=0.85\columnwidth, trim=0 0 0 1cm, clip=true]{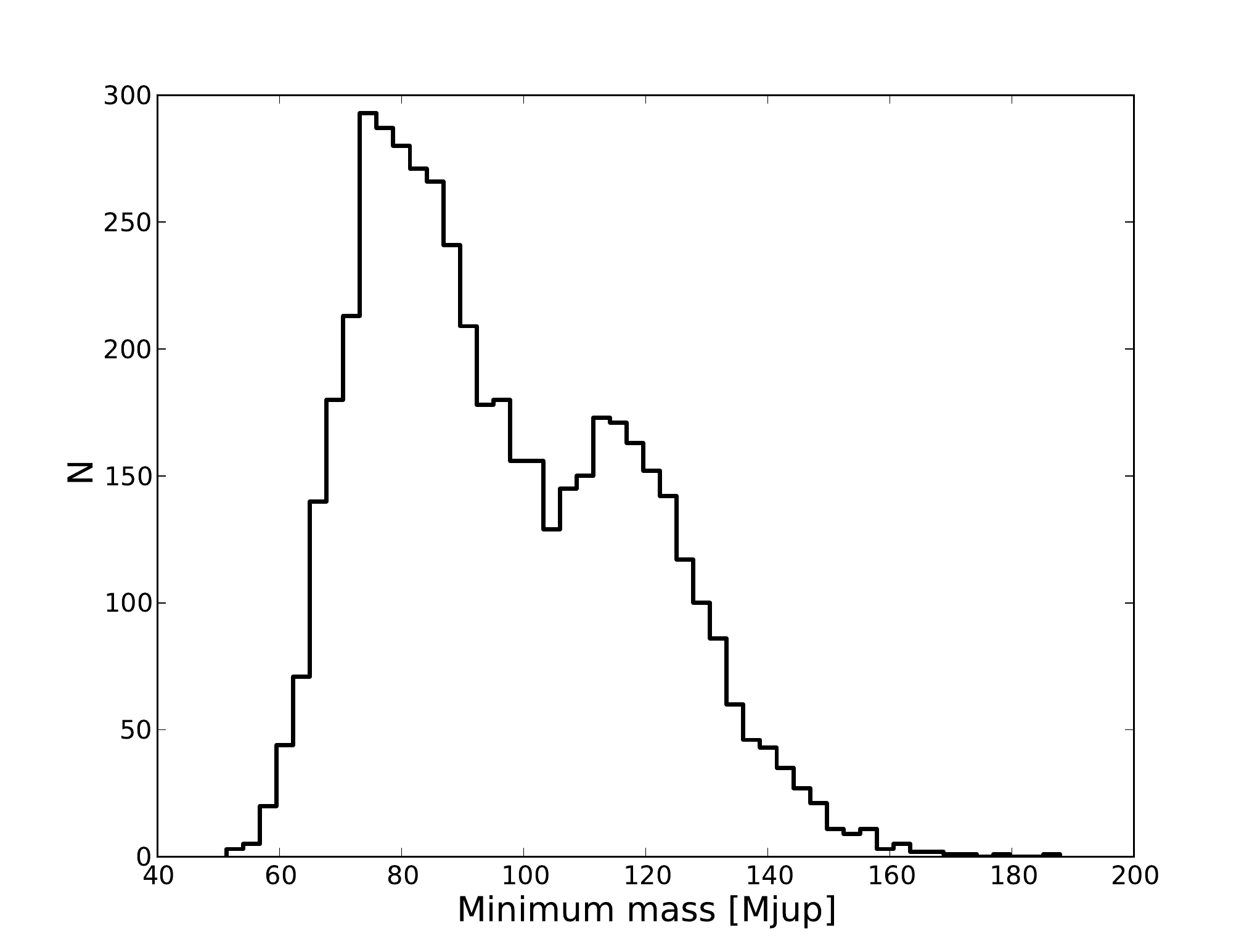}
\includegraphics[width=0.85\columnwidth, trim=0 0 0 1cm, clip=true]{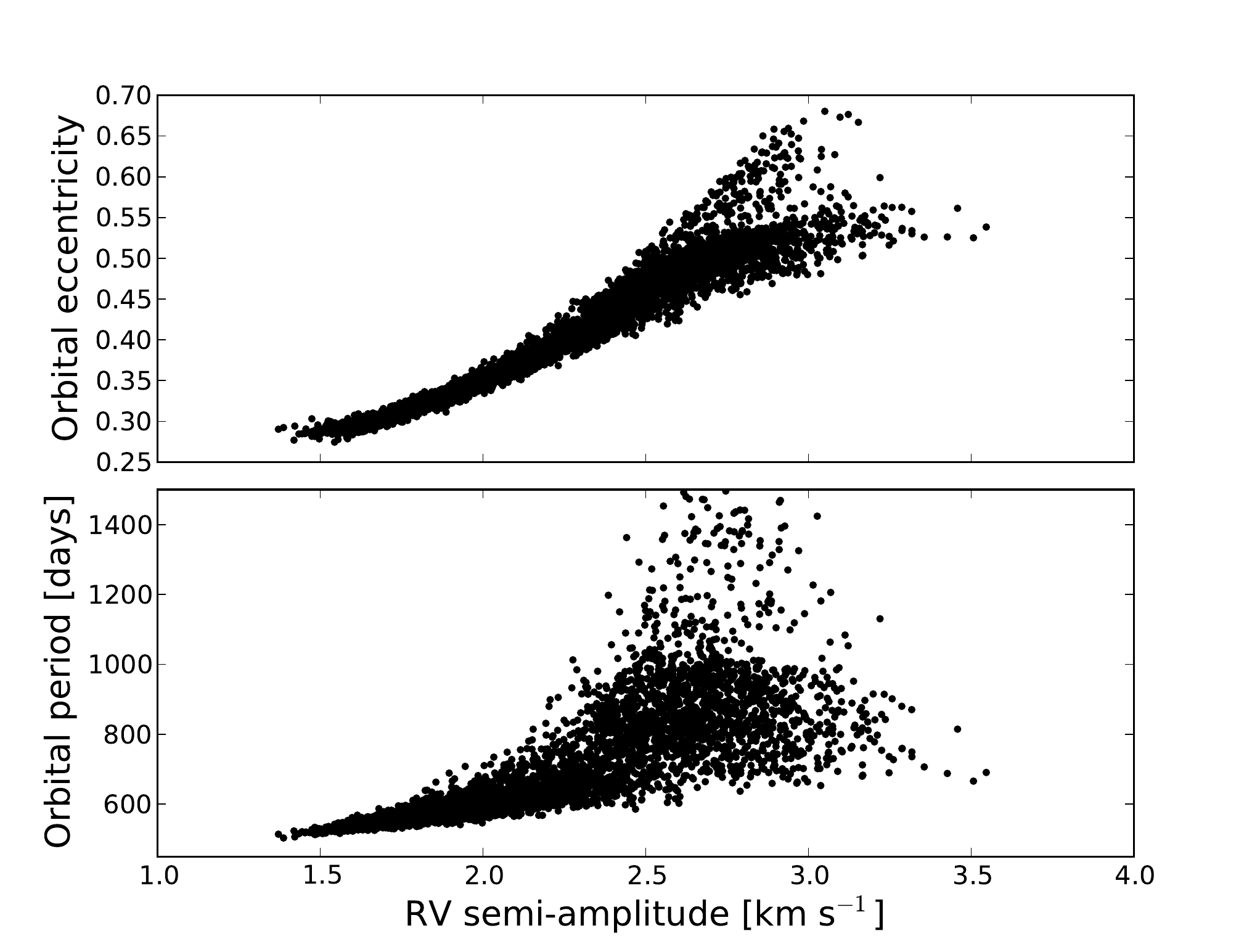}
\caption{\emph{Left:} Minimum mass distribution for HD167215, obtained from 5000 Monte Carlo simulations. Only about 25\% of the solutions lie below the 80-\MJ\ limit. \emph{Right}: Correlations between the RV variation amplitude and the orbital eccentricity (upper panel) and the orbital period (lower panel). \label{fig.bimodal167215}}
\end{center}
\end{figure*}

The Monte Carlo distribution of minimum masses (see Fig.~\ref{fig.bimodal167215}) exhibits a clear bimodal behaviour, with a lot of power on the stellar-mass range. Indeed, only about 25\% of the Monte Carlo realisations of the data are fitted to a minimum mass below 80 \MJ. Therefore, the orbiting object is most probably a low-mass star. Just as for HD30246, the high-mass solutions correspond to high eccentricity orbits. Also, the period is well correlated with the semi-amplitude $K$ (Fig.~\ref{fig.bimodal167215}).

\begin{figure} 
\begin{center}
\includegraphics[width=\columnwidth, trim=0 0 0 1cm, clip=true]{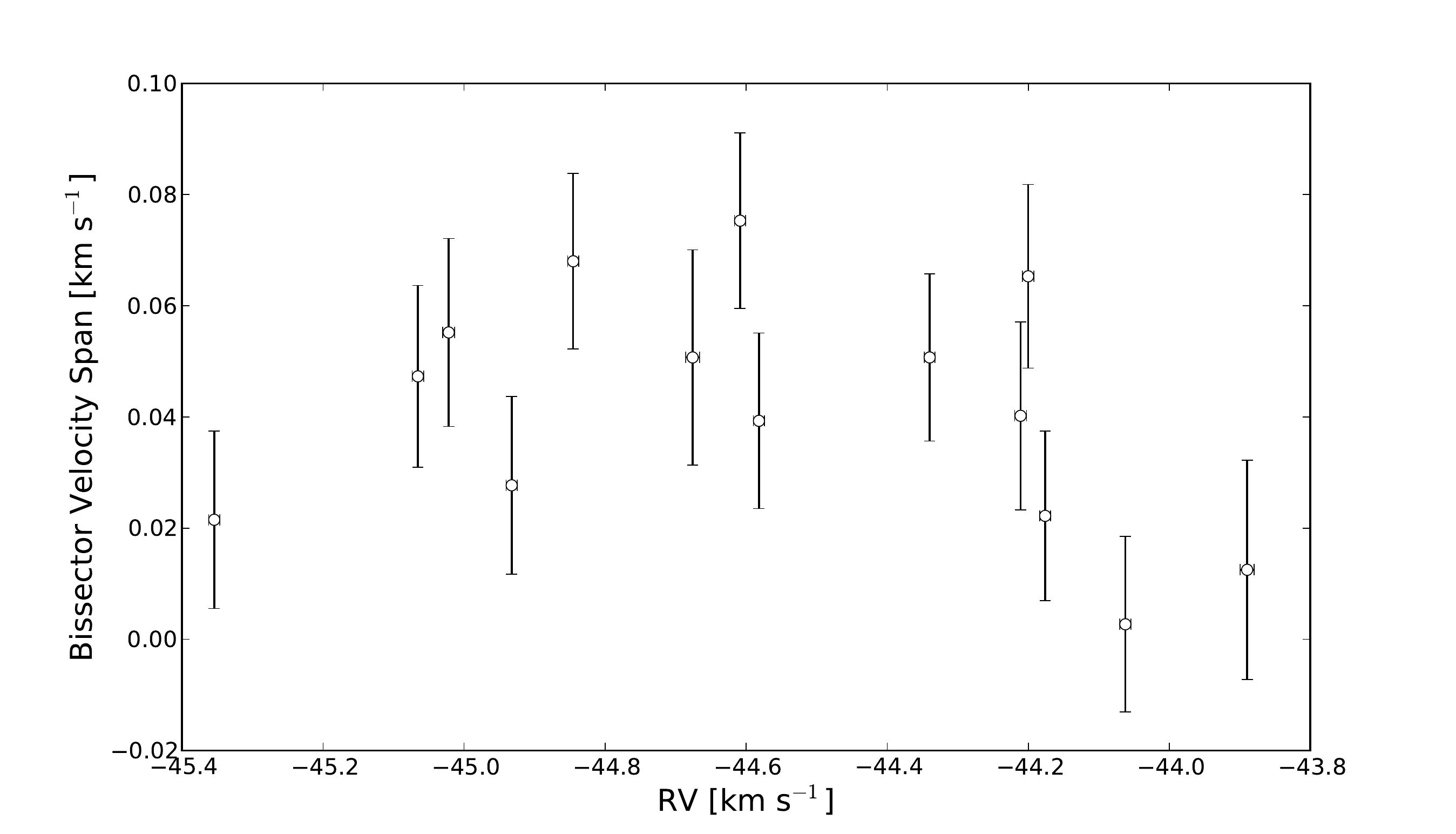}
\caption{Bisector velocity span as a function of RV measurements for HD167215. The curve is the parabolic fit to the data.\label{fig.bis167215}}
\end{center}
\end{figure}

\begin{figure} 
\begin{center}
\includegraphics[width=\columnwidth, trim=0 0 0 1cm, clip=true]{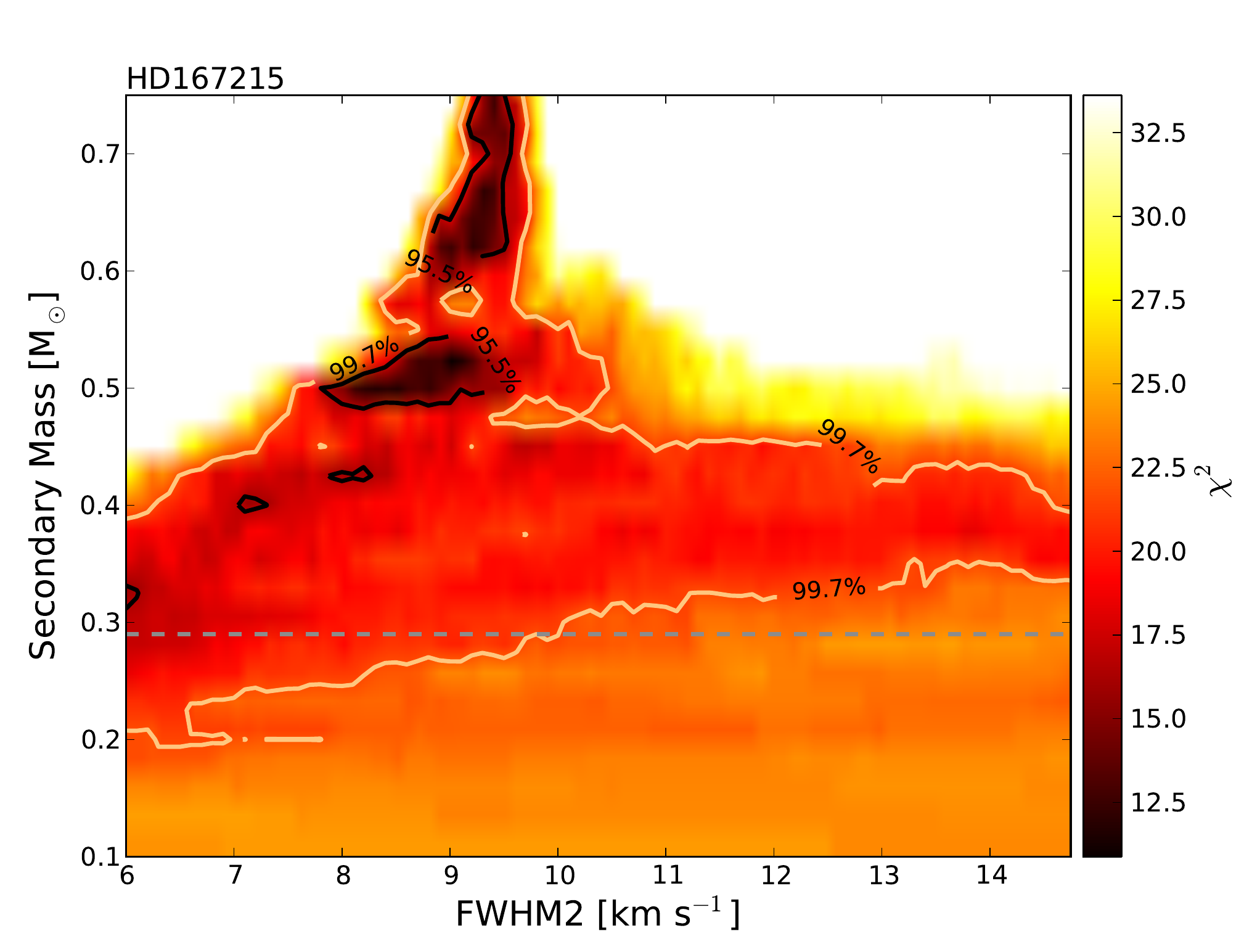}
\caption{Confidence regions for the mass of the secondary and the FWHM of its CCF, issued from numerical simulations of HD167215. When combined with the astrometric analysis, the true mass of the companion is constrained to the region below the astrometric upper limit (dashed line) and the 99.7\% contour below it. \label{fig.blendsim167215}}
\end{center}
\end{figure}

The bisector velocity span is presented as a function of RV in Fig.~\ref{fig.bis167215}. A constant fit to the data gives $\chi^2 = 23$, for 13 degrees of freedom, and a linear regression gives $\chi^2 = 21.6$, for 12 degrees of freedom. An F-test and a run test \citep[see][\S14.6.2]{frodesen}  fail at rejecting the null hypothesis at 5\% significance level. Additionally, no evidence of mask effect is detected, which means that no significant amount of light from the putative secondary star enters the spectrograph. Finally, we did not detect an orbital signature in the Hipparcos data, since the derived orbit has a significance of only $56~\%$. We were nevertheless able to set an upper limit of $0.29\,M_{\sun}$ for the mass of the companion. When the RV and astrometric analyses are combined with the CCF simulations (Fig.~\ref{fig.blendsim167215}), the mass of the secondary is confined to the wedge-like region of Fig.~\ref{fig.blendsim167215} delimited by the upper limit set by astrometry ($0.29\,M_{\sun}$, shown as a dashed line) and the 99.7 \% contour level below it. The secondary is likely to have a projected rotational velocity below $v \sin I\approx 5.3$ \kms, which is the value correspoding to FWHM $\approx 10$ \kms. The $\chi^2$ of the best fit is 11.3, for 14 observations. Regardless of the low significance of the bisector velocity span variation, the CCF simulations  place the mass of the secondary in the stellar domaine, indicating that the companion to HD167215 is likely a low-mass star, with mass ratio between $q = 0.17$ and 0.25.

\section{Discussion \label{sect.discussion}}

\begin{figure*} 
\begin{center}
\includegraphics[width=0.8\textwidth]{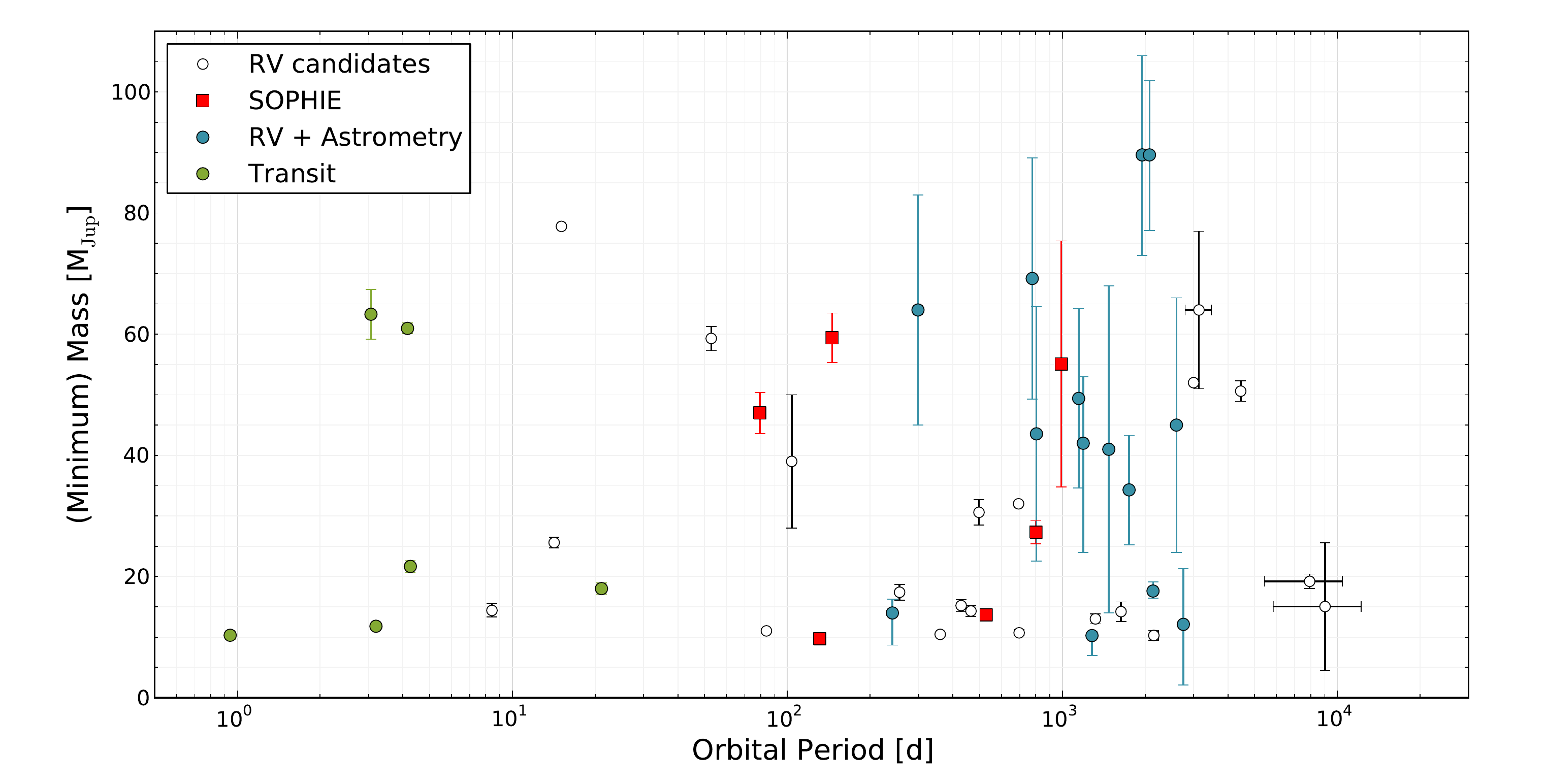}
\includegraphics[width=0.8\textwidth]{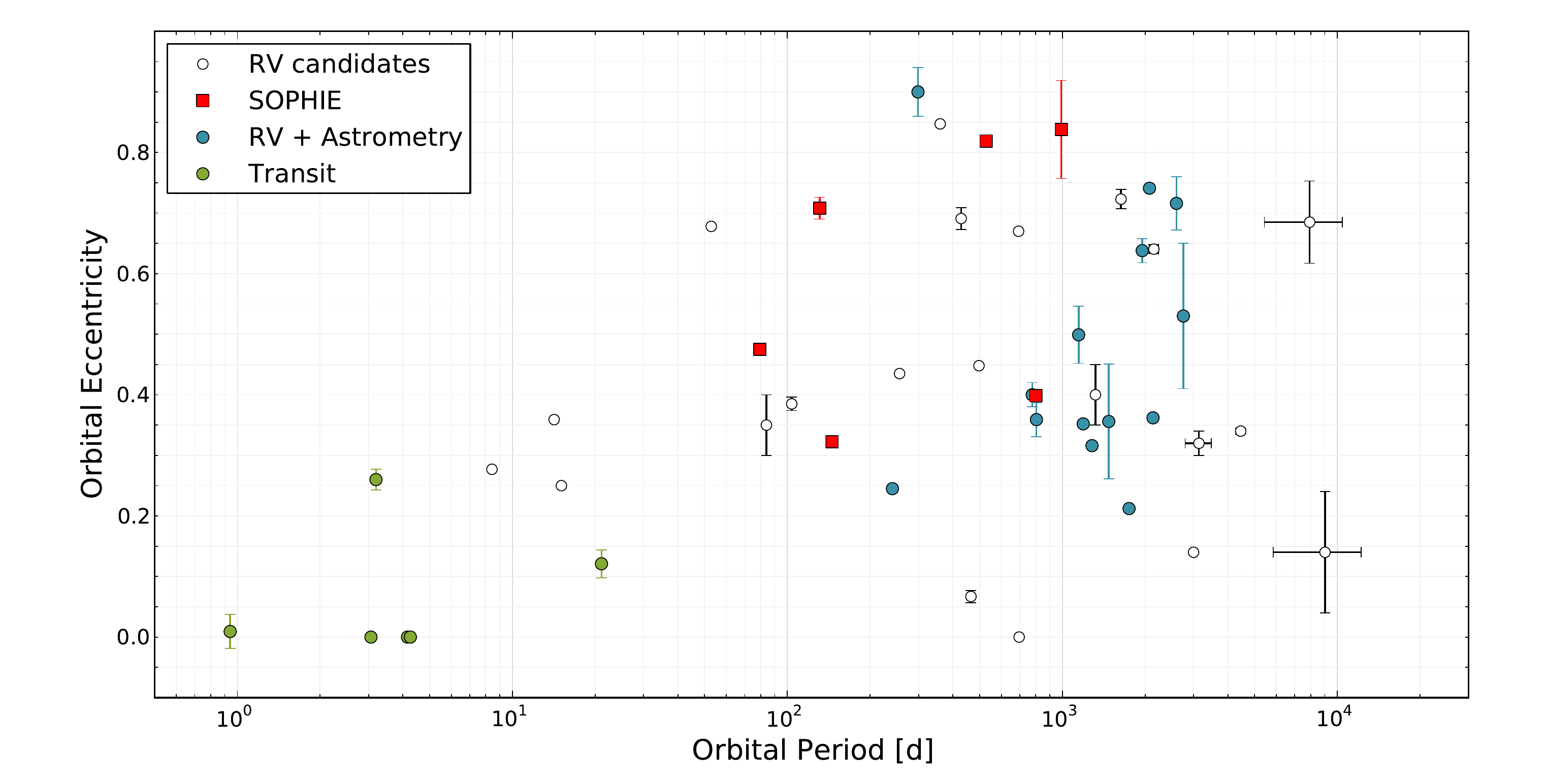}
\caption{\emph{Upper:} Mass-period diagram for all objects with (minimum) masses between 9 and 90 \MJ, orbiting solar-type stars with a period shorter than $1\times10^4$ days. \emph{Lower:} Eccentricity-period diagram for same sample of objects. In both panels,  the objects reported in this paper are shown as red squares. The other coloured symbols represent companions for which the true mass is known, either by astrometry (blue circles) or because they transit their host star (green circles). The rest of the objects are candidates detected by RV surveys (white circles), and the plotted mass should be interpreted as a lower limit. Due to the large range of objects represented, for some of them, the error bars are smaller than the symbol used in the plot. \label{fig.massperiod}}
\end{center}
\end{figure*}

\subsection{The SOPHIE objects}

The objects reported in this paper span a wide range of minimum masses, from 9.5 \MJ\ (HD156279~b) up to 92 \MJ\ (HD167215B), and are therefore useful to perform a comparative study of the properties of massive planets, brown dwarfs and low-mass stars. In particular, the object orbiting HD137510 has been established to be a brown dwarf \citep[][and this work]{endl2004, sahlmann2011}, and the companions to HD16702 and HD167215 have been identified as low-mass stars. Out of the remaining 5 objects reported here, two have minimum masses below or close to the deuterium-burning limit, and we have labelled them as planet candidates (HD22781~b and HD156279~b), while the other three are brown dwarf candidates. Except for HD92320 b, which has a minimum mass slightly larger, these candidates fall in what has been identified by \citet{gretherlineweaver2006} as the driest part of the brown dwarf desert, at $M_c = 43^{+13}_{-23}$ \MJ. This implies that these candidates are rare and therefore worthy of further observations in order to pinpoint their true mass. In Table~\ref{table.parameters} we report the minimum astrometric signal expected for the detected candidates, defined as the semi-amplitude of the periodic motion of the star projected  on the plane of the sky, assuming $i=90$ degrees. The future space mission GAIA, having an astrometric precision below 1.0$\times 10^{-5}$ arcsec for bright stars such as these \citep{prusti2011}, should easily detect the orbital signature of all three of these candidates. 

HD16702 has been identified as a stellar binary system, with a period of 72.8 days and a mass ratio $q$ lying in the interval [0.25-0.46] at the 99.7\% confidence level. Considering the mass of the companion to be $M_c = 0.35$ M$_\odot$, the magnitude differences in the $V$ and $K$ bands imply a contrast of $3\times 10^{-3}$ and $5.6\times 10^{-2}$, respectively, using the apparent $K$-band magnitude from the 2MASS catalogue \citep{2MASS}. The measured projected separation for this system is 8.6 mas. It will not be easily resolved with the current adaptive optics facilities. On the other hand, interferometric observations should be able to resolve the system.

For HD167215, we find a binary system at a period of over 600 days, and a mass ratio in the interval [0.17-0.25]. If we consider that the secondary has a mass of 0.25 M$_\odot$, inversion of the mass-luminosity relations by \citet{delfosse2000} provide the absolute magnitudes $M_V = 12.41$ and $M_K=7.28$, in the $V$ and $K$ band. The primary star has apparent magnitudes $m_V=8.1$ and $m_K=6.69$ \citep{2MASS}, which give contrasts of $2.8\times 10^{-4}$ and $8.6\times 10^{-3}$ in the $V$ and $K$ band, respectively. The measured mass for the companion implies that the inclination angle $i$ is roughly 20 degrees, and the projected separation is 26.5 mas. This system will also be difficult to resolve with current adaptive optics instruments, but it might be a suitable target for interferometric observations.

The brown dwarf orbiting HD137510 has been discovered by \citet{endl2004} and confirmed later by \citet{sahlmann2011} and \citet{reffert2011}. The minimum mass reported in this paper based on the analysis of all available datasets is compatible with previous determinations, and the analysis of the SOPHIE data alone also produces a compatible result. If we consider that the true mass of the brown dwarf is 30.9 \MJ, i.e. the expected value assuming random orientation of the system, and including the limit on the inclination provided by the astrometric analysis ($\sin i >  0.4588$), then linear interpolation of the model grids by \citet{baraffe2003} give effective temperatures between 800 K and 870 K for the two limiting values obtained for the age of this system, 3.0 Gyr and 2.3 Gyr. This identifies this object as a T-type dwarf, which should present prominent methane and water features in its near-infrared spectrum \citep[see, for example,][]{kirkpatrick2005}. However, being about 21 mas from its host star, which outshines it by over 15 magnitudes in the $H$ band, obtaining an spectrum of this object will prove challenging.

HD22781 exhibits excess variation in the residuals of the single-Keplerian fit, that might indicate the presence of a second companion in the system. A tentative estimation of the mass and period of this object results in a minimum mass of about 1 \MJ\ in a 9 years period orbit. Although this would be an unusual system, with the more massive planet in the inner orbit and a lighter companion farther away, this configuration is by no means completely unheard of. For example, HD20226 \citep{correia2005,udry2002}, present a similar configuration, with an inner massive planetary candidate in a 255-day orbit, and an outer companion close to the 5:1 resonance. Other examples of systems of this kind are the three-planet system HIP14810 \citep{wright2009,wright2007}, HD134987 \citep{jones2010}, and HD200964 \citep{johnson2011b}. 

\subsection{Formation and evolution of sub-stellar companions}

In this section, we put in context the results obtained by considering similar objects reported in the literature. We do not attempt to perform here a complete and thorough statistical analysis of the available data, which is defer to a future work, but only to present a few relevant characteristics of sub-stellar massive companions that might shed some light on mechanisms involved in their formation and evolution.

\subsubsection{Orbital period}

The SOPHIE objects cover a range of orbital periods from little over 70 days for HD16702 b up to 990 days for HD30246~b, or roughly 2 AU. The position of these objects in the mass-period diagram is shown in the upper panel of Fig.~\ref{fig.massperiod}, together with other similar objects presented in the literature\footnote{Objects with minimum mass above 9 \MJ\ and below 100 \MJ, orbiting solar-type stars (0.7 M$_\odot < M < 1.5 \mathrm{M}_\odot$), with orbital periods shorter than 10$^4$ days.} \citep[][and references therein]{sahlmann2011, sozzetti2010}. HD16702B and HD167215B are not included since their true mass is above the imposed limit of 90 \MJ. Note that none of our objects have been detected in short-period orbits. Indeed, it seems evident in Fig.~\ref{fig.massperiod} that the number of objects with masses above $\sim$10 \MJ\ increases with orbital period, even though RV and transit observations are biassed towards discovering objects in short-period orbits.  This might be explained by considering that these objects are formed far from the central star by gravitational collapse, reaching the mass needed to open a gap in the disk \citep{ward97} before start migrating inwards by interactions with the disk. Since then migration will be of type II \citep{ward97}, which has longer timescales, these objects might not have time to reach closer orbits before the disk dissipates.

\subsubsection{Orbital eccentricity}
In the lower panel of Fig.~\ref{fig.massperiod} we plot the orbital eccentricity of the SOPHIE candidates as a function of orbital period. It can be seen that all the reported objects exhibit a considerable orbital eccentricity. \citet{ribasmiraldaescude2007} report an increase in the orbital eccentricity of massive planetary objects, when compared with less massive objects. These authors argue that this might be indicative of different formation mechanisms, and propose that while low-mass objects form by core accretion in protoplanetary discs, more massive planets might form by gravitational collapse, either of a pre-stellar cloud or of a gravitationally unstable circumstellar disk. Since high-mass planets would then need to form far from their stars, the fact that those observed at shorter distances (typically a few AU at most) have higher eccentricities than low-mass planets is, according to these authors, due to the large radial migration they have undergone. Indeed, numerical simulations \citep{papaloizou2001,nelson2000} have shown that eccentricity is excited on massive ($\gtrsim$ 20 \MJ) migrating objects by means of interaction with the circumstellar disc, but not so for less massive objects. On the other hand, interactions at disk resonances can damp eccentricity \citep[see][and references therein]{tremainezakamska2004}. Therefore, although current data seem to agree with the idea that massive companions have larger eccentricities, it is not clear whether this can be explained by invoking a larger radial migration.

\subsubsection{Metallicity}

Planetary companions with masses comparable to that of Jupiter are known to occur more frequently around metal-rich stars \citep{santos2001,fischervalenti2005}, which is considered strong support for the core-accretion formation scenario \citep{pollack96,alibert2005}. On the other hand, stellar companions are more common around metal-poor primaries \citep{gretherlineweaver2007,raghavan2010}, a trend that is more evident for stars with $(B-V) > 0.625$. \citet{raghavan2010} interpret this (tentative) result by suggesting that lower-metallicity clouds might be more likely to fragment to form binary stars, as is revealed in numerical simulations. If this is so, and if brown dwarf companions form as stars, then we might expect them to show a similar trend. 

Another trend reported by  \citet{ribasmiraldaescude2007}  is that the metallicity of the host star decreases with companion mass. This is interpreted as a hint that two different mechanisms act in the formation of extrasolar planets, as discussed above, and that massive objects form as easily around metal-rich stars than around metal-poor ones. Of course, this statement assumes that the sample of target stars out of which extrasolar planets are detected is uniform in metallicity. However, since massive companions are rare and RV surveys usually have biased samples, the trend observed by \citet{ribasmiraldaescude2007} might not be real. The actual relevant measurable to study this issue is the \emph{occurrence rate} of detected objects as a function of the metallicity of their host stars. However, given the small number of detected brown dwarfs around solar-type stars, assessing the reality of any such trend is challenging.

\section{Summary and Conclusions \label{sec.summary}}

We have reported the discovery of 7 new sub-stellar candidates orbiting solar-type stars with minimum masses above $\sim 10$ \MJ. Out of these, we have managed to identify two objects as low-mass stellar companions, based on astrometric measurements and numerical simulations of their cross-correlation functions. Additionally, the orbital parameters of a known brown dwarf companion, namely HD137510 b, have been improved upon using radial velocity measurements obtained with SOPHIE. The three confirmed objects are too faint and too close to their host stars to be easily detected by current adaptive optics techniques, since they lay within the diffraction limit of the biggest available telescopes. On the other hand, the brown dwarf candidates should be promptly confirmed or rejected with GAIA's micro-arcsecond astrometry.

Different techniques were employed with the aim of discerning low-mass stars in low-inclination orbits from actual sub-stellar objects. To do this, we employed not only traditional RV diagnostics, such as the mask effect and the bisector analysis,  but we also analysed astrometric data from the Hipparcos mission, and developed a new technique to fully exploit the information present in the bisector velocity span. This latter technique has allowed us to set upper limits to the masses of all the candidates detected here, even surpassing in some cases the possibilities of current astrometric data, and has allowed the confirmation that one of the candidates was in fact a low-mass star. In some cases, the CCF simulations and the Hipparcos astrometry turn out to be complementary, as for HD167215. The effectiveness of each technique depends strongly on the orbital parameters, the number of measurements and the phase coverage, and we have therefore refrained from studying this here (see Santerne et al., in preparation).  Improvements of the CCF simulations used in this method are currently underway, and should provide further possibilities for this promising technique, such as validating planetary candidates discovered by transiting surveys that show no variation in their RV or bisector velocity span measurements, as is the case for the recently-announced CoRoT-22 b (Moutou et al., submitted).

Concerning the question of the dividing line between planets and brown dwarfs, if a mass-dependent classification is adopted, motivated by the observed mass distribution, such as the one proposed by \citet{sahlmann2011} or \citet{schneider2011}, who set the limit at around 25 \MJ, then 3 of the sub-stellar candidates are brown dwarf candidates (HD14651b, HD30246b, and HD92320b), while the remaining two are planetary candidates (HD22781b and HD156279b). A more detailed analysis of other possible classifications of the available objects is outside the scope of this paper and will be presented in a future work.

The objects reported here increase the number of similar objects present in the literature by about 12\%, contributing to establish solid observational evidence to test and constrain formation and evolution theories. At the time of writing the number of sub-stellar objects or candidates more massive than 10 \MJ\ orbiting solar-type stars in relatively short ($P<10^4$ days) orbits amounts to little less than 60 objects.  The small size of this sample currently hinders most statistical analyses. Nevertheless, these objects warrant a  rigorous statistical study, which might provide new and improved constraints for models and theories. Also, any effort to increase the number of objects in the mass range studied here, as well as all attempts to measure the actual mass of candidates,  should be encouraged.

\begin{acknowledgements}
We thank the staff of Haute-Provence Observatory for their 
support at the 1.93-m telescope and on \sophie.
We thank the ``Programme National de Plan\'etologie'' (PNP) of CNRS/INSU, 
the Swiss National Science Foundation, 
and the French National Research Agency (ANR-08-JCJC-0102-01 and ANR-NT05-4-44463) 
for their support with our planet-search programs.
D.E.\ is supported by CNES. 
I.B.\ and N.C.S.\ acknowledges the support by the European Research Council/European Community under the FP7 through Starting Grant agreement number 239953, as well as the support from Funda\c{c}\~ao para a Ci\^encia e a Tecnologia (FCT) through program Ci\^encia\,2007 funded by FCT/MCTES (Portugal) and POPH/FSE (EC), and in the form of grants reference PTDC/CTE-AST/098528/2008 and PTDC/CTE-AST/098604/2008.
A.E.\ is supported by a fellowship for advanced researchers from the Swiss National Science Foundation (grant PA00P2\_126150/1).
\end{acknowledgements}

\bibliographystyle{aa} 
\bibliography{biblio_iafe}

\Online

\onllongtab{2}{
\begin{longtable}{l l l c c c}
\caption{Radial velocity measurements \label{table.RV}}
\endfirsthead
\caption{continued.}
\endhead
\multicolumn{6}{l}{\tablefoottext{a}{Bisector Velocity Span}}\\
\multicolumn{6}{l}{\tablefoottext{\dagger}{Observations in \emph{thosimult} mode before upgrade (see \S\ref{sect.datareduc}).}}
\endlastfoot
\hline
\hline
\noalign{\smallskip}
\multicolumn{6}{l}{HD14651}\\
\noalign{\smallskip}
\hline
\noalign{\smallskip}
BJD     & RV          & $\sigma_{RV}$ & BVS\tablefootmark{a}         & Exp. time & S/N/pix    \\
-2 400 000  & (\kms) & (\kms) & (\kms) & (s)       & (at 550 nm)\\
\hline
\noalign{\smallskip}
54043.5099\tablefootmark{\dagger} & 52.243 & 0.005 & -0.052 & 500 & 70 \\
54047.4534\tablefootmark{\dagger} & 51.611 & 0.004 & -0.033 & 300 & 87 \\
54048.4953\tablefootmark{\dagger} & 51.424 & 0.004 & -0.044 & 400 & 94 \\
54049.4427\tablefootmark{\dagger} & 51.220 & 0.004 & -0.039 & 500 & 80 \\
54051.4619\tablefootmark{\dagger} & 50.745 & 0.005 & -0.034 & 400 & 64 \\
54080.3977\tablefootmark{\dagger} & 54.147 & 0.004 & -0.033 & 400 & 69 \\
54501.2668 & 53.781 & 0.005 & -0.038 & 227 & 45 \\
54507.2945 & 53.448 & 0.005 & -0.040 & 227 & 46 \\
54705.6496 & 53.075 & 0.005 & -0.036 & 341 & 49 \\
54717.6278 & 54.188 & 0.005 & -0.032 & 391 & 47 \\
54720.6097 & 54.228 & 0.005 & -0.052 & 364 & 50 \\
55240.3093 & 51.304 & 0.005 & -0.028 & 370 & 48 \\
55405.5712 & 49.679 & 0.006 & -0.046 & 227 & 36 \\
55409.5920 & 49.053 & 0.005 & -0.030 & 209 & 53 \\
55453.6099 & 53.805 & 0.004 & -0.044 & 188 & 53 \\
\noalign{\smallskip}
\hline 
%
%
%
\noalign{\bigskip}
\hline
\hline
\noalign{\smallskip}
\multicolumn{6}{l}{HD16702}\\
\noalign{\smallskip}
\hline
\noalign{\smallskip}
BJD     & RV          & $\sigma_{RV}$ & BVS\tablefootmark{a}         & Exp. time & S/N/pix    \\
-2 400 000  & (\kms) & (\kms) & (\kms) & (s)       & (at 550 nm)\\
\hline
\noalign{\smallskip}
54365.6123 & 4.844 & 0.005 & 0.005 & 200 & 54 \\
54407.4115 & 5.699 & 0.004 & 0.003 & 300 & 79 \\
54435.4567 & 5.275 & 0.004 & 0.006 & 336 & 81 \\
54461.4090 & 2.086 & 0.005 & 0.015 & 360 & 44 \\
54462.3212 & 2.091 & 0.004 & 0.009 & 221 & 60 \\
54463.3549 & 2.141 & 0.005 & 0.018 & 202 & 53 \\
55066.6291 & 6.403 & 0.005 & -0.004 & 279 & 48 \\
55067.6359 & 6.521 & 0.005 & -0.001 & 206 & 51 \\
55068.6510 & 6.663 & 0.005 & 0.003 & 424 & 47 \\
55069.6318 & 6.742 & 0.006 & -0.001 & 600 & 37 \\
55405.5953 & 2.115 & 0.005 & 0.024 & 347 & 51 \\
55425.6437 & 5.352 & 0.005 & -0.008 & 194 & 47 \\
55427.6357 & 5.811 & 0.005 & 0.004 & 159 & 50 \\
55436.6205 & 6.896 & 0.005 & -0.021 & 285 & 46 \\
55438.5971 & 6.909 & 0.005 & -0.010 & 257 & 46 \\
55444.6215 & 6.572 & 0.004 & 0.004 & 180 & 62 \\
55449.5477 & 6.089 & 0.005 & -0.004 & 202 & 46 \\
55465.6213 & 3.654 & 0.006 & 0.027 & 270 & 43 \\
55475.6539 & 2.312 & 0.005 & 0.022 & 189 & 55 \\
55479.4840 & 2.092 & 0.005 & 0.009 & 276 & 49 \\
55488.4233 & 2.816 & 0.006 & 0.031 & 200 & 42 \\
55503.4724 & 6.354 & 0.005 & -0.009 & 222 & 47 \\
\noalign{\smallskip}
\hline \pagebreak
\noalign{\bigskip}
\hline
\hline
\noalign{\smallskip}
\multicolumn{6}{l}{HD22781}\\
\noalign{\smallskip}
\hline
\noalign{\smallskip}
BJD     & RV          & $\sigma_{RV}$ & BVS\tablefootmark{a}         & Exp. time & S/N/pix    \\
-2 400 000  & (\kms) & (\kms) & (\kms) & (s)       & (at 550 nm)\\
\hline
\noalign{\smallskip}
54079.4169\tablefootmark{\dagger} & 8.154 & 0.005 & -0.005 & 600 & 61 \\
54080.4125\tablefootmark{\dagger} & 8.162 & 0.004 & 0.008 & 500 & 72 \\
54081.3843\tablefootmark{\dagger} & 8.156 & 0.004 & 0.009 & 500 & 83 \\
54087.4543 & 8.142 & 0.005 & -0.005 & 240 & 43 \\
54097.4342 & 8.134 & 0.004 & 0.011 & 300 & 56 \\
54125.3592 & 8.096 & 0.005 & 0.001 & 1000 & 42 \\
54127.3144 & 8.107 & 0.005 & 0.014 & 680 & 44 \\
54133.2899 & 8.085 & 0.004 & -0.004 & 241 & 60 \\
54138.3552 & 8.095 & 0.004 & 0.003 & 724 & 60 \\
54141.3158 & 8.085 & 0.004 & 0.009 & 483 & 75 \\
54148.3446 & 8.083 & 0.005 & 0.021 & 241 & 41 \\
54152.3596 & 8.086 & 0.006 & -0.003 & 241 & 48 \\
54173.3081 & 8.063 & 0.004 & 0.007 & 463 & 67 \\
54344.6620 & 8.297 & 0.004 & 0.014 & 367 & 64 \\
54372.6427 & 8.994 & 0.005 & 0.010 & 700 & 40 \\
54375.6321 & 8.905 & 0.003 & 0.009 & 734 & 97 \\
54376.6144 & 8.899 & 0.003 & 0.002 & 734 & 102 \\
54381.4955 & 8.816 & 0.005 & 0.009 & 194 & 47 \\
54406.4668 & 8.604 & 0.004 & -0.002 & 600 & 68 \\
54407.5038 & 8.571 & 0.004 & 0.002 & 400 & 74 \\
54409.5721 & 8.577 & 0.004 & 0.007 & 360 & 59 \\
54855.3067 & 7.990 & 0.005 & 0.009 & 831 & 47 \\
54873.2731 & 8.363 & 0.005 & 0.014 & 383 & 49 \\
54877.2718 & 8.715 & 0.005 & 0.008 & 558 & 48 \\
54878.3020 & 8.831 & 0.005 & 0.005 & 638 & 49 \\
54890.2711 & 9.297 & 0.004 & 0.017 & 288 & 52 \\
54893.3243 & 9.211 & 0.005 & 0.001 & 696 & 46 \\
55079.6628 & 8.248 & 0.007 & 0.015 & 900 & 33 \\
55088.6638 & 8.239 & 0.005 & 0.009 & 552 & 47 \\
55214.3938 & 8.101 & 0.005 & 0.006 & 350 & 49 \\
55409.5967 & 9.222 & 0.004 & 0.006 & 291 & 52 \\
55444.6436 & 8.752 & 0.004 & -0.010 & 218 & 53 \\
\noalign{\smallskip}
\hline \pagebreak
\noalign{\bigskip}
\hline
\hline
\noalign{\smallskip}
\multicolumn{6}{l}{HD30246}\\
\noalign{\smallskip}
\hline
\noalign{\smallskip}
BJD     & RV          & $\sigma_{RV}$ & BVS\tablefootmark{a}         & Exp. time & S/N/pix    \\
-2 400 000  & (\kms) & (\kms) & (\kms) & (s)       & (at 550 nm)\\
\hline
\noalign{\smallskip}
54051.5549\tablefootmark{\dagger}  & 41.300 & 0.006 & 0.020 & 400 & 61 \\
54053.5157\tablefootmark{\dagger}  & 41.301 & 0.004 & -0.002 & 350 & 78 \\
54078.4650\tablefootmark{\dagger}  & 41.231 & 0.007 & -0.011 & 500 & 53 \\
54079.4428\tablefootmark{\dagger} & 41.240 & 0.005 & -0.009 & 500 & 67 \\
54081.4249\tablefootmark{\dagger} & 41.250 & 0.004 & -0.009 & 400 & 91 \\
54088.4622\tablefootmark{\dagger} & 41.234 & 0.009 & -0.029 & 233 & 44 \\
54098.4612\tablefootmark{\dagger} & 41.198 & 0.005 & -0.012 & 300 & 67 \\
54125.3840\tablefootmark{\dagger} & 41.153 & 0.007 & 0.004 & 900 & 57 \\
54127.3579\tablefootmark{\dagger} & 41.167 & 0.006 & -0.022 & 680 & 57 \\
54133.3531\tablefootmark{\dagger} & 41.108 & 0.006 & 0.032 & 180 & 58 \\
54148.3614\tablefootmark{\dagger} & 41.067 & 0.006 & 0.004 & 180 & 53 \\
54152.3558\tablefootmark{\dagger} & 41.024 & 0.008 & -0.005 & 180 & 54 \\
54173.3335\tablefootmark{\dagger} & 41.005 & 0.005 & 0.010 & 295 & 74 \\
54855.3659 & 41.668 & 0.009 & -0.043 & 1264 & 30 \\
55096.6444 & 41.180 & 0.008 & 0.005 & 295 & 33 \\
55214.4034 & 40.810 & 0.005 & -0.015 & 216 & 48 \\
55240.3314 & 40.733 & 0.006 & 0.015 & 355 & 47 \\
55268.2826 & 40.585 & 0.005 & 0.010 & 209 & 50 \\
55283.3259 & 40.592 & 0.007 & 0.024 & 236 & 39 \\
55289.3131 & 40.635 & 0.005 & 0.002 & 367 & 60 \\
55437.6466 & 42.800 & 0.006 & 0.007 & 603 & 46 \\
55465.6651 & 42.684 & 0.005 & -0.017 & 376 & 51 \\
55488.4488 & 42.570 & 0.006 & 0.003 & 474 & 50 \\
\noalign{\smallskip}
\hline 
%
%
%
\noalign{\bigskip}
\hline
\hline
\noalign{\smallskip}
\multicolumn{6}{l}{HD92320}\\
\noalign{\smallskip}
\hline
\noalign{\smallskip}
BJD     & RV          & $\sigma_{RV}$ & BVS\tablefootmark{a}         & Exp. time & S/N/pix    \\
-2 400 000  & (\kms) & (\kms) & (\kms) & (s)       & (at 550 nm)\\
\hline
\noalign{\smallskip}
54090.6482 & 1.911 & 0.004 & -0.022 & 300 & 66 \\
54093.7075 & 1.542 & 0.004 & -0.016 & 300 & 71 \\
54096.6818 & 1.174 & 0.004 & -0.015 & 300 & 64 \\
54097.6492 & 1.071 & 0.005 & -0.013 & 300 & 52 \\
54584.4482 & -1.845 & 0.005 & -0.014 & 180 & 44 \\
54601.3449 & -1.957 & 0.005 & -0.024 & 885 & 51 \\
54642.3592 & 1.363 & 0.005 & -0.011 & 309 & 50 \\
54645.3648 & 1.939 & 0.005 & -0.028 & 295 & 47 \\
54889.4636 & -1.933 & 0.005 & -0.007 & 638 & 46 \\
54926.4396 & 0.218 & 0.005 & -0.020 & 192 & 50 \\
54927.5196 & 0.387 & 0.005 & -0.011 & 236 & 49 \\
55266.5205 & 0.467 & 0.005 & 0.003 & 288 & 47 \\
55305.3537 & -1.718 & 0.005 & -0.017 & 257 & 51 \\
55519.7222 & 2.247 & 0.005 & -0.009 & 185 & 48 \\
55525.6987 & 3.046 & 0.005 & -0.016 & 365 & 47 \\
55527.7043 & 3.184 & 0.005 & -0.016 & 270 & 45 \\
\noalign{\smallskip}
\hline \pagebreak
\noalign{\bigskip}
\hline
\hline
\noalign{\smallskip}
\multicolumn{6}{l}{HD137510}\\
\noalign{\smallskip}
\hline
\noalign{\smallskip}
BJD     & RV          & $\sigma_{RV}$ & BVS\tablefootmark{a}         & Exp. time & S/N/pix    \\
-2 400 000  & (\kms) & (\kms) & (\kms) & (s)       & (at 550 nm)\\
\hline
\noalign{\smallskip}
54549.6357 & -6.711 & 0.006 & -0.019 & 245 & 57 \\
54553.5731 & -6.707 & 0.006 & -0.003 & 128 & 56 \\
54554.5767 & -6.709 & 0.004 & -0.015 & 180 & 100 \\
54873.6698 & -6.023 & 0.005 & 0.008 & 192 & 79 \\
55051.3403 & -6.285 & 0.007 & -0.046 & 41 & 50 \\
55051.3426 & -6.267 & 0.005 & -0.009 & 181 & 80 \\
55240.7203 & -6.753 & 0.007 & -0.035 & 35 & 48 \\
55266.7010 & -6.754 & 0.007 & -0.003 & 65 & 47 \\
55268.6309 & -6.777 & 0.007 & 0.012 & 37 & 49 \\
55288.5197 & -6.766 & 0.004 & -0.013 & 194 & 102 \\
55305.3824 & -6.756 & 0.005 & -0.004 & 180 & 83 \\
55391.3744 & -6.699 & 0.006 & -0.012 & 35 & 53 \\
55436.3653 & -6.637 & 0.006 & -0.025 & 180 & 61 \\
\noalign{\smallskip}
\hline
%
%
%
\noalign{\bigskip}
\hline
\hline
\noalign{\smallskip}
\multicolumn{6}{l}{HD156279}\\
\noalign{\smallskip}
\hline
\noalign{\smallskip}
BJD     & RV          & $\sigma_{RV}$ & BVS\tablefootmark{a}         & Exp. time & S/N/pix    \\
-2 400 000  & (\kms) & (\kms) & (\kms) & (s)       & (at 550 nm)\\
\hline
\noalign{\smallskip}
55330.5522 & -20.627 & 0.004 & -0.036 & 327 & 48 \\
55361.4611 & -20.881 & 0.004 & -0.031 & 148 & 50 \\
55398.3821 & -20.163 & 0.003 & -0.037 & 377 & 90 \\
55405.3572 & -20.142 & 0.004 & -0.032 & 191 & 50 \\
55424.3514 & -20.370 & 0.004 & -0.035 & 114 & 51 \\
55425.3640 & -20.360 & 0.004 & -0.017 & 250 & 48 \\
55427.3726 & -20.376 & 0.004 & -0.028 & 176 & 48 \\
55441.4021 & -20.502 & 0.004 & -0.035 & 173 & 51 \\
55449.2932 & -20.541 & 0.004 & -0.029 & 193 & 48 \\
55479.2757 & -20.766 & 0.005 & -0.042 & 180 & 45 \\
55503.2440 & -20.999 & 0.004 & -0.042 & 182 & 49 \\
55513.2168 & -21.137 & 0.004 & -0.035 & 221 & 65 \\
55525.2393 & -20.823 & 0.005 & -0.042 & 502 & 45 \\
55540.2372 & -20.210 & 0.005 & -0.025 & 806 & 48 \\
55584.7229 & -20.592 & 0.004 & -0.031 & 180 & 54 \\
\noalign{\smallskip}
\hline 
%
%
%
\noalign{\bigskip}
\hline
\hline
\noalign{\smallskip}
\multicolumn{6}{l}{HD167215}\\
\noalign{\smallskip}
\hline
\noalign{\smallskip}
BJD     & RV          & $\sigma_{RV}$ & BVS\tablefootmark{a}         & Exp. time & S/N/pix    \\
-2 400 000  & (\kms) & (\kms) & (\kms) & (s)       & (at 550 nm)\\
\hline
\noalign{\smallskip}
55042.3613 & -44.609 & 0.008 & 0.075 & 207 & 49 \\
55051.3730 & -44.845 & 0.008 & 0.068 & 198 & 48 \\
55054.3550 & -44.932 & 0.008 & 0.028 & 181 & 50 \\
55067.4087 & -45.354 & 0.008 & 0.021 & 195 & 48 \\
55331.6227 & -45.066 & 0.008 & 0.047 & 209 & 46 \\
55333.5844 & -45.022 & 0.008 & 0.055 & 408 & 45 \\
55366.5956 & -44.676 & 0.010 & 0.051 & 192 & 39 \\
55378.5849 & -44.582 & 0.008 & 0.039 & 198 & 49 \\
55409.4168 & -44.340 & 0.008 & 0.051 & 183 & 49 \\
55423.3895 & -44.211 & 0.008 & 0.040 & 425 & 45 \\
55425.3973 & -44.200 & 0.008 & 0.065 & 161 & 46 \\
55429.3835 & -44.176 & 0.008 & 0.022 & 114 & 49 \\
55448.3524 & -44.062 & 0.008 & 0.003 & 209 & 47 \\
55479.3607 & -43.890 & 0.010 & 0.013 & 225 & 44 \\
\noalign{\smallskip}
\hline
\end{longtable}
}

\end{document}